\documentclass[a4paper,11pt]{article}
\pdfoutput=1
\usepackage{jheppub}

\usepackage{amssymb,amsmath}
\usepackage[normalem]{ulem}
\usepackage[utf8x]{inputenc}
\usepackage{slashed}
\usepackage{graphicx}
\usepackage{here}

\usepackage{csquotes} 
\usepackage{comment}
\usepackage{mathrsfs}
\usepackage{float}
\usepackage{ascmac}

\usepackage{multirow}
\usepackage{longtable}

\usepackage[hang,small,bf]{caption}
\usepackage[subrefformat=parens]{subcaption}
\newcommand{\meV}{\mbox{meV}}
\newcommand{\eV}{\mbox{eV}}
\newcommand{\MeV}{\mbox{MeV}}

\newcommand{\GeV}{\mbox{GeV}}

\def\lsim{\mathrel{\rlap{\lower4pt\hbox{\hskip1pt$\sim$}}
    \raise1pt\hbox{$<$}}}         
\def\gsim{\mathrel{\rlap{\lower4pt\hbox{\hskip1pt$\sim$}}
    \raise1pt\hbox{$>$}}}         

\numberwithin{equation}{section}

\preprint{
\begin{minipage}{5cm}
\small
\flushright
KYUSHU-HET-296
\end{minipage}} 

\title{Reinforcement learning-based statistical search strategy for an axion model from flavor}

\author{Satsuki Nishimura,} 
\author{Coh Miyao, and}
\author{Hajime Otsuka} 
\affiliation{
Department of Physics, Kyushu University, 744 Motooka, Nishi-ku, Fukuoka 819-0395, Japan\\
}
\emailAdd{nishimura.satsuki@phys.kyushu-u.ac.jp}
\emailAdd{miyao.coh@phys.kyushu-u.ac.jp}
\emailAdd{otsuka.hajime@phys.kyushu-u.ac.jp}

\abstract{
We propose a reinforcement learning-based search strategy to explore new physics beyond the Standard Model. 
The reinforcement learning, which is one of machine learning methods, is a powerful approach to find model parameters with phenomenological constraints. 
As a concrete example, we focus on a minimal axion model with a global $U(1)$ flavor symmetry. 
Agents of the learning succeed in finding $U(1)$ charge assignments of quarks and leptons solving the flavor and cosmological puzzles in the Standard Model, and find more than 150 realistic solutions for the quark sector taking renormalization effects into account. 
For the solutions found by the reinforcement learning-based analysis, we discuss the sensitivity of future experiments for the detection of an axion which is a Nambu-Goldstone boson of the spontaneously broken $U(1)$. 
We also examine how fast the reinforcement learning-based searching method finds the best discrete parameters in comparison with conventional optimization methods. 
In conclusion, the efficient parameter search based on the reinforcement learning-based strategy enables us to perform a statistical analysis of the vast parameter space associated with the axion model from flavor. 
}
\makeatletter
\gdef\@fpheader{}
\makeatother

\begin{document}

\maketitle

\section{Introduction}

The machine learning-based search strategy is of particular attention not only for collider physics but also for particle phenomenology. 
So far, the quantitative evaluation of physical observables has relied on various optimization methods to search for desirable parameters in a model and to predict new phenomena at high-energy scales. 
However, this analysis often requires computational costs due to the need to explore vast parameter spaces in the model. 
Therefore, the machine learning-based analysis is anticipated to a powerful approach for investigating physics beyond the Standard Model and reducing the computational costs. 
Among the machine learning techniques, we focus on the reinforcement learning (RL). 
In contrast to supervised learning and unsupervised learning, the RL enables us to autonomously discover solutions using a limited amount of data. 

\medskip

In this paper, we propose a new approach to search for physics beyond the Standard Model, i.e., a RL-based search strategy. 
In particular, we apply the RL-based search strategy to the minimal axion model with a global $U(1)$ flavor symmetry proposed in Refs. \cite{Ema:2016ops,Calibbi:2016hwq}\footnote{See, Refs. \cite{Davidson:1981zd,Wilczek:1982rv} for earlier works.}, as a fascinating example of solving the flavor and cosmological puzzles in the Standard Model, i.e., mass hierarchies and mixing angles of quarks and leptons, an origin of dark matter, and the inflation. 
In this model, the flavor structure of quarks and leptons can be explained by using the Froggatt-Nielsen (FN) mechanism \cite{Froggatt:1978nt}. 
Since the $U(1)$ flavor symmetry is identified with the Peccei-Quinn (PQ) symmetry, the QCD axion originated from the Nambu-Goldstone (NG) boson of the spontaneously broken $U(1)$, so-called a flaxion field, can solve the strong CP problem via the PQ mechanism \cite{PhysRevLett.38.1440,PhysRevD.16.1791}. 
Then, a proper value of the flaxion decay constant also explains the dark matter abundance in our universe. 
Furthermore, the successful inflation mechanism can be realized by the radial component of the $U(1)$ breaking field with the proper Lagrangian. 

\medskip

So far, the flavor structure of quarks and leptons was studied using the RL-based method in Ref. \cite{Nishimura:2023nre}. 
In particular, agents of RL succeeded in finding the best solutions to be consistent with the mass hierarchies and the mixing angles of quarks and leptons. 
This work motivates us to apply the RL-based strategy to explore the cosmological aspects of the minimal axion model with the global $U(1)$ flavor symmetry. 
Since couplings between the flaxion and quarks/leptons are determined by their $U(1)$ charges, 
the flaxion-photon coupling as well as the flaxion decay constant are sensitive to the assignment of these charges. 
We utilize the RL-based strategy to find the best $U(1)$ charges of quarks/leptons solving the flavor and cosmological puzzles in the Standard Model. 
By comparing traditional optimization methods with the RL-based search strategy, 
we argue that the RL-based approach for discrete parameters is more efficient than the conventional techniques. 
Since the predicted flavor structure varies significantly with different $U(1)$ charge assignments, the evaluation function of the model exhibits rapid fluctuations in response to changes in discrete parameters. 
In such a situation, the traditional numerical optimization with the gradient descent method is difficult to apply to this system, and the RL-strategy for discrete parameters has an advantage over the conventional methods.

\medskip

This paper is organized as follows. 
After briefly reviewing the flaxion model and constraints from flavor physics and cosmology in Sec. \ref{sec:FNaxion}, 
we organize the RL-based search strategy in Sec. \ref{sec:RL}. 
In Sec. \ref{sec:comparison}, we verify how fast the RL-based method finds the proper $U(1)$ charge assignments solving the problems in the Standard Model. 
In Sec. \ref{sec:flaxion_physics}, we show training results and discuss the phenomenology of the flaxion with an emphasis on the flaxion-photon coupling which will be explored in the future experiments such as DMRadio-$\mathrm{m}^3$ \cite{DMRadio:2022pkf}. 
Sec. \ref{sec:con} is devoted to conclusion and discussion. 
In Appendix \ref{app:benchmark}, we list our finding $U(1)$ charge assignments of the leptons.

\section{Flaxion model and constraints}
\label{sec:FNaxion}

In this section, we introduce a flaxion model proposed in Refs. \cite{Ema:2016ops,Calibbi:2016hwq}, which explains not only the flavor structure of quarks and leptons through the FN mechanism \cite{Froggatt:1978nt} but also the strong CP problem through the PQ mechanism \cite{PhysRevLett.38.1440,PhysRevD.16.1791}.
Moreover, it provides a good candidate for the dark matter \cite{Preskill:1982cy,Abbott:1982af,Dine:1982ah}.  

\subsection{Model}
\label{sec:model}

The model is given by the Yukawa terms as follows:
\begin{align}
    -{\cal L} \supset~ &y^u_{ij}\left(\frac{\phi}{M}\right)^{n^u_{ij}}\Bar{Q}_i H^c u_j
    + y^d_{ij}\left(\frac{\phi}{M}\right)^{n^d_{ij}}\Bar{Q}_i H d_j
    + y^l_{ij}\left(\frac{\phi}{M}\right)^{n^l_{ij}}\Bar{L}_i H l_j
    \nonumber\\
    &+ y^\nu_{ij}\left(\frac{\phi}{M}\right)^{n^\nu_{ij}}\Bar{L}_i H^c N_j
    + \frac{y^N_{ij}}{2}\left(\frac{\phi}{M}\right)^{n^N_{ij}}M\Bar{N}^{c}_i N_j
    + \text{h.c.},
\label{eq:Lagrangian}
\end{align}
where $\{Q_{i}, u_{i}, d_{i}, L_{i}, l_{i}, N_{i}, H\}$ denote the left-handed quarks, 
the right-handed up-type quarks, the right-handed down-type quarks, the left-handed leptons, 
the right-handed charged leptons, the right-handed neutrinos, and the Standard Model (SM) Higgs doublet with $H^c = i\sigma_2 H^\ast$, respectively. 
Here, a complex scalar field $\phi$, which is called a flavon field, is introduced.
We assume that a mass of the right-handed neutrinos $M$ corresponds to a cut-off scale of the model to explain the light neutrino masses through the seesaw mechanism \cite{Minkowski:1977sc,Yanagida:1979as,Gell-Mann:1979vob,Mohapatra:1979ia}. 
The Yukawa couplings $\{y^u_{ij},y^d_{ij},y^l_{ij},y^\nu_{ij},y^N_{ij}\}$ are ${\cal O}(1)$ real constants.

\medskip

We denote $U(1)$ charges of the SM fields and the flavon field by
\begin{align}
    \{q(Q_i),\,q(u_i),\,q(d_i),\,q(L_i),\,q(l_i),\,q(N_i),\,q(H),\,q(\phi)\}. \label{eq:sets_QL}
\end{align}
Since this study focuses on the charge of the quark sector primarily, the lepton sector will be omitted in the following:
\begin{align}
    \{q(Q_i),\,q(u_i),\,q(d_i),\,q(H),\,q(\phi)\}. \label{eq:sets_Q}
\end{align}
The reason for this omission is explained in Sec. \ref{sec:RL}. 
The $U(1)$ symmetry makes the indices $n_{ij}$ satisfy the following relations:
\begin{align}
    n^u_{ij} &= -\frac{q(\Bar{Q}_iH^c u_j)}{q(\phi)} = -\frac{-q(Q_i)-q(H)+q(u_j)}{q(\phi)}, \label{eq:nij_up}
    \\
    n^d_{ij} &= -\frac{q(\Bar{Q}_i H d_j)}{q(\phi)} = -\frac{-q(Q_i) + q(H) + q(d_j)}{q(\phi)}. \label{eq:nij_down}
\end{align}
In this study, $n_{ij}$ is restricted to an integer to simplify its interpretation within a field theory.

\medskip

When $\phi$ and $H$ develop vacuum expectation values (VEVs), $ \langle \phi \rangle = v_\phi$ and $\langle H\rangle = v_{\rm EW}=174\,$GeV, 
the mass matrices are given by
\begin{align}
    m_{ij}^u &= y^u_{ij} \eta^{n^u_{ij}} v_{\rm EW},\\
    m_{ij}^d &= y^d_{ij} \eta^{n^d_{ij}} v_{\rm EW},
\end{align}
with $\eta=v_{\phi}/M$.
The masses of the quarks are decided from diagonalized components as
\begin{align}
    m^u &= U^u {\rm diag}(m^u) V^{u\dagger},\\
    m^d &= U^d {\rm diag}(m^d) V^{d\dagger},
\end{align}
and the flavor mixing is defined as the difference between mass eigenstates and flavor eigenstates:
\small
\begin{align}
\begin{split}
    V_{\rm CKM} &= U^{u\dagger} U^d
    \\
    &= 
    \begin{pmatrix} c_{12} c_{13} & s_{12} c_{13} & s_{13} e^{-i \delta_{\rm CP}} \\
-s_{12} c_{23} - c_{12} s_{23} s_{13} e^{i \delta_{\text{CP}}} & c_{12} c_{23} - s_{12} s_{23} s_{13} e^{i \delta_{\text{CP}}} & s_{23} c_{13} \\
s_{12} s_{23} - c_{12} c_{23} s_{13} e^{i \delta_{\text{CP}}} & -c_{12} s_{23} - s_{12} c_{23} s_{13} e^{i \delta_{\text{CP}}} & c_{23} c_{13} 
\end{pmatrix},
\end{split}
\end{align}
\normalsize
with $c_{ij}= \cos \theta_{ij}$ and $s_{ij} = \sin \theta_{ij}$. 
Thus, the smallness of $|\eta|$ with the FN mechanism makes the hierarchical structure of flavor physics, and a similar explanation holds for the lepton sector.

\medskip

When the flavon and Higgs are expanded as 
\begin{align}
    \phi = v_{\phi} + \frac{1}{\sqrt{2}}\left(s+ia\right),
    \qquad H=\begin{pmatrix}
        0 \\
        v_{\rm EW} + \frac{h}{\sqrt 2}
	\end{pmatrix},
\end{align}
the Lagrangian for the quarks and the charged leptons is written in the following form:
\begin{align}
	\mathcal{L}_{f} = \sum_{f=u,d,l}
	\left[m_{ij}^f \left(1+\frac{h}{\sqrt{2} v_{\rm EW}}\right)
	+ \frac{m_{ij}^f n_{ij}^f (s+ia)}{\sqrt{2} v_\phi} \right] \overline {f_{Li}} f_{Rj} + {\rm h.c.}.
\end{align}
Now the Higgs Yukawa interactions can be diagonalized through rotations with unitary matrices $U^f$ and $V^f$:
\begin{align}
	f_{R_j} \equiv V^{f\dagger}_{ji} f_{R_i}',
	\qquad f_{L_i} \equiv U^{f\dagger}_{ij} f_{L_j}'.
\end{align}
Here, the fields with prime sign denote in the mass basis. 
In this basis, although the mass term is diagonalized, the Yukawa couplings relating to the flavon field are not diagonalized as follows:
\begin{align}
  \mathcal{L}_{f} = \sum_{f=u,d,l}
  \left[
    m_{i}^f \left( 1 + \frac{h}{\sqrt{2} v_{\rm EW}} \right) 
    \overline {f'_{Li}} f'_{Ri} 
    +\kappa^f_{ij} \frac{s+ia}{\sqrt{2} v_\phi}\,  
    \overline {f'_{Li}} f'_{Rj}
  \right]
  + {\rm h.c.}.
\end{align}
Here, the matrix $\kappa^f_{ij}$ is given by
\begin{align}
	\kappa^f_{ij} \equiv U^f_{ik} (m_{kn}^f n_{kn}^f) V^{f\dagger}_{nj}.
\end{align}
In particular, the off-diagonal interactions of the pseudo-scalar field $a$ can be written as
\begin{align}
    \mathcal{L}_{a} = \frac{ia}{\sqrt 2 v_\phi} \sum_{f=u',d',l'}\left[ \left(\kappa^f_{\rm H}\right)_{ij}  \overline f_i \gamma_5 f_j 
    + \left(\kappa^f_{\rm AH}\right)_{ij} \overline f_i f_j  \right], \label{eq:interaction_aff}
\end{align}
with $\kappa^f_{\rm H} = (\kappa^f + \kappa^{f\dagger})/2$ and $\kappa^f_{\rm AH} = (\kappa^f - \kappa^{f\dagger})/2$, which mean Hermitian part and anti-Hermitian part of $\kappa^f$ respectively. 
After diagonalizing the quark masses, we found useful formulas for $\kappa_H^f$ and $\kappa_{AH}^f$ as following forms:
\begin{align}
    (\kappa_H^f)_{ij} &= \frac{1}{2} ({V^f}^\dagger \hat{q}_Q V^f - {U^f}^\dagger \hat{q}_f U^f)_{ij} (m_j^f + m_i^f), \\
    (\kappa_{AH}^f)_{ij} &= \frac{1}{2} ({V^f}^\dagger \hat{q}_Q V^f + {U^f}^\dagger \hat{q}_f U^f)_{ij} (m_j^f - m_i^f), \label{eq:kappa_AH}
\end{align}
where $f=u,d$ and $(\hat{q}_X)_{ij} = q_{X_i}\delta_{ij}$. 

\medskip

The pseudo-scalar $a$ can play the same role as the QCD axion, so it is called the flaxion.
The effective Lagrangian of flaxion-gluon-gluon interaction is given as follows:
\begin{align}
    \mathcal{L}_{\mathrm{eff}}
    = \frac{g_s^2}{32\pi^2} \frac{a}{f_a} G_{\mu\nu}^a \widetilde G^{\mu\nu a}.   \label{eq:interaction_aGG}
\end{align}
Here, $g_{s}$ is a coupling constant of strong interaction and $f_{a}$ is a decay constant of the flaxion given by
\begin{align}
  f_{a} \equiv \frac{\sqrt{2}|v_\phi|}{N_{\mathrm{DW}}}
  = \frac{\sqrt{2}|\eta| M}{N_{\mathrm{DW}}}\, ,
\label{eq:fa}
\end{align}
with a domain wall number
\begin{align}
    N_{\mathrm{DW}} 
    = {\rm Tr} \left[ n^u +  n^d\right]\, .
\end{align}
When the flaxion works as the QCD axion, a flaxion mass is related to the PQ scale as
\begin{align}
m_a \simeq 6\times 10^{-6}\,\eV \left( \frac{10^{12}\,\GeV}{f_a}\right).
\end{align}

\medskip

For the phenomenological study, it is also useful to consider an effective coupling between the flaxion and the photon.
The effective Lagrangian of the interaction is given by
\begin{align}
    \mathcal{L} =  \frac{g_{a\gamma}}{4} a F_{\mu \nu} \tilde{F}^{\mu \nu},
    \label{eq:int_agg}
\end{align}
where $g_{a\gamma}$ is given in Ref. \cite{KIM19871} as follows:
\begin{align}
    g_{a\gamma} = \frac{8 \pi^2 f_a}{e^2}\left[ \frac{2}{N_{\rm DW}}\sum_{f=u,d,l}\left[ N_f {\rm Tr}(n^f) \left( q^{(\rm em)}_f \right)^2 \right] - \frac{2(4+z)}{3(1+z)}\right],
\end{align}
with $z=m_{u}/m_{d}$. 
Here, $q^{(\rm em)}_f$ is the electromagnetic charge of the quarks and the leptons. 
In addition, $N_f$ is 3 for the quarks and 1 for the leptons.

\medskip

In general, $N_{\mathrm{DW}}$ corresponds to the number of minima of a potential.
Based on the discussion of the QCD instanton, the strong CP phase is canceled at a minimum of the flaxion potential. 
Thus, the flaxion $a$ solves the strong CP problem. 

\medskip

Incidentally, domain walls are two-dimensional phase defects that arise from symmetry breaking in the early universe, and the number $N_{\mathrm{DW}}$ governs whether the domain walls are stable or not.
When $N_{\mathrm{DW}}=1$, the domain wall is unstable because it undergoes cleavage and contraction on its own. 
Numerical simulations have confirmed that such domain wall disappears quickly after its formation \cite{Hiramatsu:2012gg}. 
On the other hand, when $N_{\mathrm{DW}}>1$, the domain walls are stable and their energy density decays very slowly. 
This means that the dominant energy of the universe is composed of the domain walls even at present. 
However, that scenario violates uniform isotropy and contradicts cosmological observations, so it is called the domain wall problem.
In the case of $N_{\mathrm{DW}}>1$, the problem is avoided by considering an additional term in the Lagrangian and breaking the symmetry before the inflation, or by adding a new scalar field during the inflation \cite{Ibe:2019yew}. 

\subsection{Constraints from flavor physics}

The flaxion $a$ has flavor-changing neutral current interactions with the quarks and the leptons. 
The strictest bound on $f_a$ is given by $K^+\to\pi^+ a$ process. 
In the previous work \cite{Ema:2016ops}, a decay rate is calculated as follows:
\begin{align}
  \Gamma(K^+\to\pi^+ a) =
  \frac{m_K^3}{32\pi v_\phi^2}
  \left(1-\frac{m_\pi^2}{m_K^2} \right)^3 
  \left|\frac{(\kappa^d_{\rm AH})_{12}}{m_s-m_d} \right|^2.
\end{align}
By using $\left(m_{K^{+}}, m_{\pi}, m_{d}, m_{s}\right)\sim \left(493.677, 139.570, 4.67, 93.4\right)\MeV$, the rate gives following braching ratio:
\begin{align}
  {\rm Br}(K^+\to \pi^+ a) \simeq 
  \frac{3.5\times 10^{13}\,\GeV}{f_{a}^{2} N_{\mathrm{DW}}^{2}} 
  \left|\frac{(\kappa^d_{\rm AH})_{12}}{m_s-m_d} \right|^2.
\end{align}
Comparing with a current experimental bound from NA62 collaboration \cite{NA62:2020pwi,NA62:2021zjw}, Br$(K^+\to \pi^+ a)
\lesssim (3-6)\times 10^{-11}$ at $90\%$ C.L., a lower bound on $f_a$ is given by
\begin{align}
	f_a \gtrsim 2.5\times 10^{10}\,\GeV \frac{30}{N_{\mathrm{DW}}}\left|\frac{(\kappa^d_{\rm AH})_{12}}{m_s-m_d}\right|.
\end{align}
This bound can be evaluated by determining appropriate parameters of the FN model. 
The specific ranges of $\eta, M$ and $q$ to be examined in this study are described later, but in that domain, this inequality is almost automatically satisfied.

\subsection{Constraints from cosmology}
\label{sec:Cosmology}

The flaxion is related to some cosmological topics (e.x., dark matter, isocurvature perturbation, and inflation).
Here, we review constraints from these topics.

\paragraph{Dark Matter}

When the flaxion is regarded as the dark matter, the density parameter of the flaxion is given in Ref. \cite{PhysRevD.33.889} as
\begin{align}
    \Omega_a h^2 = 0.18\times \theta_i^2\ \left( \frac{f_a}{10^{12}\,\GeV}\right)^{1.19},
    \label{eq:DM_axion}
\end{align}
where $\Omega_a$ is the energy density parameter of flaxion, $h$ is the dimensionless Hubble constant and $\theta_i$ is the misalignment angle, respectively.
By fixing $\Omega_a h^2 = 0.12$, we arrive at a relation between the misalignment angle and the PQ scale as
\begin{align}
    \theta_{i} = 0.82 \times \left( \frac{10^{12}\,\GeV}{f_a}\right)^{0.595}.
    \label{eq:alignment-angle}
\end{align}
From this relation, the misalignment angle $\theta_{i}$ is determined by searching results of RL through Eq. \eqref{eq:fa}.

\paragraph{Isocurvature Perturbation}

The Planck 2018 result \cite{Planck:2018jri} gives the latest constraint on a power spectrum of the isocurvature perturbation.
This constraint provides an upper bound of the inflationary scale $H_{\rm inf}$ as
\begin{align}
    H_{\rm inf} \lesssim 3\times 10^{7}\,\GeV\ \theta_i^{-1} \left( \frac{10^{12}\,\GeV}{f_a}\right)^{0.19}.
    \label{eq:inf-scale_iso}
\end{align}
As discussed later, the misalignment angle is fixed through Eq. \eqref{eq:alignment-angle} for each model found by RL, so the upper bound of $H_{\rm inf}$ is given for each of the models.

\paragraph{Inflation}

Let us identify the flavon field as an inflaton field. 
Following Ref. \cite{Ema:2016ops}, we adopt the following Lagrangian for the flavon field:
\begin{align}
    \mathcal{L} = -\frac{|\partial \phi|^2}{\left( 1-\frac{|\phi|^2}{\Lambda^2}\right)} - (|\phi|^2-v_\phi^2)^2,
\end{align}
leading to the successful slow-roll inflation with the inflaton $\varphi=\sqrt{2}{\rm Re}(\phi)$. 
It is consistent with the Planck data \cite{Planck:2018jri}. 
Note that when the inflaton $\varphi$ develops a large field value during the inflation $\varphi \gg v_{\phi}$, there exists the domain wall problem after the QCD phase transition. 
Since the inflaton oscillates around the origin during the reheating epoch, it leads to the parametric resonant enhancement of the flaxion and the symmetry is restored \cite{Kofman:1995fi}. 
However, such a problem does not occur for the small-field inflation in which the inflation starts around the origin.
In our inflation model, 
the non-trivial kinetic term of the flavon field makes the potential flatten at the large field value. 
Furthermore, if $v_{\phi} < \Lambda < \sqrt{2}v_{\phi}$, the flavon does not reach the origin after the inflation, and the symmetry is not restored, i.e., there is no domain wall problem as discussed in Ref. \cite{Ema:2016ops}. 

\medskip

The inflationary scale is given by
\begin{align}
    H_{\rm inf} &\simeq 5\times 10^{8}\,\GeV \left( \frac{\Lambda}{10^{14}\,\GeV} \right) \nonumber \\
    &\gtrsim 5\times 10^{7}\,\GeV.
    \label{eq:inf-scale}
\end{align}
In the last line, we apply the lower bound of the inflationary scale as $\Lambda\gtrsim 10^{13}\,\GeV$.
In summary, if the value of the upper bound Eq. \eqref{eq:inf-scale_iso} for a certain model is less than the lower bound Eq. \eqref{eq:inf-scale}, that model is excluded.

\section{Reinforcement learning}
\label{sec:RL}

In RL, the subject of learning is called an agent, and the problem to be solved is called an environment. 
The learning process is constructed from three stages: observing the environment, choosing an action, and getting rewards.
By repeating these procedures, it is known that the agent autonomously acquires a principle of actions that maximizes the sum of rewards. Details of RL and Q-learning, which are basic ingredients of RL, are explained in Ref. \cite{RL}. 
In this paper, we adopt Deep Q-Network (DQN) as a RL algorithm, and use the package ``gym'' developed by OpenAI.
Since the architecture in this work is similar to our previous work \cite{Nishimura:2023nre}, 
we briefly review the RL, and setups of the learning will be explained by focusing on the key points. 
For more details about the architecture, see Ref. \cite{Nishimura:2023nre}.

\medskip

The flavor structure of quarks and leptons is determined by the charge shown in Eq. \eqref{eq:sets_QL}.
In Ref. \cite{Nishimura:2023nre}, many realistic models of the lepton sector are found in contrast to the quark sector. 
In other words, there are many varieties of the $U(1)$ charges of leptons leading to successful mass hierarchies and the flavor structure. 
On the other hand, viable solutions for the quark sector have proven difficult to obtain, even in the absence of renormalization effects. 
This fact raises concerns that the existence of renormalization effects will obstruct the agents from reaching valid solutions and make the learning of RL unstable. 
Moreover, the benchmarks established in the previous work, which neglected renormalization effects, are not directly applicable to the quark sector in a precise sense. 
Consequently, those solutions require modification to account for the influence of radiative corrections.
Hence, at the moment, we will perform RL on the quark sector to clarify the effectiveness of RL and to compare the different energy scales.

\subsection{The environment}
\label{sec:env}

We impose two constraints on the environment. First, the indices $n_{ij}$ are considered positive integers. Second, to ensure the value of top quark mass, the Yukawa term $\Bar{Q}_3H^c u_3$ is irrelevant to $q(\phi)$:
\begin{align}
    q(\Bar{Q}_3H^c u_3) = 0 
    \Leftrightarrow
    q(H) = q(u_3) - q(Q_3),
\label{eq:Higgscharge}
\end{align}
from which the $U(1)$ charge of the Higgs is specified by the quarks. 
Based on this equation, the flavor structure of the quarks is specified by the following charge vector:
\begin{align}
    {\cal Q}_a := \{q(Q_i),\,q(u_i),\,q(d_i),\,q(\phi)\},
\end{align}
with $i=1,2,3$, so the vector has 10 elements. 
This is the input data for neural networks which decide an action of the agent. 
To avoid a hierarchy of $U(1)$ charges, we take the value of each charge within:
\begin{align}
    -9 \leq {\cal Q}_a \leq 9,
\end{align}
corresponding to total $19^{10}\,\sim\,10^{12}$ possibilities for only the quark sector.
In a general $U(1)$ charge pair, non-integers $n_{ij}$ can be realized. 
To avoid this situation, we initialize the set of charges as $q(\phi)=+1$ with 50\% probability and $q(\phi)=-1$ with 50\% probability.
This makes it easier to generate pairs of charges such that the denominator of Eq. \eqref{eq:nij_up} and Eq. \eqref{eq:nij_down} is $\pm 1$, so integers $n_{ij}$ are preferentially realized.

\subsection{Neural Network}

We utilize neural networks to determine an action of the agent. 
Their architecture is summarized in Table \ref{tab:network}. 
The activation function is chosen as the SELU function for hidden layers and the softmax function for an output layer. 
We employ the ADAM optimizer in TensorFlow \cite{DBLP:journals/corr/AbadiABBCCCDDDG16}, and the Huber function is used as a loss function:
\begin{align}
    \begin{split}
        L_{\rm Huber}(x, y) = 
        \left\{
        \begin{array}{l}
             \frac{1}{2} (x_i - y_i)^2\qquad \qquad\,\,\,\, \text{if}\ |x_i - y_i| \leq \delta   \\
             \delta \cdot |x_i - y_i| - \frac{1}{2}\delta^2\qquad \text{if}\ |x_i - y_i| > \delta
        \end{array}
        \right.
        ,
    \end{split}
    \label{eq:Huber}
\end{align}
with $\delta = 1$.
This function combining the mean squared error and the mean absolute error will lead to the efficient learning by taking advantage of the features of both.

\begin{table}[H]
    \centering
    \begin{tabular}{|c||c|c|c|c|c|}\hline
       layer  &  Input & Hidden 1 & Hidden 2 & Hidden 3 & Output\\
       \hline
       Dimension  & $\mathbb{Z}^{10}$ & $\mathbb{R}^{64}$ & $\mathbb{R}^{64}$ & $\mathbb{R}^{64}$ & $\mathbb{R}^{20}$
       \\
       \hline
    \end{tabular}
    \caption{In the neural networks, the input is the charge assignment ${\cal Q}_a$. 
    The activation functions are the SELU function for the hidden layers $\{1, 2, 3\}$, and the softmax function for the output layer.}
    \label{tab:network}
\end{table}

\subsection{Agent}
\label{sec:agent}

The action $\mathfrak{a}$ of the agent is realized in the following way at each step:
\begin{align}
    \mathfrak{a}\,:\, {\cal Q}_a \rightarrow {\cal Q}_a \pm 1\,\,(a \in A),
    \label{eq:action_agent}
\end{align}
where $A$ corresponds to $\{Q_i,u_i,d_i,\phi\}$. 
At an initial stage of the learning, for each environment, the ${\cal O}(1)$ coefficients in Yukawa terms Eq. \eqref{eq:Lagrangian} are picked up from the two Gaussian distribution with an average $\pm 1$ and a standard deviation 0.25. 
After the training with the neural networks introduced in the previous subsection, the coefficients are optimized to proper values by the Monte-Carlo simulation. 
Note that the coefficient $y^{u}_{33}$ was fixed at 0.4 in both the initialization and the optimization to ensure that the mass of the top quark is reproduced.

\medskip

The agent chooses an action based on $\varepsilon$-greedy method. 
In this method, $\mathfrak{a}$ is determined by the neural network with probability $1-\varepsilon$ and randomly with probability $\varepsilon$ as follows:
\begin{align}
    \mathfrak{a} = 
    \left\{
    \begin{array}{l}
         \mathfrak{b}\quad (\text{with}\,1-\varepsilon)  \\
         \mathfrak{c}\quad (\text{with}\,\varepsilon) 
    \end{array}
    \right.
    ,
\end{align}
where $\mathfrak{b}$ is called a greedy action, and $\mathfrak{c}$ represents a random action.
By repeating choosing an action, a sequence of charges is defined as follows: 
\begin{align}
    {\cal Q}^{\,\left(1\right)} \xrightarrow{\mathfrak{a}_{1}}
    {\cal Q}^{\,\left(2\right)} \xrightarrow{\mathfrak{a}_{2}}
    {\cal Q}^{\,\left(3\right)} \xrightarrow{\mathfrak{a}_{3}} \cdots,
\end{align}
and this chain is called an episode. The initial set of charges ${\cal Q}^{\,\left(1\right)}$ is determined randomly.
The number of actions is specified by $N_{\rm step}$ for one episode and the agent repeats the steps $N_{\rm ep}$ times.
We took $\varepsilon$ as the following form to ensure that the agent gradually takes the greedy action: 
\begin{align}
    \varepsilon = \max \left( \varepsilon_0 r^{k-1},\  \varepsilon_{\rm min}\right),
\end{align}
with $k=1,2,...,N_{\rm ep}$. 
Following the same way as in Ref. \cite{Nishimura:2023nre}, we adopt $\varepsilon_0 = 1,\ r=0.99999$ and $\varepsilon_{\rm min}=0.01$. 
Other hyper-parameters are set as $N_{\rm ep}=10^5$ for an episode number and $N_{\rm step}=32$ for a step number.
Moreover, a batch size is 32, an epoch number is 32, and a learning rate is $\alpha= 2.5\times 10^{-4}$, respectively
\footnote{For simplicity, we use the same hyper-parameters as in the previous study, but the choice of hyper-parameters is also important as a factor that determines the success or failure of the learning. 
There are various optimization methods for hyper-parameters. Then, ``Hyperopt'' \cite{Bergstra_2015} and ``Optuna'' \cite{10.1145/3292500.3330701} are known as libraries that provide tuning methods based on Bayesian optimization. 
In general, a ``black box function'' refers to a function whose functional form cannot be revealed in advance and whose value is determined only after the arguments are given. 
Bayesian optimization and genetic algorithms are effective methods for solving optimization problems for black box functions.}. 

\medskip

Then, to evaluate how well the values calculated from the FN model reproduce reference values such as renormalized masses, 
an intrinsic value is defined as follows:
\begin{align}
    {\cal V}({\cal Q}) = -{\rm min}_{\eta}\bigl[{\cal M}_{\rm quark} + {\cal C} \bigl],
\label{eq:intrinsic_value}
\end{align}
whose components will be defined below. 
The flavon VEV is defined to maximize the intrinsic value, and we search for the VEV within
\begin{align}
    0.01 \leq |\eta| \leq 0.3,\qquad -\pi \leq {\rm arg}(\eta) \leq \pi. \label{eq:eta_range}
\end{align}

\begin{enumerate}
    \item Masses:

${\cal M}_{\rm quark}$ evaluates errors of quark masses:
\begin{align}
    {\cal M}_{\rm quark} = \sum_{\alpha = u,d}E_\alpha,
\end{align}
with
\begin{align}
    E_\alpha = \biggl| \log_{10}\left(\frac{|m_\alpha|}{|m_{\alpha,{\rm RG}}|}\right) \biggl|.
\end{align}
Here, $m_{\alpha}$ is a predicted mass by the agent, and $m_{\alpha,\mathrm{RG}}$ is a renormalized mass. 
$m_{\alpha,\mathrm{RG}}$ at energy scale $M$ is extrapolated from calculated masses in Ref. \cite{Huang:2020hdv}.
Specifically, we consider cases of $M=10^{14},10^{15},10^{16},10^{17}\ \GeV$, and the referred values are shown in Table \ref{tab:QuarkMassRatio_in_HighEnergy}.

\item Mixing angles:

$\mathcal{C}$ evaluates errors of quark mixings:
\begin{align}
    {\cal C} = \sum_{i,j} E_{\cal C}^{ij},
\end{align}
with
\begin{align}
    E_{\cal C}^{ij} = \biggl| \log_{10}\left(\frac{|V_{\rm CKM}^{ij}|}{|V_{\rm CKM,\,exp}^{ij}|}\right) \biggl|.
\end{align}
From Table \ref{tab:data_quark}, the CKM matrix is given by the following form:
\begin{align}
|V_{\mathrm{CKM}}| = \begin{pmatrix}
0.97435 \pm 0.00016 & 0.22501 \pm 0.00068 & 0.003732^{+0.000090}_{-0.000085} \\
0.22487 \pm 0.00068 & 0.97349 \pm 0.00016 & 0.04183^{+0.00079}_{-0.00069} \\
0.00858^{+0.00019}_{-0.00017} & 0.04111^{+0.00077}_{-0.00068} & 0.999118^{+0.000029}_{-0.000034}
\end{pmatrix}.
\end{align}

For the mixing angles, we focused on reproducing the characteristics of weak mixing rather than the exact experimental values. 
Following the previous studies utilizing RL, the ratios of the best-fit values are used to train the agent for the quark sector simply, and the uncertainties associated with the CKM matrix are not taken into account.
In addition, only one flavon field is introduced in the Lagrangian Eq. \eqref{eq:Lagrangian}.
Thus, $\mathcal{C}$ does not include the evaluation of the CP phase.

\end{enumerate}

\begin{table}[H]
\centering
\scalebox{0.9}{
\begin{tabular}{|c|ccc|}
    \hline
    Energy scale & $m_{u,\mathrm{RG}}/\MeV$ & $m_{c,\mathrm{RG}}/\GeV$ & $m_{t,\mathrm{RG}}/\GeV$ \\
    \hline
    $10^{8}\,\GeV$ & $0.69\pm0.12$ & $0.350\pm0.011$ & $102.49\pm0.89$ \\
    $10^{12}\,\GeV$ & $0.56\pm0.10$ & $0.283\pm0.009$ & $85.07\pm0.89$ \\
    $10^{14}\,\GeV$ (ext) & $0.49$ & $0.248$ & $77.10$ \\
    $10^{15}\,\GeV$ (ext) & $0.46$ & $0.235$ & $73.91$ \\
    $10^{16}\,\GeV$ (ext) & $0.44$ & $0.223$ & $70.93$ \\
    $10^{17}\,\GeV$ (ext) & $0.42$ & $0.211$ & $68.13$ \\
    \hline
    \hline
    Energy scale & $m_{d,\mathrm{RG}}/\MeV$ & $m_{s,\mathrm{RG}}/\MeV$ & $m_{b,\mathrm{RG}}/\GeV$ \\
    \hline
    $10^{8}\,\GeV$ & $1.52\pm0.11$ & $30.34\pm2.65$ & $1.502\pm0.018$ \\
    $10^{12}\,\GeV$ & $1.24\pm0.09$ & $24.76\pm2.17$ & $1.194\pm0.015$ \\
    $10^{14}\,\GeV$ (ext) & $1.09$ & $21.76$ & $1.009$ \\
    $10^{15}\,\GeV$ (ext) & $1.03$ & $20.67$ & $0.945$ \\
    $10^{16}\,\GeV$ (ext) & $0.98$ & $19.65$ & $0.886$ \\
    $10^{17}\,\GeV$ (ext) & $0.93$ & $18.68$ & $0.830$ \\
    \hline
\end{tabular}
}
\caption{This shows the renormalized masses and the extrapolated masses (ext) at each energy scale for the quarks. The renormalized masses are evaluated based on Ref. \cite{Huang:2020hdv}.}
\label{tab:QuarkMassRatio_in_HighEnergy}
\end{table}

\begin{table}[H]
\renewcommand{\arraystretch}{1.25}
\centering
\scalebox{0.88}{
   \begin{tabular}{|c|c|c|c|}\hline
          $s_{12}$ & $s_{13}$ & $s_{23}$ & $\delta_{\rm CP}$ \\\hline
          $0.22501 \pm 0.0006$ &
          $0.003732^{+0.000090}_{-0.00008}$ &
          $0.04183^{+0.00079}_{-0.0006}$ & 
          $1.147 \pm 0.02$ \\
         \hline
      \end{tabular}
      }
  \caption{Mixing angles and CP phase in the quark sector \cite{ParticleDataGroup:2024prd}.}
    \label{tab:data_quark}  
\end{table}
     \renewcommand{\arraystretch}{1}

\medskip

A large intrinsic value indicates that an obtained charge assignment well reproduces the renormalized masses and mixings
\footnote{In the present study, the intrinsic value is defined the same as in Ref. \cite{Nishimura:2023nre} to be based on the findings of the previous study.
If the $\chi^2$ value is adopted for intrinsic value, the search can be conducted with more statistical significance.}.  
Such charge assignment is called a {\it terminal state}. 
Specifically, the terminal state is defined to satisfy the following requirements:
\begin{align}
    |{\cal V}({\cal Q})| < V_0,\qquad
    E_\alpha <V_1\quad ({\rm for}\,\forall\alpha),\qquad
    E_{{\cal C}}^{ij}<V_2\quad ({\rm for}\,\forall i,j).
\end{align}
In this paper, we adopt $V_0 = 10.0$, $V_1 = 1.75$, and $V_2 = 0.2$. 
Here, $V_1=1.75$ ($V_2=0.2)$ means that a ratio of the calculated masses (mixings) 
to the renormalized masses (observed mixings) satisfies $1.78\times10^{-2} \leq r_{\rm mass} \leq 56.2$ ($0.63 \leq r_{\rm mixings} \leq 1.58$).
The difference between $V_1$ and $V_2$ occurs from relative errors associated with the observed values of the masses and mixing matrices. 
While the error for the mass is ${\cal O}(1)$-${\cal O}(10)$\%, the elements of the CKM matrix contain ${\cal O}(10^{-3})$-${\cal O}(1)$\% error. 
Therefore, it is reasonable to search for a mixing matrix with a narrower tolerance than that of the masses. 
Note that the detailed values of $\{V_0, V_1, V_2\}$ are not completely determined by any specific principle, as these values are hyperparameters in our architecture.

\medskip

We change the threshold value from $V_{1}=1.0$ used in Ref. \cite{Nishimura:2023nre} to $V_1=1.75$. 
That work used experimental masses at the electroweak scale, which are shown in Particle Data Group \cite{ParticleDataGroup:2022pth}. 
On the other hand, this study considers renormalized masses that are smaller than those used in the previous research. 
For example, the up-quark mass shown in Ref. \cite{ParticleDataGroup:2024prd} is $m_{u,\mathrm{exp}}=2.16\,\MeV$, and an allowed range for a terminal state in the case of $V_{1}=1.0$ is $0.216\,\MeV \leq m_{u} \leq 21.6\,\MeV$. 
However, $m_{u,\mathrm{RG}}=0.49\,\MeV$ for $M=10^{14}\,\GeV$ as shown in Table \ref{tab:QuarkMassRatio_in_HighEnergy}, so an allowed range with the same $V_{1}$ is $0 .049 \,\MeV \leq m_{u} \leq 4.9\,\MeV$.
This variation is quantified as a tolerance $\Delta m_{u}$ in the following way:
\begin{align}
    \Delta m_{u}^{\mathrm{exp}}\left(V_1=1.0\right)&=\left(10^{V_1}-10^{-V_1}\right)m_{u,\mathrm{exp}}=21.4\nonumber\\
    \rightarrow \Delta m_{u}^{\mathrm{RG}}\left(V_1=1.0\right)&=\left(10^{V_1}-10^{-V_1}\right)m_{u,\mathrm{RG}}=4.9.
\end{align}
For the agent, the smaller $\Delta m_{u}$ becomes, the harder it is to reach a terminal state and the less experience it has to receive $\mathcal{R}_{\rm term}$ which is mentioned below. 
This means that the agent cannot learn the appropriate behavior to reach a terminal state.
Thus, we adopt $V_1 = 1.75$ to promote the learning, and this allows the tolerance to remain comparable to the previous work:
\begin{align}
    \Delta m_{u}^{\mathrm{RG}}\left(V_1=1.75\right)=27.5.
\end{align}
Since changes of the renormalized masses from $10^{14}\,\GeV$ to $10^{17}\,\GeV$ are relatively small, the same $V_1$ is adopted for all energy scales.
Then, after optimizing the ${\cal O}(1)$ coefficients, we manually extract terminal states where $n_{ij}$ is positive. 
Moreover, at this time, the reproducibility of the masses is ensured by extracting models that satisfy $V_1 = 1.75\,(0.1 \leq r_{\rm mass} \leq 10.0)$.

\medskip

Let us denote the charge assignment ${\cal Q}$ as observed by the agent, and ${\cal Q}'$ followed by the action $\mathfrak{a}$.
For a pair of the assignment and the action $({\cal Q}, \mathfrak{a})$, we give the agent rewards $\mathcal{R}$.
The rewarding procedure is exactly the same as in Ref. \cite{Nishimura:2023nre} in which the design of the rewards is described by:

\medskip

\begin{screen}
\begin{enumerate}
    \item Give the basic point $\mathcal{R}_{\rm base}$, depending on the value of intrinsic value:
    \begin{align}
        \mathcal{R}_{\rm base}=
        \left\{
        \begin{array}{ll}
             {\cal V}({\cal Q}') -{\cal V}({\cal Q})&\qquad {\rm if}\,{\cal V}({\cal Q}') -{\cal V}({\cal Q})>0 \\
             \mathcal{R}_{\rm offset}&\qquad {\rm if}\,{\cal V}({\cal Q}') -{\cal V}({\cal Q})\leq 0         
        \end{array}
        \right.
        ,
    \end{align}
    where $\mathcal{R}_{\rm offset}$ corresponds to a penalty, chosen as $\mathcal{R}_{\rm offset}=-10$. 

    \item When the ${\cal Q}'$ lies outside $-9\leq {\cal Q}' \leq 9$ or the flavon charge satisfies $q(\phi)=0$, we give the penalty ${\cal R}_{\rm offset}$ and the environment comes back to the original charge assignment ${\cal Q}$. 

    \item When the ${\cal Q}'$ is turned out to be a terminal state, we give the bonus point ${\cal R}_{\rm term}$, chosen as $\mathcal{R}_{\rm term}=100$. 

    \item Summing up the above points, we define the reward ${\cal R}({\cal Q},\mathfrak{a})$.
\end{enumerate}
\end{screen}

The success or failure of RL depends greatly on the design of the reward\footnote{In many cases, the appropriate design of rewards requires heuristic construction depending on the environment and action space. 
In contrast, it has been pointed out that the design may be systematized by combining Bayesian statistics with RL \cite{levine2018reinforcementlearningcontrolprobabilistic} 
Alternatively, by learning based on models in Bayesian statistics, various algorithms have been developed that can perform learning even with small amounts of data. 
``PILCO'' \cite{10.5555/3104482.3104541} is one example. 
We will leave these comprehensive studies for future work.}. 
If the intrinsic value is drastically changed, it may become difficult to find FN charges. 
In order to reduce such uncertainties in our research, we do not include cosmological restrictions in the intrinsic value.

\section{Comparison with traditional methods}
\label{sec:comparison}

In this section, we show the distribution of terminal states for the quark sector, which is found by the RL. 
We made 20 agents for each energy scale, and in total, it took six days to train the neural networks on a single CPU.
The loss functions tend to be minimized, and we also check for an increase in the sum of rewards.
Thus, overfittings did not occur. 

\medskip

By performing the Monte-Carlo search with the Gaussian distribution described in Sec. \ref{sec:agent}, the ${\cal O}(1)$ coefficients $y_{ij}$ are optimized to more realistic ones.
In other words, the intrinsic values are surely optimized. 
We summarize the number of terminal states found by the agents in Table \ref{tab:num_of_terminal}. 
From this, the terminal states are relatively easy to find at $10^{14},10^{15}\,\GeV$.

\begin{table}[H]
\renewcommand{\arraystretch}{1.25}
\centering
\scalebox{0.88}{
   \begin{tabular}{|c|c|c|c|c|c|}\hline
          Energy scale $M$ & $10^{14}\,\GeV$ & $10^{15}\,\GeV$ & $10^{16}\,\GeV$ & $10^{17}\,\GeV$ & Total \\\hline
          Terminal states & 710 & 555 & 374 & 546 & 2,185 \\
          Positive $n_{ij}$ & 434 & 323 & 236 & 323 & 1,316 \\
          Extracted in $V_{1}=1.0$ & 44 & 52 & 24 & 36 & 156 \\
         \hline
      \end{tabular}
      }
  \caption{The number of models at each scale. The terminal states consist of non-integer $n_{ij}$, so we should narrow down according to the value of $n_{ij}$. Then, since they are searched in $V_{1}=1.75$, we extract models that satisfy $V_{1}=1.0$ so that the errors of masses are within single-digit finally.}
    \label{tab:num_of_terminal}  
\end{table}
     \renewcommand{\arraystretch}{1}

\medskip

The predicted flavor structure is strongly dependent on the assignments of $U(1)$ charges. 
In terms of intrinsic values, the evaluation of each model fluctuates dramatically with changes in discrete parameters.
Therefore, the usual gradient descent method cannot be used to obtain good charges that reproduce the flavor structure, and we must rely on brute force calculations or random searches. 
With this background, our statistical approach with RL is practically more efficient than brute force computation.
Let $p$ be the fraction of charge pairs that satisfy $V_{1}=1.0$ in the parameter space generated by the charge combinations. 
In our learning, we generated the following number of random sets of charges as initial states of each episode:
\begin{align}
    4\,\mathrm{scales}
    \times 20\,\mathrm{agents}
    \times 10^5\,\mathrm{episodes}
    = 8\times10^6\,\mathrm{sets}.
\end{align}
Since no terminal state is found in those sets, it is reasonable to estimate $p<1/\left(8\times10^6\right)$.
Let $t$ be the time required to find one terminal state that satisfies $V_{1}=1.75$ using a method other than RL. 
Then, the total time $T$, which is required to find realistic models with a similar number of charges as those found in this work, can be estimated as $T=2{,}185\cdot t/p$.

\medskip

We tested various optimization methods of ``SciPy'', which is a library of Python, to find $\eta$ satisfying Eq. \eqref{eq:eta_range}. 
In Table \ref{tab:calc_time}, we show the results of actual measurements using the same computer as in the case of RL.
It turns out that 0.274 ms is required at least to find the optimal value of $\eta$ such that maximize the intrinsic value ${\cal V}$ under a particular set of FN charges.
Thus, the total time $T$ is estimated as follows:
\begin{align}
    T > 2{,}185 \times 0.274\,\mathrm{ms} \times \left(8\times10^6\right)
    \sim 55\,\mathrm{days}.
\end{align}
In practice, optimization of couplings $y_{ij}$ must also be performed, so the efficiency of the search by RL has a very large impact.
Note that the SLSQP method is also used in our RL algorithm, and sufficiently fast optimization of $\eta$ has been performed.

\begin{table}[H]
\renewcommand{\arraystretch}{1.25}
\centering
\scalebox{0.88}{
   \begin{tabular}{|c|c||c|c|}\hline
        Method & Calculation time (ms) & Method & Calculation time (ms) \\
        \hline
        Nelder-Mead & $7.225 \pm 0.063$ & TNC & $0.737 \pm 0.012$ \\
        Powell & $5.132 \pm 0.033$ & COBYLA & $1.687 \pm 0.011$ \\
        L-BFGS-B & $0.597 \pm 0.012$ & SLSQP & $0.274 \pm 0.005$ \\
        \hline
    \end{tabular}
    }
  \caption{The scipy.optimize.minimize is used for searching an appropriate $\eta$.}
    \label{tab:calc_time}  
\end{table}
     \renewcommand{\arraystretch}{1}

\medskip

\begin{figure}[H]
\begin{minipage}{0.49\hsize}
  \begin{center}
  \includegraphics[height=60mm]{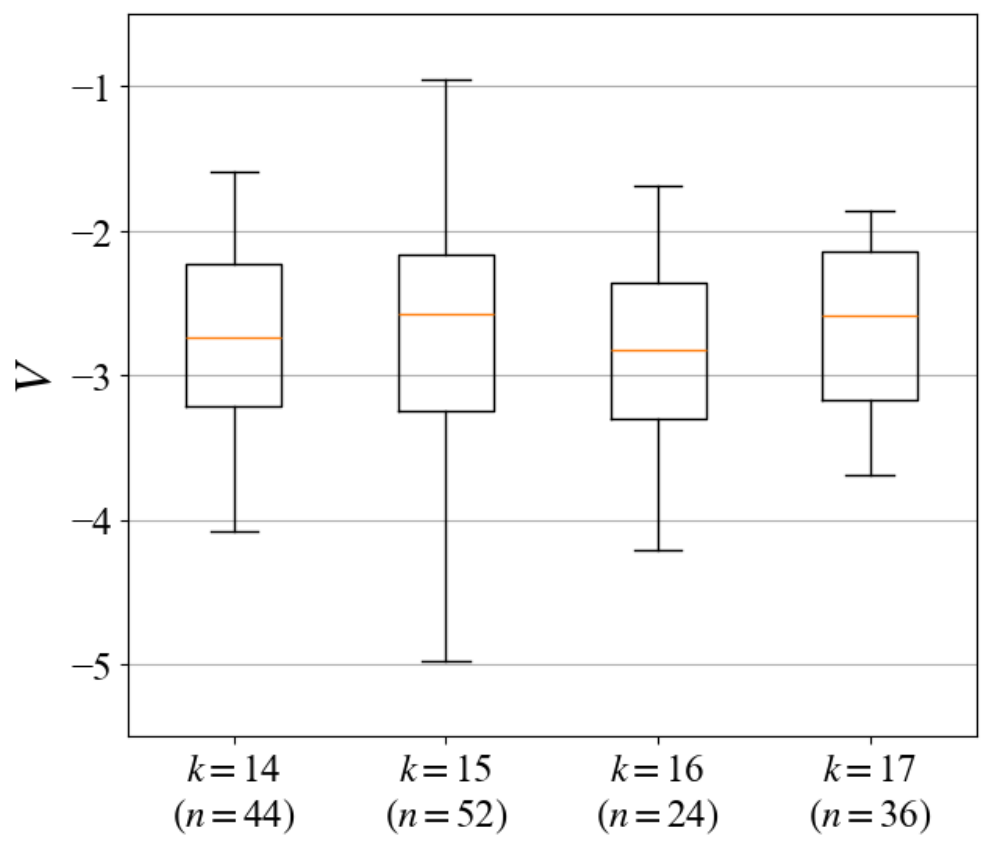}
  \end{center}
 \end{minipage}
 \begin{minipage}{0.49\hsize}
  \begin{center}
   \includegraphics[height=60mm]{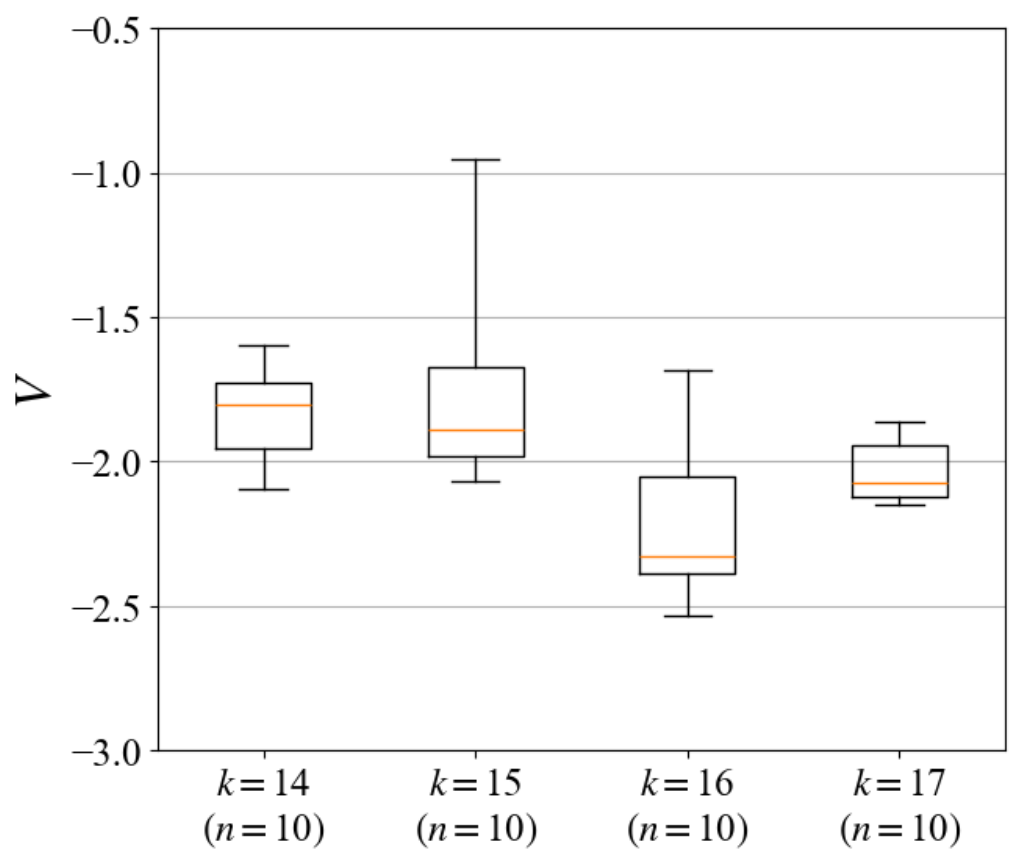}
  \end{center}
 \end{minipage}
  \caption{$k=\log_{10} M$ means the energy scale and $n$ is the number of models found in our numerical analysis. 
  The left figure shows the distribution of intrinsic values $V$ for the whole terminal states. On the other hand, we extract 10 models with high intrinsic values at each $k$ in the right figure. Realistic models are numerous for $k=14,15$. In addition, their intrinsic values tend to be high. Note that the range of the vertical axis is different in each graph.}
\label{fig:intrinsic_value}
\end{figure}

We show the distribution for all terminal states in Fig. \ref{fig:intrinsic_value}. 
In addition, the distribution of the top 10 models with the highest intrinsic value is also shown to compare with an aligned number of terminal states at each energy scale.
In Fig. \ref{fig:intrinsic_value}, the quartile ranges are represented by boxes, and the orange lines represent the medians.
The upper and lower ends of the error bars refer to the maximum and minimum intrinsic values of the models, respectively. 
Taking into account all terminal states, there is no significant difference in the median and interquartile range of intrinsic values for each energy scale. 
This suggests that the training of the agents is not biased at each scale and stable learning is realized. 
In fact, the behavior of the loss function and rewards also showed no occurrence of overfitting.

\medskip

On the other hand, the number of terminal states found by RL varies across different energy scales. 
Once sufficiently trained, agents tend to seek out terminal states as quickly as possible within their environment in order to avoid negative rewards (punishments) associated with reductions in intrinsic value. 
However, if no terminal state exists near the randomly assigned initial charge, the agents fail to reach a solution and simply proceed to the next episode.
Therefore, while the learning process remains stable, the variation in the number of terminal states results in differing ratios of solutions corresponding to these terminal states.
Moreover, terminal states have high intrinsic values in the case of $M=10^{14}$ and $10^{15}\,\GeV$.
Taken together, it is inferred that there are relatively many solutions that can reproduce the experimentally confirmed flavor structure of quarks at $M=10^{14}, 10^{15}\,\GeV$.
The two perspectives presented above enable us to consider the differences between the energy scales. 
However, future large-scale searches will be necessary to provide additional statistical support for each piece of evidence.

\medskip

\begin{figure}[H]
    \centering
    \includegraphics[width=65mm]{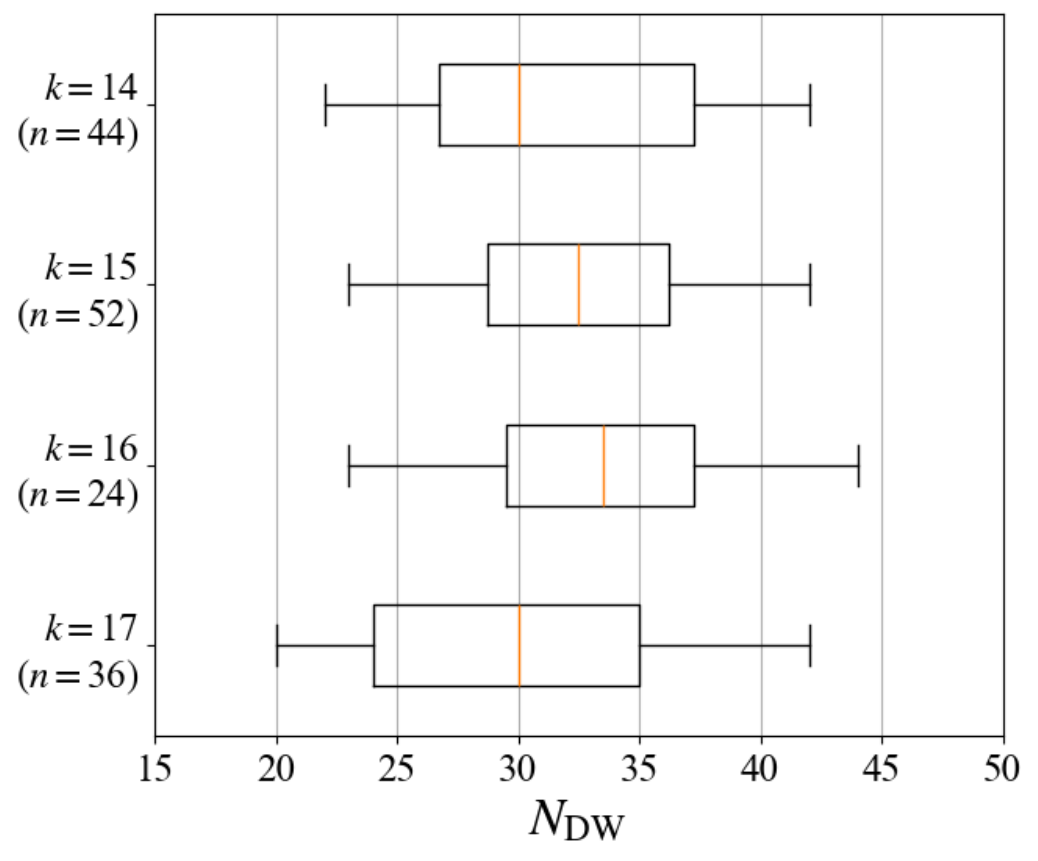}
    \caption{$k=\log_{10} M$ means the energy scale and $n$ is the number of models found in our numerical analysis. Our finding models are located around $N_{\mathrm{DW}}=30$, and a minimum value is $N_{\mathrm{DW}}=20$ at $k=17$.}
    \label{fig:N_DW}
\end{figure}

The distribution of the domain wall number $N_{\mathrm{DW}}$ at each energy scale is shown in Fig. \ref{fig:N_DW}.
For all scales, the models are located around $N_{\mathrm{DW}}=30$. 
In addition, Table \ref{tab:IV_NDW} shows the average of intrinsic values ${\cal V}$ for each $N_{\mathrm{DW}}$ category at each scale.
The table shows that the distribution of ${\cal V}$ varies from scale to scale. 
For example, the intrinsic value tends to be highest in the region of $N_{\mathrm{DW}} \geq 36$ at $M=10^{14}, 10^{17}\,\GeV$, so some cases with small $N_{\mathrm{DW}}$ are not expected to reproduce the renormalized masses. 
The scale of $M=10^{15}\,\GeV$ also tends to have large ${\cal V}$ at about $N_{\mathrm{DW}}=30$ even considering the $1\sigma$ error range. 
Since the model with the highest intrinsic value among the 156 terminal states is located at $\left(M/\GeV,N_{\mathrm{DW}}\right)=\left(10^{15},36\right)$, it is still expected to be difficult to reproduce renormalized masses with small $N_{\mathrm{DW}}$. 
On the other hand, for $M=10^{16}\,\GeV$, the category with smaller $N_{\mathrm{DW}}$ tends to have higher ${\cal V}$. 
Although the analysis with $M=10^{16}\,\GeV$ includes the uncertainty due to the relatively large error, it would be possible that there are even smaller $N_{\mathrm{DW}}$ terminal states at $M=10^{16}\,\GeV$.
The potential for discovering nontrivial relationships through RL is promising, but more precise investigation is required in the future.

\medskip

In a previous study \cite{Ema:2016ops}, an example with $N_{\mathrm{DW}}=26$ was treated as a benchmark for roughly reproducing the flavor structure. 
By statistical analysis with RL as described above, the lower limit of $N_{\mathrm{DW}}$ reproducing the flavor structure may be around $N_{\mathrm{DW}}=20$, summarized in Fig. \ref{fig:N_DW} and Table \ref{tab:IV_NDW}. 
In any case, RL is useful as a flexible method to explore discrete parameters such as FN charges and to consider the constraints from cosmology and so on.

\vspace{\stretch{1}}

\vspace{\stretch{1}}
\newpage

\vspace{\stretch{1}}

\begin{table}[H]
\renewcommand{\arraystretch}{1.25}
\centering
\scalebox{0.88}{
   \begin{tabular}{|c|c|c|c|}\hline
         & $N_{\mathrm{DW}} \in \left[20,28\right)$ & $N_{\mathrm{DW}} \in \left[28,36\right)$ & $N_{\mathrm{DW}} \in \left[36,44\right]$ \\
        \hline
        $k=14$ & $-2.78 \pm 0.13$ & $-2.71 \pm 0.19$ & $\mathbf{-2.61} \pm 0.20$ \\
        $k=15$ & $-2.89 \pm 0.20$ & $\mathbf{-2.55} \pm 0.15$ & $-2.91 \pm 0.24$ \\
        $k=16$ & $\mathbf{-2.47} \pm 0.27$ & $-2.77 \pm 0.17$ & $-3.12 \pm 0.26$ \\
        $k=17$ & $-2.69 \pm 0.16$ & $-2.86 \pm 0.17$ & $\mathbf{-2.44} \pm 0.10$ \\
        \hline
    \end{tabular}
    }
  \caption{The terminal states at each energy scale are divided into three categories according to $N_{\mathrm{DW}}$, and the average intrinsic value in each category is shown with $k=\log_{10} M$. The largest averages among the three categories are written in bold. The distribution of intrinsic value is different at each energy scale. Especially at $k=16$, the smaller the $N_{\mathrm{DW}}$, the higher the intrinsic value.}
    \label{tab:IV_NDW}  
\end{table}
     \renewcommand{\arraystretch}{1}

\vspace{\stretch{1}}

\section{Exploring the flaxion physics}
\label{sec:flaxion_physics}

\begin{figure}[H]
    \centering
    \includegraphics[width=70mm]{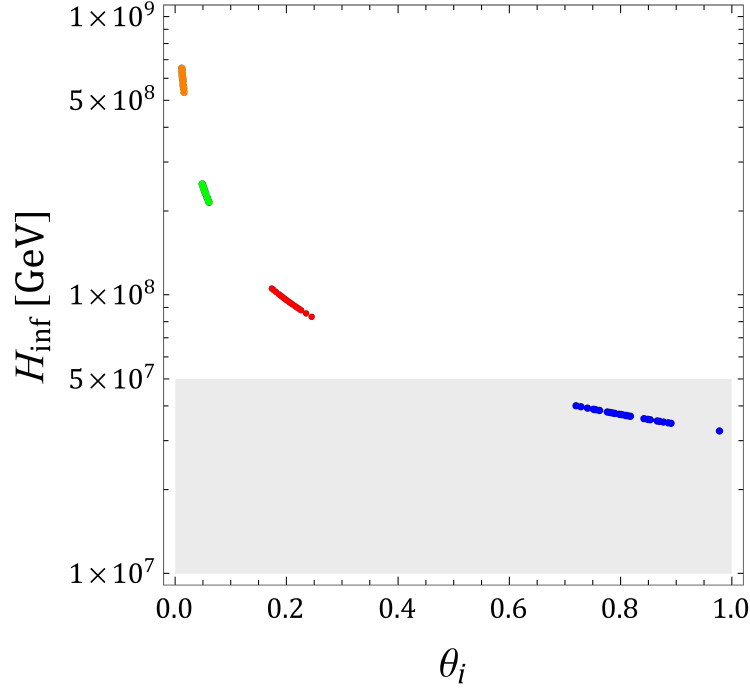}
    \caption{The point clouds are formed at 
    $M=10^{14}\,\GeV$ (blue), $10^{15}\,\GeV$ (red), $10^{16}\,\GeV$ (green) and $10^{17}\,\GeV$ (orange) in ascending order of $H_{\mathrm{inf}}$. The gray area is the excluded region from the inflation scenario. Thus, all of the terminal states found at $M=10^{14}\,\GeV$ are unsuitable due to cosmological constraints.}
    \label{fig:H-theta_log}
\end{figure}

Each of the models found by the agents takes a different charge configuration. 
Depending on their charges, the flaxion decay constant $f_a$ as well as the misalignment angle $\theta_{i}$ given by Eq. \eqref{eq:alignment-angle} are determined by results discovered by the agent. 
Then, based on the results of RL, we can set the upper bound for the inflationary scale $H_{\mathrm{inf}}$ through Eq. \eqref{eq:inf-scale_iso}. 
Fig. \ref{fig:H-theta_log} shows the correlation between inflationary scales and the misalignment angles for each model. 
The colors of the data classify energy scales. 
The blue, red, green, and orange mean $M = 10^{14},\ 10^{15},\ 10^{16},$ and $10^{17}\,\GeV$ respectively. 
The gray shaded region is excluded by Eq. \eqref{eq:inf-scale}.
Therefore, we can state that $M\gtrsim 10^{15}\,\GeV$ models are favored by cosmology.

\vspace{\stretch{1}}
\newpage
\medskip

In this section, we discuss the flaxion phenomenology.
To calculate the flaxion-photon coupling, we should determine the FN charges for both quarks and leptons.
Thus, we run some training for the lepton sector.
Two benchmark points of the quark sector are shown in Table \ref{tab:benchmark_quark15} and Table \ref{tab:benchmark_quark16}.
Benchmark 1 has the highest intrinsic value in the case of $M=10^{15}\,\GeV$.
On the other hand, benchmark 2 has the highest intrinsic value in the case of $M=10^{16}\,\GeV$.
By definition, models with such high intrinsic values tend to successfully reproduce the renormalized masses and mixings. 
Therefore, these benchmarks can be regarded as providing the most suitable Lagrangian for exploring flaxion physics. 
In the quark sector, since the Yukawa couplings are treated as ordinary continuous parameters, it is possible to reproduce the flavor structure with even greater accuracy by optimizing the parameters using conventional methods, such as Markov Chain Monte Carlo. 
However, this approach requires a long computational time due to the large number of coupling constants involved.

\vspace{\stretch{1}}

\vspace{\stretch{1}}
\newpage
\vspace*{\stretch{1}}

\begin{table}[H]
\centering
\scalebox{0.9}{
\begin{tabular}{l|c}
    \hline
    \begin{tabular}{l} Charges \end{tabular} &
    ${\cal Q}=\left(\begin{array}{ccc|ccc|ccc|cc}
        Q_{1} & Q_{2} & Q_{3} & u_{1} & u_{2} & u_{3} & d_{1} & d_{2} & d_{3} & H & \phi \\ 
        \hline
        5 & 5 & 2 & 1 & -3 & 4 & -4 & -6 & -4 & 2 & 1 \\ 
    \end{array}\right)$ \\ 
    \hline
    \begin{tabular}{l} $\mathcal{O}\left(1\right)$ coeff. \end{tabular} & $y^{u} \simeq
    \left(\begin{array}{rrr}
        -1.483 & -1.021 & -0.996 \\ 
        -1.513 & 0.960 & 0.866 \\ 
        -1.219 & -0.500 & 0.400
    \end{array}\right) \ ,\ 
    y^{d} \simeq
    \left(\begin{array}{rrr}
        -1.117 & 0.793 & 1.291 \\ 
        -1.087 & 0.507 & -1.313 \\ 
        -0.868 & -1.618 & -0.929
    \end{array}\right)$ \\ 
    \hline
    \begin{tabular}{l} VEV\end{tabular} & $\eta \simeq 0.255\cdot e^{1.533i}$ \\ 
    \hline
    \begin{tabular}{l} Intrinsic value \end{tabular} & $\mathcal{V}_{\mathrm{opt}}\simeq-0.953$ \\ 
    \hline
    \begin{tabular}{l} Masses\\(output) \end{tabular} & $\left(\begin{array}{lll}
    m_{u}/\MeV & m_{c}/\GeV & m_{t}/\GeV\\
    m_{d}/\MeV & m_{s}/\MeV & m_{b}/\GeV
    \end{array}\right)
    \simeq \left(\begin{array}{lll}
    0.289 & 0.222 & 69.79 \\
    2.13 & 20.70 & 0.938
    \end{array}\right)$\\
    \hline
    \begin{tabular}{l} Ratios\\(masses) \end{tabular}
    & $\left(\begin{array}{lll}
    E_{u} & E_{c} & E_{t}\\
    E_{d} & E_{s} & E_{b}
    \end{array}\right)
    \simeq \left(\begin{array}{lll}
    0.205 & 0.025 & 0.025 \\
    0.313 & 0.001 & 0.004
    \end{array}\right)$ \\
    \hline
    \begin{tabular}{l} CKM matrix\\(output) \end{tabular} & $\left|V_{\mathrm{CKM}}\right| \simeq
    \left(\begin{array}{lll}
    0.983 & 0.185 & 0.004 \\
    0.184 & 0.982 & 0.041 \\
    0.012 & 0.039 & 0.999
    \end{array}\right)$
    \\
    \hline
    \begin{tabular}{l} Ratios\\(mixings) \end{tabular}
    &
    $E_{\cal C} \simeq
    \left(\begin{array}{lll}
    0.004 & 0.086 & 0.041 \\
    0.086 & 0.004 & 0.010 \\
    0.128 & 0.019 & 0.000
    \end{array}\right)$ \\
    \hline
\end{tabular}
}
\caption{Benchmark 1 for the quark sector at $M=10^{15}\,\GeV$.}
\label{tab:benchmark_quark15}
\end{table}

\vspace{\stretch{1}}

\begin{table}[H]
\centering
\scalebox{0.9}{
\begin{tabular}{l|c}
    \hline
    \begin{tabular}{l} Charges \end{tabular} &
    ${\cal Q}=\left(\begin{array}{ccc|ccc|ccc|cc}
        Q_{1} & Q_{2} & Q_{3} & u_{1} & u_{2} & u_{3} & d_{1} & d_{2} & d_{3} & H & \phi \\ 
        \hline
        -2 & -1 & 2 & 1 & 1 & -2 & 7 & 7 & 8 & -3 & -1 \\ 
    \end{array}\right)$ \\ 
    \hline
    \begin{tabular}{l} $\mathcal{O}\left(1\right)$ coeff. \end{tabular} & $y^{u} \simeq
    \left(\begin{array}{rrr}
        -0.980 & -1.125 & 1.128 \\ 
        -1.004 & -0.660 & 1.076 \\ 
        1.535 & 1.148 & 0.400
    \end{array}\right) \ ,\ 
    y^{d} \simeq
    \left(\begin{array}{rrr}
        1.138 & -1.178 & -1.184 \\ 
        -1.281 & -0.899 & 1.612 \\ 
        -1.106 & 1.243 & -1.421
    \end{array}\right)$ \\ 
    \hline
    \begin{tabular}{l} VEV\end{tabular} & $\eta \simeq 0.148\cdot e^{1.505i}$ \\ 
    \hline
    \begin{tabular}{l} Intrinsic value \end{tabular} & $\mathcal{V}_{\mathrm{opt}}\simeq-1.684$ \\ 
    \hline
    \begin{tabular}{l} Masses\\(output) \end{tabular} & $\left(\begin{array}{lll}
    m_{u}/\MeV & m_{c}/\GeV & m_{t}/\GeV\\
    m_{d}/\MeV & m_{s}/\MeV & m_{b}/\GeV
    \end{array}\right)
    \simeq \left(\begin{array}{lll}
    0.718 & 0.080 & 697.33 \\
    0.67 & 19.70 & 0.952
    \end{array}\right)$\\
    \hline
    \begin{tabular}{l} Ratios\\(masses) \end{tabular}
    & $\left(\begin{array}{lll}
    E_{u} & E_{c} & E_{t}\\
    E_{d} & E_{s} & E_{b}
    \end{array}\right)
    \simeq \left(\begin{array}{lll}
    0.214 & 0.443 & 0.007 \\
    0.163 & 0.001 & 0.032
    \end{array}\right)$ \\
    \hline
    \begin{tabular}{l} CKM matrix\\(output) \end{tabular} & $\left|V_{\mathrm{CKM}}\right| \simeq
    \left(\begin{array}{lll}
    0.985 & 0.170 & 0.003 \\
    0.170 & 0.984 & 0.058 \\
    0.013 & 0.057 & 0.998
    \end{array}\right)$
    \\
    \hline
    \begin{tabular}{l} Ratios\\(mixings) \end{tabular}
    &
    $E_{\cal C} \simeq
    \left(\begin{array}{lll}
    0.005 & 0.121 & 0.074 \\
    0.123 & 0.005 & 0.145 \\
    0.179 & 0.142 & 0.000
    \end{array}\right)$ \\
    \hline
\end{tabular}
}
\caption{Benchmark 2 for the quark sector at $M=10^{16}\,\GeV$.}
\label{tab:benchmark_quark16}
\end{table}

\vspace{\stretch{1}}
\newpage

\medskip

We make the agent learn the lepton sector for these benchmarks.
We adopt $N_{\rm ep}=6\times10^4$ and the other hyper-parameters are the same as the learning of the quark sector.
We considered renormalized masses of the charged leptons shown in Table \ref{tab:LeptonMassRatio_in_HighEnergy} and a constraint for the sum of neutrino masses $\Sigma m_{\nu}<85\,\meV$ based on the $\Lambda \mathrm{CDM}$ model in Ref. \cite{Brieden:2022lsd}. 
Moreover, we also check constraints for the mixing angles based on NuFIT v5.3.

\medskip

An intrinsic value for the lepton sector is defined as
\begin{align}
    {\cal V}({\cal Q}) = {\cal M}_{\rm lepton} + {\cal P},
\end{align}
with
\begin{align}
    {\cal M}_{\rm lepton} &= \sum_{\alpha = l}E_\alpha,\\
    {\cal P} &= \sum_{i,j} E_{\cal P}^{ij}.
\end{align}
The errors $E_{\alpha,{\cal P}}$ are calculated in following ways:
\begin{align}
    E_\alpha &= \biggl| \log_{10}\left(\frac{|m_\alpha|}{|m_{\alpha,{\rm RG}}|}\right) \biggl|,\\
\begin{split}
    E_{\cal P}^{ij} &= 
        \left\{
        \begin{array}{ll}    
    0.0&\quad ({\rm in\,3\sigma\,CL\,ranges\,of\,Eq.\,\eqref{eq:PMNS_mat}})
    \\
    \biggl| \log_{10}\left(\frac{|V_{\rm PMNS}^{ij}|}{|V_{\rm PMNS,\,exp}^{ij}|}\right) \biggl|&\quad 
    ({\rm other})
        \end{array}
        \right.
\end{split}
.
\label{eq:intrinsic_value}
    \end{align}
From NuFIT v5.3 and Ref. \cite{Esteban:2020cvm}, $3\sigma$ CL ranges of the PMNS matrix are given by
\begin{align}
|V_{\mathrm{PMNS}}| = \begin{pmatrix}
0.801 \rightarrow 0.842, & 0.518 \rightarrow 0.580, & 0.142 \rightarrow 0.155 \\
0.236 \rightarrow 0.507, & 0.458 \rightarrow 0.691, & 0.630 \rightarrow 0.779 \\
0.264 \rightarrow 0.527, & 0.471 \rightarrow 0.700, & 0.610 \rightarrow 0.762
\end{pmatrix}.
\label{eq:PMNS_mat}
\end{align}

\vspace{\stretch{1}}

\begin{table}[H]
\centering
\scalebox{0.9}{
\begin{tabular}{|c|ccc|}
    \hline
    energy & $m_{e,\mathrm{RG}}/\MeV$ & $m_{\mu,\mathrm{RG}}/\GeV$ & $m_{\tau,\mathrm{RG}}/\GeV$ \\
    \hline
    $10^{8}\,\GeV$ & $0.49691\pm0.00098$ & $0.104681\pm0.000183$ & $1.77852\pm0.00308$ \\
    $10^{12}\,\GeV$ & $0.48388\pm0.00139$ & $0.101936\pm0.000277$ & $1.73194\pm0.00466$ \\
    $10^{14}\,\GeV$ (ext) & $0.47856$ & $0.100813$ & $1.71303$ \\
    $10^{15}\,\GeV$ (ext) & $0.47733$ & $0.100550$ & $1.70871$ \\
    $10^{16}\,\GeV$ (ext) & $0.47747$ & $0.100577$ & $1.70936$ \\
    $10^{17}\,\GeV$ (ext) & $0.47931$ & $0.100961$ & $1.71614$ \\
    \hline
\end{tabular}
}
\caption{This shows the renormalized masses and the extrapolated masses (ext) at each energy scale for the charged leptons. The renormalized masses are evaluated based on Ref. \cite{Huang:2020hdv}.}
\label{tab:LeptonMassRatio_in_HighEnergy}
\end{table}

\vspace{\stretch{1}}
\newpage
\medskip

We make six agents for each of the benchmarks and need 15 hours on a single CPU for the whole learning process. 
In the training, we choose the normal ordering of neutrino masses. 
The number of terminal states found by the agent is summarized in Table \ref{tab:num_of_terminal_lepton}.
For $M=10^{15}\,\GeV$, we find 4,732 terminal states, and among them, 23 models satisfy the $3\sigma$ constraints for the mixing angles $\theta_{12}, \theta_{13}, \theta_{23}$. 
On the other hand, for $M=10^{16}\,\GeV$, we find few terminal states with $V_{2} = 0.2$, and no state satisfies all of the constraints for the mixing angles. 
Thus, similar to the evaluation of renormalized masses, we repeat the learning process with $V_{2}=0.3$ and then narrow down terminal states to $V_{2}=0.2$. 
We then obtain 6,219 terminal states, and among them, 7 models satisfy the $3\sigma$ constraints for the mixing angles. 
Finally, 30 models were extracted by considering the mixing constraints, which were chosen as benchmarks for the lepton sector.
The details of these benchmarks for the lepton sector are summarized in Appendix \ref{app:benchmark}. 

\medskip

The tendency for the number of terminal states at $M=10^{15}, 10^{16}\,\GeV$ are different due to the value of $M$ in the neutrino mass matrix. 
In other words, the change in $M$ complicates the reproduction of the mixing matrix. 
The appropriate value of $V_{2}$ at each energy scale needs to be analyzed, but we leave this issue for future work.

\medskip

\begin{figure}[H]
    \centering
    \includegraphics[width=70mm]{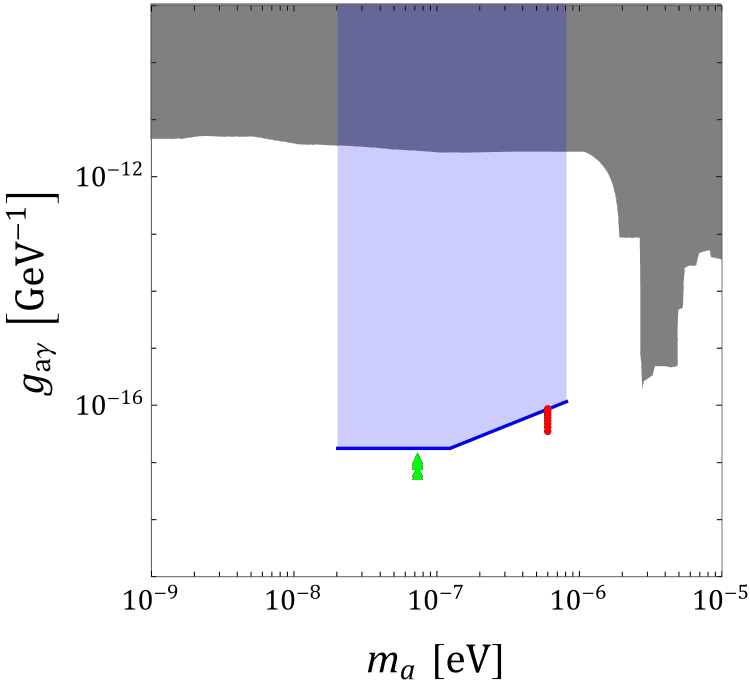}
    \caption{Right circles (Red) and left triangles (green) are the results calculated by using the FN charges of the lepton sector corresponding to Benchmarks 1 and 2, respectively. The gray area indicates the region excluded by existing experiments \cite{PhysRevLett.104.041301,ADMX:2018ogs,ADMX:2019uok,Salemi:2021gck,PhysRevLett.127.261803,Noordhuis:2022ljw,Benabou:2025jcv} summarized in Ref. \cite{AxionLimits}. On the other hand, the blue area indicates the sensitivity of DMRadio-$\mathrm{m}^3$ proposed in Ref. \cite{DMRadio:2022pkf}.}
    \label{fig:ma-gag}
\end{figure}

The flaxion-photon coupling is given by using the charges obtained from the results of RL. 
Note that the charges of both quarks and leptons are needed to calculate the coupling. 
Therefore, we adopt two benchmark charge assignments: benchmark 1 with the quark and lepton charges listed in Tables \ref{tab:benchmark_quark15} and \ref{tab:lepton_15_1}, respectively, and benchmark 2 with the corresponding charges given in Tables \ref{tab:benchmark_quark16} and \ref{tab:lepton_16_1}.
Fig. \ref{fig:ma-gag} shows the effective coupling constant based on the charges of the benchmark models.
This mass range will be explored in the future experiment, ex., DMRadio-$\mathrm{m}^3$ in Ref. \cite{DMRadio:2022pkf}.
In this figure, only the results at $M=10^{15}\,\GeV$ are within the sensitivity region of the DMRadio-$\mathrm{m}^3$.
However, the calculated values at $M=10^{16}\,\GeV$ are also close to the sensitivity region.
Therefore, it is expected that the RL with the appropriate value of $V_{2}$ will lead to the discovery of models that overlap in the same sensitivity region in a short time.


\begin{table}[H]
\renewcommand{\arraystretch}{1.25}
\centering
\scalebox{0.88}{
   \begin{tabular}{|c|c|c|}\hline
         & Benchmark 1 & Benchmark 2 \\
        Energy scale $M$ & $10^{15}\,\GeV$ & $10^{16}\,\GeV$ \\
        \hline
        Terminal states & 4,732 & 6,219 \\
        Positive $n_{ij}$ & 575 & 451 \\
        Extracted in $\left(V_{1}, V_{2}\right)=\left(1.0, 0.2\right)$ & 364 & 197 \\
        Extracted in $\Sigma m_{\nu}<85\,\meV$ & 361 & 197 \\
        Extracted in mixing angles & 23 & 7 \\
        \hline
    \end{tabular}
    }
  \caption{This shows the number of models for the lepton sector based on the benchmark point of the quark sector. The terminal states include non-integer $n_{ij}$, so we should narrow down according to the value of $n_{ij}$. Since they are limited by $V_{1}=1.75$, we extract models that satisfy $V_{1}=1.0$ so that the errors of masses are within single-digit. Finally, the constraint $\Sigma m_{\nu}<85\,\meV$ from Ref. \cite{Brieden:2022lsd} and the $3\sigma$ constraints for the mixing angles from NuFIT v5.3 are considered.}
    \label{tab:num_of_terminal_lepton}  
\end{table}
     \renewcommand{\arraystretch}{1}

\newpage
\vspace*{\stretch{1}}

\begin{table}[H]
\centering
\scalebox{0.82}{
\begin{tabular}{l|c}
    \hline
    \begin{tabular}{l} Charges \end{tabular} &
    ${\cal Q}=\left(\begin{array}{ccc|ccc|ccc|cc}
        L_{1} & L_{2} & L_{3} & N_{1} & N_{2} & N_{3} & l_{1} & l_{2} & l_{3} & H & \phi \\ 
        \hline
        3 & 3 & 4 & -4 & -4 & -5 & -8 & -4 & -2 & 2 & 1 \\ 
    \end{array}\right)$ \\ 
    \hline
    \begin{tabular}{l} $\mathcal{O}\left(1\right)$ coeff. \end{tabular} & $y^{l} \simeq
    \left(\begin{array}{rrr}
        -0.810 & -1.032 & -0.321 \\
        1.169 & -0.925 & -0.784 \\
        -0.910 & -0.714 & -0.939
    \end{array}\right) \ ,\ 
    y^{\nu} \simeq
    \left(\begin{array}{rrr}
        -0.928 & 0.674 & -1.224 \\
        0.743 & 1.088 & 1.250 \\
        1.152 & 0.613 & -1.377
    \end{array}\right)$ \\ 
    & $y^{N} \simeq
    \left(\begin{array}{rrr}
        -1.095 & -1.146 & 0.987 \\
        -1.146 & -1.352 & 0.721 \\
        0.987 & 0.721 & 1.209
    \end{array}\right)$ \\ 
    \hline
    \begin{tabular}{l} Intrinsic value \end{tabular} & $\mathcal{V}_{\mathrm{opt}}\simeq-0.321$ \\ 
    \hline
    \begin{tabular}{l} Masses\\(output) \end{tabular} & \begin{tabular}{l} $\left(\begin{array}{lll}
    m_{e}/\MeV & m_{\mu}/\GeV & m_{\tau}/\GeV\\
    m_{\nu_{1}}/\meV & m_{\nu_{2}}/\meV & m_{\nu_{3}}/\meV
    \end{array}\right)
    \simeq \left(\begin{array}{lll}
    0.57815 & 0.116738 & 2.54709 \\
    4.475\times 10^{-7} & 8.178\times 10^{-7} & 3.761\times 10^{-6}
    \end{array}\right)$ \end{tabular} \\
    \hline
    \begin{tabular}{l} PMNS matrix\\(output) \end{tabular} & $\left|V_{\mathrm{PMNS}}\right| \simeq
    \left(\begin{array}{lll}
    0.825 & 0.544 & 0.151 \\
    0.298 & 0.647 & 0.702 \\
    0.479 & 0.534 & 0.696
    \end{array}\right)$
    \\
    \hline
    \begin{tabular}{l} Mixing angles\\(output) \end{tabular}
    & $\left(\begin{array}{lll}
    \theta_{12} & \theta_{13} & \theta_{23}
    \end{array}\right)
    \simeq \left(\begin{array}{lll}
    0.185\pi & 0.048\pi & 0.251\pi
    \end{array}\right)$ \\
    \hline
\end{tabular}
}
\caption{Benchmark point for the lepton sector. The VEV is same as Table \ref{tab:benchmark_quark15}, and this has the highest intrinsic value among the models which are found by the agent.}
\label{tab:lepton_15_1}
\end{table}

\vspace{\stretch{1}}

\begin{table}[H]
\centering
\scalebox{0.82}{
\begin{tabular}{l|c}
    \hline
    \begin{tabular}{l} Charges \end{tabular} &
    ${\cal Q}=\left(\begin{array}{ccc|ccc|ccc|cc}
        L_{1} & L_{2} & L_{3} & N_{1} & N_{2} & N_{3} & l_{1} & l_{2} & l_{3} & H & \phi \\ 
        \hline
        -4 & -4 & -4 & 3 & 8 & 1 & 6 & 3 & 2 & -3 & -1 \\ 
    \end{array}\right)$ \\ 
    \hline
    \begin{tabular}{l} $\mathcal{O}\left(1\right)$ coeff. \end{tabular} & $y^{l} \simeq
    \left(\begin{array}{rrr}
        -0.652 & -1.396 & 1.040 \\
        -1.175 & -1.376 & -1.075 \\
        1.171 & -1.295 & -0.925
    \end{array}\right) \ ,\ 
    y^{\nu} \simeq
    \left(\begin{array}{rrr}
        0.815 & 0.553 & -0.837 \\
        0.903 & -1.349 & 1.040 \\
        1.326 & -1.310 & -0.467

    \end{array}\right)$ \\ 
    & $y^{N} \simeq
    \left(\begin{array}{rrr}
        1.068 & -1.147 & 0.874 \\
        -1.147 & 1.003 & 1.487 \\
        0.874 & 1.487 & -0.885

    \end{array}\right)$ \\ 
    \hline
    \begin{tabular}{l} Intrinsic value \end{tabular} & $\mathcal{V}_{\mathrm{opt}}\simeq-0.595$ \\ 
    \hline
    \begin{tabular}{l} Masses\\(output) \end{tabular} & \begin{tabular}{l} $\left(\begin{array}{lll}
    m_{e}/\MeV & m_{\mu}/\GeV & m_{\tau}/\GeV\\
    m_{\nu_{1}}/\meV & m_{\nu_{2}}/\meV & m_{\nu_{3}}/\meV
    \end{array}\right)
    \simeq \left(\begin{array}{lll}
    0.46065 & 0.188688 & 1.00064 \\
    1.724\times 10^{-12} & 9.254\times 10^{-12} & 2.056\times 10^{-11}
    \end{array}\right)$ \end{tabular} \\
    \hline
    \begin{tabular}{l} PMNS matrix\\(output) \end{tabular} & $\left|V_{\mathrm{PMNS}}\right| \simeq
    \left(\begin{array}{lll}
    0.834 & 0.531 & 0.151 \\
    0.456 & 0.508 & 0.731 \\
    0.311 & 0.679 & 0.665
    \end{array}\right)$
    \\
    \hline
    \begin{tabular}{l} Mixing angles\\(output) \end{tabular}
    & $\left(\begin{array}{lll}
    \theta_{12} & \theta_{13} & \theta_{23}
    \end{array}\right)
    \simeq \left(\begin{array}{lll}
    0.180\pi & 0.048\pi & 0.265\pi
    \end{array}\right)$ \\
    \hline
\end{tabular}
}
\caption{Benchmark point for the lepton sector. The VEV is the same as Table \ref{tab:benchmark_quark16}, and this has the highest intrinsic value among the models which are found by the agent.}
\label{tab:lepton_16_1}
\end{table}

\vspace{\stretch{1}}
\newpage

\section{Conclusion}
\label{sec:con}

Flavor symmetries give many insights into new physics beyond the Standard Model. 
Recently, numerous studies have analyzed theoretical models using machine learning techniques. 
Under these circumstances, RL can be a new statistical analysis method to capture features of the models that are difficult to observe in sequential calculations. 
In particular, RL is useful for solving issues which are difficult within conventional descent approaches.

\medskip

In this paper, we constructed a new algorithm to search for the model parameters in the minimal axion model from flavor, utilizing the RL-based search strategy. The flavor structure has been previously analyzed in Refs. \cite{Nishimura:2023nre} and \cite{Harvey:2021oue}.
In Sec. \ref{sec:FNaxion}, we introduced the $U(1)$ flavor model with the FN mechanism, and a pseudo-scalar field arising from the imaginary component of the flavon.
Then, in Sec. \ref{sec:RL}, we developed an RL to find $U(1)$ charge assignments of the quark sector and considered renormalizing effects on masses. 
In Sec. \ref{sec:comparison}, we studied the statistical differences at the four energy scales, i.e., $M=10^{14}, 10^{15}, 10^{16}, 10^{17}$ GeV, by training 20 agents at each scale. 
Through six days of neural network training, the agents found a total of 156 terminal states that had errors within single-digit from the renormalized masses. 
In addition, we estimated the time required to discover the same number of terminal states without using RL. 
Various conventional optimization methods would require at least 55 days to perform the same task.
This indicates that the RL-based search method for discrete parameters is faster than conventional computational methods. 
We also discussed the phenomenology of the flaxion in Sec. \ref{sec:flaxion_physics}, taking the constraints from the dark matter, the isocurvature perturbation and the inflation into account. 
It turned out that $U(1)$ charge assignments with high intrinsic values are realized around $M=10^{14}, 10^{15}$ GeV, although the isocurvature constraint puts a lower bound on $M$, i.e., $M\gtrsim 10^{15}$ GeV. 
Note that the reproducibility of renormalized masses depends on the energy scales. 
Furthermore, we ran additional training for the lepton sector for the benchmarks of quarks. 
As a result, we found some terminal states for the leptons that satisfy the constraints on the sum of neutrino masses and mixing angles. 
From those states, we showed the distribution of the flaxion mass and flaxion-photon coupling at $M=10^{15}, 10^{16}$ GeV, which will be explored in future experiments such as DMRadio-$\mathrm{m}^3$ \cite{DMRadio:2022pkf}.
Our proposed RL-based efficient analysis may provide a statistical perspective not only for our discussion of the cosmological axion model but also for other areas of particle phenomenology and cosmology. 

\medskip

Before closing our paper, we will mention possible future works:

\begin{itemize}
    \item In this work, since only one flavon field is introduced, we are not able to derive the CP-symmetry breaking in the quark sector. 
    One way to improve this situation is to introduce two flavons and consider the phase difference between them. 
    In this case, one of the two flaxions will be regarded as the QCD axion, and the mechanism for mass generation of the other flaxion in the orthogonal direction should be discussed. 
    Alternatively, another way to derive CP-symmetry breaking is to consider the complex Yukawa couplings. 
    In this case, it is nontrivial whether the RL can find many terminal states immediately. 
    That is because the parameter space of Yukawa couplings generated in the first step of learning is expanded. 
    Then, it is also useful to discuss how the degree of the FN charge space has been explored by the agent when no terminal state is found during a single training session with fixed Yukawa couplings. 

    \item 
    We focused on the specific axion model with $U(1)$ flavor symmetry, but our neural network architecture can be applied to other axion models as well as flavors by replacing the input data with the appropriate ones. 
    Furthermore, it would be interesting to verify the axion through other experiments and cosmological observations using the RL-based strategy, but we leave comprehensive studies for future works. 

    \item Our architecture of the neural network follows a previous study \cite{Nishimura:2023nre}. 
    This network has about 10,000 parameters, so in total, one agent has 20,000 parameters to be optimized since two types of neural networks are used in DQN (Q network and target network). 
    In this study, the parameters of both networks were initialized in the first episode.
    Since each agent works through a different learning process independently, the parameters of the networks differ for each agent. 
    On the other hand, neural networks are known to exhibit the phenomenon of ``generalization''.
    This is the ability to make appropriate classifications or predictions for unknown data that have never been observed.
    Our networks output optimal actions even for sets of FN charges that have not been inputted before.
    This attempt itself is established based on the generalization ability. 
    However, since the generalization holds only for specific fixed values of the Yukawa couplings, this study does not fully exploit the distinctive capabilities of neural networks.
    Thus, preparing a larger-scale architecture is a stepping stone to realizing a more generalized search strategy by RL. 
    If we construct neural networks that can output the optimal action for arbitrary Yukawa couplings, the learning itself can be done only once, and further reduction of computational time is expected.

\end{itemize}

\acknowledgments

This work was supported in part by Kyushu University’s Innovator Fellowship Program (S. N., C. M.), JSPS KAKENHI Grant Numbers JP25KJ1927 (S.N.), JP23H04512 (H.O.).

\appendix

\section{Benchmark points for lepton sector}
\label{app:benchmark}

\subsection{The case of $10^{15}\,\GeV$}
\label{app:15GeV}

We show benchmarks for the leptons at $M=10^{15}\,\GeV$.
Note that the VEV is the same as Table \ref{tab:benchmark_quark15}.
Following tables are ordered by the intrinsic values.

\vspace{\stretch{1}}

\begin{table}[H]
\centering
\scalebox{0.82}{
\begin{tabular}{l|c}
    \hline
    \begin{tabular}{l} Charges \end{tabular} &
    ${\cal Q}=\left(\begin{array}{ccc|ccc|ccc|cc}
        L_{1} & L_{2} & L_{3} & N_{1} & N_{2} & N_{3} & l_{1} & l_{2} & l_{3} & H & \phi \\ 
        \hline
        3 & 3 & 4 & -4 & -4 & -5 & -8 & -4 & -2 & 2 & 1 \\ 
    \end{array}\right)$ \\ 
    \hline
    \begin{tabular}{l} $\mathcal{O}\left(1\right)$ coeff. \end{tabular} & $y^{l} \simeq
    \left(\begin{array}{rrr}
        -0.810 & -1.032 & -0.321 \\
        1.169 & -0.925 & -0.784 \\
        -0.910 & -0.714 & -0.939
    \end{array}\right) \ ,\ 
    y^{\nu} \simeq
    \left(\begin{array}{rrr}
        -0.928 & 0.674 & -1.224 \\
        0.743 & 1.088 & 1.250 \\
        1.152 & 0.613 & -1.377
    \end{array}\right)$ \\ 
    & $y^{N} \simeq
    \left(\begin{array}{rrr}
        -1.095 & -1.146 & 0.987 \\
        -1.146 & -1.352 & 0.721 \\
        0.987 & 0.721 & 1.209
    \end{array}\right)$ \\ 
    \hline
    \begin{tabular}{l} Intrinsic value \end{tabular} & $\mathcal{V}_{\mathrm{opt}}\simeq-0.321$ \\ 
    \hline
    \begin{tabular}{l} Masses\\(output) \end{tabular} & \begin{tabular}{l} $\left(\begin{array}{lll}
    m_{e}/\MeV & m_{\mu}/\GeV & m_{\tau}/\GeV\\
    m_{\nu_{1}}/\meV & m_{\nu_{2}}/\meV & m_{\nu_{3}}/\meV
    \end{array}\right)
    \simeq \left(\begin{array}{lll}
    0.57815 & 0.116738 & 2.54709 \\
    4.475\times 10^{-6} & 8.178\times 10^{-5} & 3.761\times 10^{-4}
    \end{array}\right)$ \end{tabular} \\
    \hline
    \begin{tabular}{l} PMNS matrix\\(output) \end{tabular} & $\left|V_{\mathrm{PMNS}}\right| \simeq
    \left(\begin{array}{lll}
    0.825 & 0.544 & 0.151 \\
    0.298 & 0.647 & 0.702 \\
    0.479 & 0.534 & 0.696
    \end{array}\right)$
    \\
    \hline
    \begin{tabular}{l} Mixing angles\\(output) \end{tabular}
    & $\left(\begin{array}{lll}
    \theta_{12} & \theta_{13} & \theta_{23}
    \end{array}\right)
    \simeq \left(\begin{array}{lll}
    0.185\pi & 0.048\pi & 0.251\pi
    \end{array}\right)$ \\
    \hline
\end{tabular}
}
\label{tab:lepton_15_16}
\end{table}

\begin{table}[H]
\centering
\scalebox{0.82}{
\begin{tabular}{l|c}
    \hline
    \begin{tabular}{l} Charges \end{tabular} &
    ${\cal Q}=\left(\begin{array}{ccc|ccc|ccc|cc}
        L_{1} & L_{2} & L_{3} & N_{1} & N_{2} & N_{3} & l_{1} & l_{2} & l_{3} & H & \phi \\ 
        \hline
        4 & 4 & 5 & -2 & -9 & -7 & -5 & -2 & -4 & 2 & 1 \\ 
    \end{array}\right)$ \\ 
    \hline
    \begin{tabular}{l} $\mathcal{O}\left(1\right)$ coeff. \end{tabular} & $y^{l} \simeq
    \left(\begin{array}{rrr}
        0.963 & 0.894 & 0.873 \\
        1.020 & 1.320 & -1.311 \\
        0.908 & 0.616 & 1.089
    \end{array}\right) \ ,\ 
    y^{\nu} \simeq
    \left(\begin{array}{rrr}
        0.536 & -0.883 & 0.921 \\
        -1.371 & -0.904 & -0.975 \\
        -0.959 & -1.103 & 1.298
    \end{array}\right)$ \\ 
    & $y^{N} \simeq
    \left(\begin{array}{rrr}
        1.143 & -1.011 & 1.444 \\
        -1.011 & 1.287 & 0.649 \\
        1.444 & 0.649 & -1.082
    \end{array}\right)$ \\ 
    \hline
    \begin{tabular}{l} Intrinsic value \end{tabular} & $\mathcal{V}_{\mathrm{opt}}\simeq-0.335$ \\ 
    \hline
    \begin{tabular}{l} Masses\\(output) \end{tabular} & \begin{tabular}{l} $\left(\begin{array}{lll}
    m_{e}/\MeV & m_{\mu}/\GeV & m_{\tau}/\GeV\\
    m_{\nu_{1}}/\meV & m_{\nu_{2}}/\meV & m_{\nu_{3}}/\meV
    \end{array}\right)
    \simeq \left(\begin{array}{lll}
    0.50593 & 0.071445 & 1.17743 \\
    3.316\times 10^{-7} & 5.354\times 10^{-7} & 7.069\times 10^{-6}
    \end{array}\right)$ \end{tabular} \\
    \hline
    \begin{tabular}{l} PMNS matrix\\(output) \end{tabular} & $\left|V_{\mathrm{PMNS}}\right| \simeq
    \left(\begin{array}{lll}
    0.825 & 0.544 & 0.153 \\
    0.491 & 0.556 & 0.671 \\
    0.280 & 0.628 & 0.726
    \end{array}\right)$
    \\
    \hline
    \begin{tabular}{l} Mixing angles\\(output) \end{tabular}
    & $\left(\begin{array}{lll}
    \theta_{12} & \theta_{13} & \theta_{23}
    \end{array}\right)
    \simeq \left(\begin{array}{lll}
    0.186\pi & 0.049\pi & 0.237\pi
    \end{array}\right)$ \\
    \hline
\end{tabular}
}
\label{tab:lepton_15_20}
\end{table}

\vspace{\stretch{1}}
\newpage
\vspace*{\stretch{1}}

\begin{table}[H]
\centering
\scalebox{0.82}{
\begin{tabular}{l|c}
    \hline
    \begin{tabular}{l} Charges \end{tabular} &
    ${\cal Q}=\left(\begin{array}{ccc|ccc|ccc|cc}
        L_{1} & L_{2} & L_{3} & N_{1} & N_{2} & N_{3} & l_{1} & l_{2} & l_{3} & H & \phi \\ 
        \hline
        8 & 8 & 9 & -2 & -8 & 0 & 2 & -2 & 0 & 2 & 1 \\ 
    \end{array}\right)$ \\ 
    \hline
    \begin{tabular}{l} $\mathcal{O}\left(1\right)$ coeff. \end{tabular} & $y^{l} \simeq
    \left(\begin{array}{rrr}
        1.398 & 1.025 & -1.205 \\
        1.179 & -0.946 & 0.882 \\
        -0.973 & -1.312 & 1.064
    \end{array}\right) \ ,\ 
    y^{\nu} \simeq
    \left(\begin{array}{rrr}
        -1.013 & 1.028 & 0.860 \\
        -1.250 & -1.370 & 1.168 \\
        1.408 & 1.130 & 0.635
    \end{array}\right)$ \\ 
    & $y^{N} \simeq
    \left(\begin{array}{rrr}
        1.197 & -0.612 & 1.020 \\
        -0.612 & -0.735 & -0.890 \\
        1.020 & -0.890 & -0.984
    \end{array}\right)$ \\ 
    \hline
    \begin{tabular}{l} Intrinsic value \end{tabular} & $\mathcal{V}_{\mathrm{opt}}\simeq-0.463$ \\ 
    \hline
    \begin{tabular}{l} Masses\\(output) \end{tabular} & \begin{tabular}{l} $\left(\begin{array}{lll}
    m_{e}/\MeV & m_{\mu}/\GeV & m_{\tau}/\GeV\\
    m_{\nu_{1}}/\meV & m_{\nu_{2}}/\meV & m_{\nu_{3}}/\meV
    \end{array}\right)
    \simeq \left(\begin{array}{lll}
    0.29641 & 0.070178 & 1.35618 \\
    4.884\times 10^{-12} & 3.498\times 10^{-11} & 3.151\times 10^{-10}
    \end{array}\right)$ \end{tabular} \\
    \hline
    \begin{tabular}{l} PMNS matrix\\(output) \end{tabular} & $\left|V_{\mathrm{PMNS}}\right| \simeq
    \left(\begin{array}{lll}
    0.821 & 0.551 & 0.146 \\
    0.484 & 0.538 & 0.690 \\
    0.302 & 0.638 & 0.709
    \end{array}\right)$
    \\
    \hline
    \begin{tabular}{l} Mixing angles\\(output) \end{tabular}
    & $\left(\begin{array}{lll}
    \theta_{12} & \theta_{13} & \theta_{23}
    \end{array}\right)
    \simeq \left(\begin{array}{lll}
    0.188\pi & 0.047\pi & 0.246\pi
    \end{array}\right)$ \\
    \hline
\end{tabular}
}
\label{tab:lepton_15_17}
\end{table}

\begin{table}[H]
\centering
\scalebox{0.82}{
\begin{tabular}{l|c}
    \hline
    \begin{tabular}{l} Charges \end{tabular} &
    ${\cal Q}=\left(\begin{array}{ccc|ccc|ccc|cc}
        L_{1} & L_{2} & L_{3} & N_{1} & N_{2} & N_{3} & l_{1} & l_{2} & l_{3} & H & \phi \\ 
        \hline
        2 & 2 & 3 & -8 & -5 & -7 & -3 & -8 & -6 & 2 & 1 \\ 
    \end{array}\right)$ \\ 
    \hline
    \begin{tabular}{l} $\mathcal{O}\left(1\right)$ coeff. \end{tabular} & $y^{l} \simeq
    \left(\begin{array}{rrr}
        0.406 & 0.979 & 0.625 \\
        0.816 & 1.028 & -1.222 \\
        -0.890 & -1.556 & -1.123
    \end{array}\right) \ ,\ 
    y^{\nu} \simeq
    \left(\begin{array}{rrr}
        -1.140 & -0.935 & -1.551 \\
        -0.897 & 0.761 & 1.083 \\
        -0.902 & -1.061 & 1.135
    \end{array}\right)$ \\ 
    & $y^{N} \simeq
    \left(\begin{array}{rrr}
        -1.397 & -1.127 & -1.309 \\
        -1.127 & 0.516 & -1.077 \\
        -1.309 & -1.077 & -1.393
    \end{array}\right)$ \\ 
    \hline
    \begin{tabular}{l} Intrinsic value \end{tabular} & $\mathcal{V}_{\mathrm{opt}}\simeq-0.578$ \\ 
    \hline
    \begin{tabular}{l} Masses\\(output) \end{tabular} & \begin{tabular}{l} $\left(\begin{array}{lll}
    m_{e}/\MeV & m_{\mu}/\GeV & m_{\tau}/\GeV\\
    m_{\nu_{1}}/\meV & m_{\nu_{2}}/\meV & m_{\nu_{3}}/\meV
    \end{array}\right)
    \simeq \left(\begin{array}{lll}
    0.35035 & 0.057377 & 2.70682 \\
    8.581\times 10^{-5} & 6.419\times 10^{-4} & 1.336\times 10^{-2}
    \end{array}\right)$ \end{tabular} \\
    \hline
    \begin{tabular}{l} PMNS matrix\\(output) \end{tabular} & $\left|V_{\mathrm{PMNS}}\right| \simeq
    \left(\begin{array}{lll}
    0.818 & 0.554 & 0.154 \\
    0.278 & 0.616 & 0.737 \\
    0.503 & 0.561 & 0.658
    \end{array}\right)$
    \\
    \hline
    \begin{tabular}{l} Mixing angles\\(output) \end{tabular}
    & $\left(\begin{array}{lll}
    \theta_{12} & \theta_{13} & \theta_{23}
    \end{array}\right)
    \simeq \left(\begin{array}{lll}
    0.189\pi & 0.049\pi & 0.268\pi
    \end{array}\right)$ \\
    \hline
\end{tabular}
}
\label{tab:lepton_15_15}
\end{table}

\vspace{\stretch{1}}
\newpage
\vspace*{\stretch{1}}

\begin{table}[H]
\centering
\scalebox{0.82}{
\begin{tabular}{l|c}
    \hline
    \begin{tabular}{l} Charges \end{tabular} &
    ${\cal Q}=\left(\begin{array}{ccc|ccc|ccc|cc}
        L_{1} & L_{2} & L_{3} & N_{1} & N_{2} & N_{3} & l_{1} & l_{2} & l_{3} & H & \phi \\ 
        \hline
        5 & 4 & 5 & -7 & -5 & -8 & -5 & -1 & -4 & 2 & 1 \\ 
    \end{array}\right)$ \\ 
    \hline
    \begin{tabular}{l} $\mathcal{O}\left(1\right)$ coeff. \end{tabular} & $y^{l} \simeq
    \left(\begin{array}{rrr}
        0.850 & -0.895 & 1.098 \\
        1.145 & 0.584 & 0.945 \\
        -1.057 & -0.914 & -0.950
    \end{array}\right) \ ,\ 
    y^{\nu} \simeq
    \left(\begin{array}{rrr}
        0.966 & -0.676 & -0.774 \\
        0.427 & -1.293 & 1.260 \\
        -1.075 & 0.965 & -0.950
    \end{array}\right)$ \\ 
    & $y^{N} \simeq
    \left(\begin{array}{rrr}
        1.116 & 0.968 & 1.203 \\
        0.968 & 1.136 & 1.075 \\
        1.203 & 1.075 & 1.059
    \end{array}\right)$ \\ 
    \hline
    \begin{tabular}{l} Intrinsic value \end{tabular} & $\mathcal{V}_{\mathrm{opt}}\simeq-0.637$ \\ 
    \hline
    \begin{tabular}{l} Masses\\(output) \end{tabular} & \begin{tabular}{l} $\left(\begin{array}{lll}
    m_{e}/\MeV & m_{\mu}/\GeV & m_{\tau}/\GeV\\
    m_{\nu_{1}}/\meV & m_{\nu_{2}}/\meV & m_{\nu_{3}}/\meV
    \end{array}\right)
    \simeq \left(\begin{array}{lll}
    0.54572 & 0.029910 & 1.92809 \\
    1.164\times 10^{-8} & 5.437\times 10^{-6} & 1.966\times 10^{-5}
    \end{array}\right)$ \end{tabular} \\
    \hline
    \begin{tabular}{l} PMNS matrix\\(output) \end{tabular} & $\left|V_{\mathrm{PMNS}}\right| \simeq
    \left(\begin{array}{lll}
    0.814 & 0.560 & 0.153 \\
    0.499 & 0.541 & 0.677 \\
    0.297 & 0.628 & 0.720
    \end{array}\right)$
    \\
    \hline
    \begin{tabular}{l} Mixing angles\\(output) \end{tabular}
    & $\left(\begin{array}{lll}
    \theta_{12} & \theta_{13} & \theta_{23}
    \end{array}\right)
    \simeq \left(\begin{array}{lll}
    0.192\pi & 0.049\pi & 0.240\pi
    \end{array}\right)$ \\
    \hline
\end{tabular}
}
\end{table}

\begin{table}[H]
\centering
\scalebox{0.82}{
\begin{tabular}{l|c}
    \hline
    \begin{tabular}{l} Charges \end{tabular} &
    ${\cal Q}=\left(\begin{array}{ccc|ccc|ccc|cc}
        L_{1} & L_{2} & L_{3} & N_{1} & N_{2} & N_{3} & l_{1} & l_{2} & l_{3} & H & \phi \\ 
        \hline
        4 & 3 & 4 & -1 & -3 & -2 & -7 & -3 & -4 & 2 & 1 \\ 
    \end{array}\right)$ \\ 
    \hline
    \begin{tabular}{l} $\mathcal{O}\left(1\right)$ coeff. \end{tabular} & $y^{l} \simeq
    \left(\begin{array}{rrr}
        -0.717 & 0.716 & 0.960 \\
        -1.300 & -1.192 & 1.586 \\
        -1.240 & -0.578 & 1.225
    \end{array}\right) \ ,\ 
    y^{\nu} \simeq
    \left(\begin{array}{rrr}
        -1.022 & -1.462 & 0.763 \\
        -0.515 & -0.946 & -0.931 \\
        -0.924 & 1.167 & 0.668
    \end{array}\right)$ \\ 
    & $y^{N} \simeq
    \left(\begin{array}{rrr}
        1.081 & 1.196 & 1.299 \\
        1.196 & 1.172 & -0.455 \\
        1.299 & -0.455 & -0.887
    \end{array}\right)$ \\ 
    \hline
    \begin{tabular}{l} Intrinsic value \end{tabular} & $\mathcal{V}_{\mathrm{opt}}\simeq-0.710$ \\ 
    \hline
    \begin{tabular}{l} Masses\\(output) \end{tabular} & \begin{tabular}{l} $\left(\begin{array}{lll}
    m_{e}/\MeV & m_{\mu}/\GeV & m_{\tau}/\GeV\\
    m_{\nu_{1}}/\meV & m_{\nu_{2}}/\meV & m_{\nu_{3}}/\meV
    \end{array}\right)
    \simeq \left(\begin{array}{lll}
    0.19203 & 0.088546 & 0.93953 \\
    5.174\times 10^{-6} & 7.689\times 10^{-6} & 1.482\times 10^{-5}
    \end{array}\right)$ \end{tabular} \\
    \hline
    \begin{tabular}{l} PMNS matrix\\(output) \end{tabular} & $\left|V_{\mathrm{PMNS}}\right| \simeq
    \left(\begin{array}{lll}
    0.809 & 0.569 & 0.145 \\
    0.338 & 0.654 & 0.676 \\
    0.480 & 0.498 & 0.722
    \end{array}\right)$
    \\
    \hline
    \begin{tabular}{l} Mixing angles\\(output) \end{tabular}
    & $\left(\begin{array}{lll}
    \theta_{12} & \theta_{13} & \theta_{23}
    \end{array}\right)
    \simeq \left(\begin{array}{lll}
    0.195\pi & 0.046\pi & 0.240\pi
    \end{array}\right)$ \\
    \hline
\end{tabular}
}
\label{tab:lepton_15_2}
\end{table}

\vspace{\stretch{1}}
\newpage
\vspace*{\stretch{1}}

\begin{table}[H]
\centering
\scalebox{0.82}{
\begin{tabular}{l|c}
    \hline
    \begin{tabular}{l} Charges \end{tabular} &
    ${\cal Q}=\left(\begin{array}{ccc|ccc|ccc|cc}
        L_{1} & L_{2} & L_{3} & N_{1} & N_{2} & N_{3} & l_{1} & l_{2} & l_{3} & H & \phi \\ 
        \hline
        8 & 8 & 9 & -3 & -8 & -6 & 2 & -1 & 1 & 2 & 1 \\ 
    \end{array}\right)$ \\ 
    \hline
    \begin{tabular}{l} $\mathcal{O}\left(1\right)$ coeff. \end{tabular} & $y^{l} \simeq
    \left(\begin{array}{rrr}
        0.948 & -1.089 & -0.508 \\
        1.817 & -1.052 & 1.125 \\
        1.488 & -1.128 & 1.333
    \end{array}\right) \ ,\ 
    y^{\nu} \simeq
    \left(\begin{array}{rrr}
        -0.804 & -1.092 & -0.790 \\
        -0.644 & 0.620 & -0.793 \\
        1.113 & -0.928 & 1.219
    \end{array}\right)$ \\ 
    & $y^{N} \simeq
    \left(\begin{array}{rrr}
        1.005 & -1.203 & -1.475 \\
        -1.203 & 0.550 & -1.382 \\
        -1.475 & -1.382 & -0.796
    \end{array}\right)$ \\ 
    \hline
    \begin{tabular}{l} Intrinsic value \end{tabular} & $\mathcal{V}_{\mathrm{opt}}\simeq-0.778$ \\ 
    \hline
    \begin{tabular}{l} Masses\\(output) \end{tabular} & \begin{tabular}{l} $\left(\begin{array}{lll}
    m_{e}/\MeV & m_{\mu}/\GeV & m_{\tau}/\GeV\\
    m_{\nu_{1}}/\meV & m_{\nu_{2}}/\meV & m_{\nu_{3}}/\meV
    \end{array}\right)
    \simeq \left(\begin{array}{lll}
    1.40507 & 0.184622 & 1.53861 \\
    1.124\times 10^{-14} & 3.755\times 10^{-11} & 3.824\times 10^{-11}
    \end{array}\right)$ \end{tabular} \\
    \hline
    \begin{tabular}{l} PMNS matrix\\(output) \end{tabular} & $\left|V_{\mathrm{PMNS}}\right| \simeq
    \left(\begin{array}{lll}
    0.806 & 0.574 & 0.144 \\
    0.323 & 0.632 & 0.705 \\
    0.495 & 0.522 & 0.695
    \end{array}\right)$
    \\
    \hline
    \begin{tabular}{l} Mixing angles\\(output) \end{tabular}
    & $\left(\begin{array}{lll}
    \theta_{12} & \theta_{13} & \theta_{23}
    \end{array}\right)
    \simeq \left(\begin{array}{lll}
    0.197\pi & 0.046\pi & 0.252\pi
    \end{array}\right)$ \\
    \hline
\end{tabular}
}
\label{tab:lepton_15_12}
\end{table}

\begin{table}[H]
\centering
\scalebox{0.82}{
\begin{tabular}{l|c}
    \hline
    \begin{tabular}{l} Charges \end{tabular} &
    ${\cal Q}=\left(\begin{array}{ccc|ccc|ccc|cc}
        L_{1} & L_{2} & L_{3} & N_{1} & N_{2} & N_{3} & l_{1} & l_{2} & l_{3} & H & \phi \\ 
        \hline
        4 & 4 & 5 & 0 & -9 & -2 & -2 & -5 & -3 & 2 & 1 \\ 
    \end{array}\right)$ \\ 
    \hline
    \begin{tabular}{l} $\mathcal{O}\left(1\right)$ coeff. \end{tabular} & $y^{l} \simeq
    \left(\begin{array}{rrr}
        0.996 & 0.700 & 0.367 \\
        1.067 & 0.802 & 1.142 \\
        0.964 & 1.077 & -1.064
    \end{array}\right) \ ,\ 
    y^{\nu} \simeq
    \left(\begin{array}{rrr}
        -0.947 & -0.389 & -0.819 \\
        -1.113 & 0.951 & 0.777 \\
        1.197 & -0.949 & -0.942
    \end{array}\right)$ \\ 
    & $y^{N} \simeq
    \left(\begin{array}{rrr}
        -1.015 & 0.769 & -0.747 \\
        0.769 & 0.765 & -0.894 \\
        -0.747 & -0.894 & -0.946
    \end{array}\right)$ \\ 
    \hline
    \begin{tabular}{l} Intrinsic value \end{tabular} & $\mathcal{V}_{\mathrm{opt}}\simeq-0.822$ \\ 
    \hline
    \begin{tabular}{l} Masses\\(output) \end{tabular} & \begin{tabular}{l} $\left(\begin{array}{lll}
    m_{e}/\MeV & m_{\mu}/\GeV & m_{\tau}/\GeV\\
    m_{\nu_{1}}/\meV & m_{\nu_{2}}/\meV & m_{\nu_{3}}/\meV
    \end{array}\right)
    \simeq \left(\begin{array}{lll}
    1.14824 & 0.125244 & 1.10475 \\
    1.517\times 10^{-9} & 1.953\times 10^{-6} & 6.610\times 10^{-6}
    \end{array}\right)$ \end{tabular} \\
    \hline
    \begin{tabular}{l} PMNS matrix\\(output) \end{tabular} & $\left|V_{\mathrm{PMNS}}\right| \simeq
    \left(\begin{array}{lll}
    0.823 & 0.547 & 0.154 \\
    0.465 & 0.491 & 0.737 \\
    0.328 & 0.678 & 0.658
    \end{array}\right)$
    \\
    \hline
    \begin{tabular}{l} Mixing angles\\(output) \end{tabular}
    & $\left(\begin{array}{lll}
    \theta_{12} & \theta_{13} & \theta_{23}
    \end{array}\right)
    \simeq \left(\begin{array}{lll}
    0.187\pi & 0.049\pi & 0.268\pi
    \end{array}\right)$ \\
    \hline
\end{tabular}
}
\label{tab:lepton_15_10}
\end{table}

\vspace{\stretch{1}}
\newpage
\vspace*{\stretch{1}}

\begin{table}[H]
\centering
\scalebox{0.82}{
\begin{tabular}{l|c}
    \hline
    \begin{tabular}{l} Charges \end{tabular} &
    ${\cal Q}=\left(\begin{array}{ccc|ccc|ccc|cc}
        L_{1} & L_{2} & L_{3} & N_{1} & N_{2} & N_{3} & l_{1} & l_{2} & l_{3} & H & \phi \\ 
        \hline
        1 & 1 & 2 & -1 & -7 & -1 & -5 & -7 & -6 & 2 & 1 \\ 
    \end{array}\right)$ \\ 
    \hline
    \begin{tabular}{l} $\mathcal{O}\left(1\right)$ coeff. \end{tabular} & $y^{l} \simeq
    \left(\begin{array}{rrr}
        0.369 & 0.760 & -1.134 \\
        -0.968 & -0.850 & -0.777 \\
        -1.004 & -0.839 & -0.776
    \end{array}\right) \ ,\ 
    y^{\nu} \simeq
    \left(\begin{array}{rrr}
        1.286 & -0.867 & 1.232 \\
        -0.895 & 1.418 & 0.894 \\
        -1.083 & -0.940 & 1.211
    \end{array}\right)$ \\ 
    & $y^{N} \simeq
    \left(\begin{array}{rrr}
        -0.494 & 0.849 & 1.207 \\
        0.849 & 1.357 & 0.957 \\
        1.207 & 0.957 & -1.132
    \end{array}\right)$ \\ 
    \hline
    \begin{tabular}{l} Intrinsic value \end{tabular} & $\mathcal{V}_{\mathrm{opt}}\simeq-0.830$ \\ 
    \hline
    \begin{tabular}{l} Masses\\(output) \end{tabular} & \begin{tabular}{l} $\left(\begin{array}{lll}
    m_{e}/\MeV & m_{\mu}/\GeV & m_{\tau}/\GeV\\
    m_{\nu_{1}}/\meV & m_{\nu_{2}}/\meV & m_{\nu_{3}}/\meV
    \end{array}\right)
    \simeq \left(\begin{array}{lll}
    0.59596 & 0.251105 & 0.78891 \\
    2.244\times 10^{-3} & 3.333\times 10^{-3} & 9.258\times 10^{-2}
    \end{array}\right)$ \end{tabular} \\
    \hline
    \begin{tabular}{l} PMNS matrix\\(output) \end{tabular} & $\left|V_{\mathrm{PMNS}}\right| \simeq
    \left(\begin{array}{lll}
    0.829 & 0.538 & 0.152 \\
    0.479 & 0.543 & 0.690 \\
    0.288 & 0.645 & 0.708
    \end{array}\right)$
    \\
    \hline
    \begin{tabular}{l} Mixing angles\\(output) \end{tabular}
    & $\left(\begin{array}{lll}
    \theta_{12} & \theta_{13} & \theta_{23}
    \end{array}\right)
    \simeq \left(\begin{array}{lll}
    0.183\pi & 0.049\pi & 0.246\pi
    \end{array}\right)$ \\
    \hline
\end{tabular}
}
\label{tab:lepton_15_22}
\end{table}

\begin{table}[H]
\centering
\scalebox{0.82}{
\begin{tabular}{l|c}
    \hline
    \begin{tabular}{l} Charges \end{tabular} &
    ${\cal Q}=\left(\begin{array}{ccc|ccc|ccc|cc}
        L_{1} & L_{2} & L_{3} & N_{1} & N_{2} & N_{3} & l_{1} & l_{2} & l_{3} & H & \phi \\ 
        \hline
        4 & 4 & 5 & -2 & -3 & -3 & -7 & -3 & -4 & 2 & 1 \\ 
    \end{array}\right)$ \\ 
    \hline
    \begin{tabular}{l} $\mathcal{O}\left(1\right)$ coeff. \end{tabular} & $y^{l} \simeq
    \left(\begin{array}{rrr}
        1.035& -0.908 & -1.208 \\
        1.088 & 1.092 & -1.579 \\
        -1.249 & 0.405 & -1.881
    \end{array}\right) \ ,\ 
    y^{\nu} \simeq
    \left(\begin{array}{rrr}
        0.946 & -0.940 & 1.418 \\
        0.888 & 1.071 & 1.011 \\
        -0.955 & -1.163 & 1.186
    \end{array}\right)$ \\ 
    & $y^{N} \simeq
    \left(\begin{array}{rrr}
        -0.942 & -1.342 & -1.523 \\
        -1.342 & 1.276 & 1.198 \\
        -1.523 & 1.198 & 0.919
    \end{array}\right)$ \\ 
    \hline
    \begin{tabular}{l} Intrinsic value \end{tabular} & $\mathcal{V}_{\mathrm{opt}}\simeq-0.868$ \\ 
    \hline
    \begin{tabular}{l} Masses\\(output) \end{tabular} & \begin{tabular}{l} $\left(\begin{array}{lll}
    m_{e}/\MeV & m_{\mu}/\GeV & m_{\tau}/\GeV\\
    m_{\nu_{1}}/\meV & m_{\nu_{2}}/\meV & m_{\nu_{3}}/\meV
    \end{array}\right)
    \simeq \left(\begin{array}{lll}
    0.51886 & 0.094567 & 0.26772 \\
    1.401\times 10^{-7} & 2.510\times 10^{-6} & 7.317\times 10^{-5}
    \end{array}\right)$ \end{tabular} \\
    \hline
    \begin{tabular}{l} PMNS matrix\\(output) \end{tabular} & $\left|V_{\mathrm{PMNS}}\right| \simeq
    \left(\begin{array}{lll}
    0.821 & 0.550 & 0.155 \\
    0.253 & 0.592 & 0.766 \\
    0.513 & 0.589 & 0.624
    \end{array}\right)$
    \\
    \hline
    \begin{tabular}{l} Mixing angles\\(output) \end{tabular}
    & $\left(\begin{array}{lll}
    \theta_{12} & \theta_{13} & \theta_{23}
    \end{array}\right)
    \simeq \left(\begin{array}{lll}
    0.188\pi & 0.049\pi & 0.282\pi
    \end{array}\right)$ \\
    \hline
\end{tabular}
}
\label{tab:lepton_15_6}
\end{table}

\vspace{\stretch{1}}
\newpage
\vspace*{\stretch{1}}

\begin{table}[H]
\centering
\scalebox{0.82}{
\begin{tabular}{l|c}
    \hline
    \begin{tabular}{l} Charges \end{tabular} &
    ${\cal Q}=\left(\begin{array}{ccc|ccc|ccc|cc}
        L_{1} & L_{2} & L_{3} & N_{1} & N_{2} & N_{3} & l_{1} & l_{2} & l_{3} & H & \phi \\ 
        \hline
        3 & 4 & 4 & -1 & -4 & -8 & -2 & -6 & -7 & 2 & 1 \\ 
    \end{array}\right)$ \\ 
    \hline
    \begin{tabular}{l} $\mathcal{O}\left(1\right)$ coeff. \end{tabular} & $y^{l} \simeq
    \left(\begin{array}{rrr}
        0.390 & -1.245 & -0.905 \\
        1.056 & 1.442 & 0.743 \\
        1.159 & 1.087 & 1.182
    \end{array}\right) \ ,\ 
    y^{\nu} \simeq
    \left(\begin{array}{rrr}
        1.001 & 1.276 & 1.324 \\
        -1.020 & -0.978 & -0.749 \\
        -0.818 & 1.201 & 0.999
    \end{array}\right)$ \\ 
    & $y^{N} \simeq
    \left(\begin{array}{rrr}
        0.858 & -0.914 & -0.839 \\
        -0.914 & 0.854 & -0.988 \\
        -0.839 & -0.988 & 1.293
    \end{array}\right)$ \\ 
    \hline
    \begin{tabular}{l} Intrinsic value \end{tabular} & $\mathcal{V}_{\mathrm{opt}}\simeq-0.949$ \\ 
    \hline
    \begin{tabular}{l} Masses\\(output) \end{tabular} & \begin{tabular}{l} $\left(\begin{array}{lll}
    m_{e}/\MeV & m_{\mu}/\GeV & m_{\tau}/\GeV\\
    m_{\nu_{1}}/\meV & m_{\nu_{2}}/\meV & m_{\nu_{3}}/\meV
    \end{array}\right)
    \simeq \left(\begin{array}{lll}
    0.38264 & 0.014974 & 1.60993 \\
    4.912\times 10^{-8} & 2.981\times 10^{-6} & 1.844\times 10^{-4}
    \end{array}\right)$ \end{tabular} \\
    \hline
    \begin{tabular}{l} PMNS matrix\\(output) \end{tabular} & $\left|V_{\mathrm{PMNS}}\right| \simeq
    \left(\begin{array}{lll}
    0.823 & 0.548 & 0.151 \\
    0.258 & 0.597 & 0.760 \\
    0.506 & 0.587 & 0.632
    \end{array}\right)$
    \\
    \hline
    \begin{tabular}{l} Mixing angles\\(output) \end{tabular}
    & $\left(\begin{array}{lll}
    \theta_{12} & \theta_{13} & \theta_{23}
    \end{array}\right)
    \simeq \left(\begin{array}{lll}
    0.187\pi & 0.048\pi & 0.279\pi
    \end{array}\right)$ \\
    \hline
\end{tabular}
}
\label{tab:lepton_15_18}
\end{table}

\begin{table}[H]
\centering
\scalebox{0.82}{
\begin{tabular}{l|c}
    \hline
    \begin{tabular}{l} Charges \end{tabular} &
    ${\cal Q}=\left(\begin{array}{ccc|ccc|ccc|cc}
        L_{1} & L_{2} & L_{3} & N_{1} & N_{2} & N_{3} & l_{1} & l_{2} & l_{3} & H & \phi \\ 
        \hline
        6 & 6 & 7 & -5 & -9 & 0 & -3 & -1 & -1 & 2 & 1 \\ 
    \end{array}\right)$ \\ 
    \hline
    \begin{tabular}{l} $\mathcal{O}\left(1\right)$ coeff. \end{tabular} & $y^{l} \simeq
    \left(\begin{array}{rrr}
        -0.546 & 1.157 & -0.515 \\
        0.874 & -1.025 & 1.105 \\
        -0.883 & 0.700 & -1.044
    \end{array}\right) \ ,\ 
    y^{\nu} \simeq
    \left(\begin{array}{rrr}
        -1.006 & 0.713 & -1.098 \\
        0.847 & 0.822 & 1.051 \\
        1.290 & 1.127 & -1.126
    \end{array}\right)$ \\ 
    & $y^{N} \simeq
    \left(\begin{array}{rrr}
        -0.985 & -1.079 & -1.145 \\
        -1.079 & -1.029 & -1.071 \\
        -1.145 & -1.071 & -0.604
    \end{array}\right)$ \\ 
    \hline
    \begin{tabular}{l} Intrinsic value \end{tabular} & $\mathcal{V}_{\mathrm{opt}}\simeq-0.954$ \\ 
    \hline
    \begin{tabular}{l} Masses\\(output) \end{tabular} & \begin{tabular}{l} $\left(\begin{array}{lll}
    m_{e}/\MeV & m_{\mu}/\GeV & m_{\tau}/\GeV\\
    m_{\nu_{1}}/\meV & m_{\nu_{2}}/\meV & m_{\nu_{3}}/\meV
    \end{array}\right)
    \simeq \left(\begin{array}{lll}
    0.33117 & 0.075216 & 0.36599 \\
    5.619\times 10^{-9} & 7.186\times 10^{-9} & 2.842\times 10^{-7}
    \end{array}\right)$ \end{tabular} \\
    \hline
    \begin{tabular}{l} PMNS matrix\\(output) \end{tabular} & $\left|V_{\mathrm{PMNS}}\right| \simeq
    \left(\begin{array}{lll}
    0.823 & 0.548 & 0.148 \\
    0.466 & 0.502 & 0.729 \\
    0.326 & 0.669 & 0.668
    \end{array}\right)$
    \\
    \hline
    \begin{tabular}{l} Mixing angles\\(output) \end{tabular}
    & $\left(\begin{array}{lll}
    \theta_{12} & \theta_{13} & \theta_{23}
    \end{array}\right)
    \simeq \left(\begin{array}{lll}
    0.187\pi & 0.047\pi & 0.264\pi
    \end{array}\right)$ \\
    \hline
\end{tabular}
}
\label{tab:lepton_15_13}
\end{table}

\vspace{\stretch{1}}
\newpage
\vspace*{\stretch{1}}

\begin{table}[H]
\centering
\scalebox{0.82}{
\begin{tabular}{l|c}
    \hline
    \begin{tabular}{l} Charges \end{tabular} &
    ${\cal Q}=\left(\begin{array}{ccc|ccc|ccc|cc}
        L_{1} & L_{2} & L_{3} & N_{1} & N_{2} & N_{3} & l_{1} & l_{2} & l_{3} & H & \phi \\ 
        \hline
        4 & 4 & 5 & -7 & -7 & -6 & -6 & -3 & -4 & 2 & 1 \\ 
    \end{array}\right)$ \\ 
    \hline
    \begin{tabular}{l} $\mathcal{O}\left(1\right)$ coeff. \end{tabular} & $y^{l} \simeq
    \left(\begin{array}{rrr}
        -1.156 & -1.212 & -0.915 \\
        -0.706 & 0.674 & -1.390 \\
        -1.084 & 1.150 & -1.238
    \end{array}\right) \ ,\ 
    y^{\nu} \simeq
    \left(\begin{array}{rrr}
        -0.885 & 1.215 & -1.136 \\
        -0.995 & -0.873 & -1.130 \\
        0.847 & -1.302 & -0.920
    \end{array}\right)$ \\ 
    & $y^{N} \simeq
    \left(\begin{array}{rrr}
        0.907 & 0.839 & 1.212 \\
        0.839 & -1.634 & -0.799 \\
        1.212 & -0.799 & 0.696
    \end{array}\right)$ \\ 
    \hline
    \begin{tabular}{l} Intrinsic value \end{tabular} & $\mathcal{V}_{\mathrm{opt}}\simeq-0.958$ \\ 
    \hline
    \begin{tabular}{l} Masses\\(output) \end{tabular} & \begin{tabular}{l} $\left(\begin{array}{lll}
    m_{e}/\MeV & m_{\mu}/\GeV & m_{\tau}/\GeV\\
    m_{\nu_{1}}/\meV & m_{\nu_{2}}/\meV & m_{\nu_{3}}/\meV
    \end{array}\right)
    \simeq \left(\begin{array}{lll}
    0.51912 & 0.080900 & 0.26552 \\
    4.688\times 10^{-7} & 2.551\times 10^{-6} & 7.757\times 10^{-6}
    \end{array}\right)$ \end{tabular} \\
    \hline
    \begin{tabular}{l} PMNS matrix\\(output) \end{tabular} & $\left|V_{\mathrm{PMNS}}\right| \simeq
    \left(\begin{array}{lll}
    0.833 & 0.534 & 0.143 \\
    0.303 & 0.657 & 0.690 \\
    0.462 & 0.532 & 0.710
    \end{array}\right)$
    \\
    \hline
    \begin{tabular}{l} Mixing angles\\(output) \end{tabular}
    & $\left(\begin{array}{lll}
    \theta_{12} & \theta_{13} & \theta_{23}
    \end{array}\right)
    \simeq \left(\begin{array}{lll}
    0.181\pi & 0.046\pi & 0.246\pi
    \end{array}\right)$ \\
    \hline
\end{tabular}
}
\label{tab:lepton_15_8}
\end{table}

\begin{table}[H]
\centering
\scalebox{0.82}{
\begin{tabular}{l|c}
    \hline
    \begin{tabular}{l} Charges \end{tabular} &
    ${\cal Q}=\left(\begin{array}{ccc|ccc|ccc|cc}
        L_{1} & L_{2} & L_{3} & N_{1} & N_{2} & N_{3} & l_{1} & l_{2} & l_{3} & H & \phi \\ 
        \hline
        4 & 4 & 5 & -4 & 0 & -7 & -7 & -4 & -3 & 2 & 1 \\ 
    \end{array}\right)$ \\ 
    \hline
    \begin{tabular}{l} $\mathcal{O}\left(1\right)$ coeff. \end{tabular} & $y^{l} \simeq
    \left(\begin{array}{rrr}
        -1.166 & 1.411 & -1.216 \\
        -0.939 & 0.645 & 1.066 \\
        -1.001 & -0.892 & -1.028
    \end{array}\right) \ ,\ 
    y^{\nu} \simeq
    \left(\begin{array}{rrr}
        1.091 & -1.437 & -0.851 \\
        -0.971 & -0.766 & 1.027 \\
        -0.585 & -1.023 & 1.345
    \end{array}\right)$ \\ 
    & $y^{N} \simeq
    \left(\begin{array}{rrr}
        1.048 & -0.507 & 1.165 \\
        -0.507 & 1.182 & 1.298 \\
        1.165 & 1.298 & -1.406
    \end{array}\right)$ \\ 
    \hline
    \begin{tabular}{l} Intrinsic value \end{tabular} & $\mathcal{V}_{\mathrm{opt}}\simeq-0.958$ \\ 
    \hline
    \begin{tabular}{l} Masses\\(output) \end{tabular} & \begin{tabular}{l} $\left(\begin{array}{lll}
    m_{e}/\MeV & m_{\mu}/\GeV & m_{\tau}/\GeV\\
    m_{\nu_{1}}/\meV & m_{\nu_{2}}/\meV & m_{\nu_{3}}/\meV
    \end{array}\right)
    \simeq \left(\begin{array}{lll}
    0.42420 & 0.069048 & 0.30810 \\
    2.754\times 10^{-8} & 1.989\times 10^{-6} & 4.726\times 10^{-6}
    \end{array}\right)$ \end{tabular} \\
    \hline
    \begin{tabular}{l} PMNS matrix\\(output) \end{tabular} & $\left|V_{\mathrm{PMNS}}\right| \simeq
    \left(\begin{array}{lll}
    0.807 & 0.573 & 0.145 \\
    0.485 & 0.502 & 0.716 \\
    0.338 & 0.648 & 0.683
    \end{array}\right)$
    \\
    \hline
    \begin{tabular}{l} Mixing angles\\(output) \end{tabular}
    & $\left(\begin{array}{lll}
    \theta_{12} & \theta_{13} & \theta_{23}
    \end{array}\right)
    \simeq \left(\begin{array}{lll}
    0.197\pi & 0.046\pi & 0.258\pi
    \end{array}\right)$ \\
    \hline
\end{tabular}
}
\label{tab:lepton_15_11}
\end{table}

\vspace{\stretch{1}}
\newpage
\vspace*{\stretch{1}}

\begin{table}[H]
\centering
\scalebox{0.82}{
\begin{tabular}{l|c}
    \hline
    \begin{tabular}{l} Charges \end{tabular} &
    ${\cal Q}=\left(\begin{array}{ccc|ccc|ccc|cc}
        L_{1} & L_{2} & L_{3} & N_{1} & N_{2} & N_{3} & l_{1} & l_{2} & l_{3} & H & \phi \\ 
        \hline
        7 & 6 & 7 & -4 & -7 & -7 & 0 & 2 & -5 & 2 & 1 \\ 
    \end{array}\right)$ \\ 
    \hline
    \begin{tabular}{l} $\mathcal{O}\left(1\right)$ coeff. \end{tabular} & $y^{l} \simeq
    \left(\begin{array}{rrr}
        -0.615 & 1.039 & -1.143 \\
        0.477 & 0.677 & -1.155 \\
        -1.099 & -1.004 & -0.832
    \end{array}\right) \ ,\ 
    y^{\nu} \simeq
    \left(\begin{array}{rrr}
        -0.985 & -1.396 & -0.960 \\
        0.657 & 1.157 & -1.294 \\
        -0.703 & -0.948 & -1.041
    \end{array}\right)$ \\ 
    & $y^{N} \simeq
    \left(\begin{array}{rrr}
        -1.006 & 0.882 & 0.558 \\
        0.882 & -1.056 & -0.847 \\
        0.558 & -0.847 & 0.589
    \end{array}\right)$ \\ 
    \hline
    \begin{tabular}{l} Intrinsic value \end{tabular} & $\mathcal{V}_{\mathrm{opt}}\simeq-1.112$ \\ 
    \hline
    \begin{tabular}{l} Masses\\(output) \end{tabular} & \begin{tabular}{l} $\left(\begin{array}{lll}
    m_{e}/\MeV & m_{\mu}/\GeV & m_{\tau}/\GeV\\
    m_{\nu_{1}}/\meV & m_{\nu_{2}}/\meV & m_{\nu_{3}}/\meV
    \end{array}\right)
    \simeq \left(\begin{array}{lll}
    0.47722 & 0.255129 & 8.71773 \\
    3.531\times 10^{-15} & 9.909\times 10^{-9} & 6.345\times 10^{-8}
    \end{array}\right)$ \end{tabular} \\
    \hline
    \begin{tabular}{l} PMNS matrix\\(output) \end{tabular} & $\left|V_{\mathrm{PMNS}}\right| \simeq
    \left(\begin{array}{lll}
    0.812 & 0.565 & 0.147 \\
    0.290 & 0.609 & 0.739 \\
    0.507 & 0.557 & 0.658
    \end{array}\right)$
    \\
    \hline
    \begin{tabular}{l} Mixing angles\\(output) \end{tabular}
    & $\left(\begin{array}{lll}
    \theta_{12} & \theta_{13} & \theta_{23}
    \end{array}\right)
    \simeq \left(\begin{array}{lll}
    0.193\pi & 0.047\pi & 0.268\pi
    \end{array}\right)$ \\
    \hline
\end{tabular}
}
\label{tab:lepton_15_3}
\end{table}

\begin{table}[H]
\centering
\scalebox{0.82}{
\begin{tabular}{l|c}
    \hline
    \begin{tabular}{l} Charges \end{tabular} &
    ${\cal Q}=\left(\begin{array}{ccc|ccc|ccc|cc}
        L_{1} & L_{2} & L_{3} & N_{1} & N_{2} & N_{3} & l_{1} & l_{2} & l_{3} & H & \phi \\ 
        \hline
        6 & 7 & 7 & -6 & -8 & -8 & 2 & 0 & -4 & 2 & 1 \\ 
    \end{array}\right)$ \\ 
    \hline
    \begin{tabular}{l} $\mathcal{O}\left(1\right)$ coeff. \end{tabular} & $y^{l} \simeq
    \left(\begin{array}{rrr}
        0.695 & 0.703 & 0.702 \\
        -1.072 & -1.293 & 0.816 \\
        1.013 & 0.934 & 1.073
    \end{array}\right) \ ,\ 
    y^{\nu} \simeq
    \left(\begin{array}{rrr}
        1.355 & -0.705 & 1.423 \\
        0.986 & -1.266 & 0.730 \\
        0.788 & -0.984 & 1.567
    \end{array}\right)$ \\ 
    & $y^{N} \simeq
    \left(\begin{array}{rrr}
        1.103 & 1.093 & 0.772 \\
        1.093 & -1.115 & 0.905 \\
        0.772 & 0.905 & -1.038
    \end{array}\right)$ \\ 
    \hline
    \begin{tabular}{l} Intrinsic value \end{tabular} & $\mathcal{V}_{\mathrm{opt}}\simeq-1.122$ \\ 
    \hline
    \begin{tabular}{l} Masses\\(output) \end{tabular} & \begin{tabular}{l} $\left(\begin{array}{lll}
    m_{e}/\MeV & m_{\mu}/\GeV & m_{\tau}/\GeV\\
    m_{\nu_{1}}/\meV & m_{\nu_{2}}/\meV & m_{\nu_{3}}/\meV
    \end{array}\right)
    \simeq \left(\begin{array}{lll}
    0.50280 & 0.041922 & 8.95755 \\
    2.236\times 10^{-10} & 8.061\times 10^{-10} & 5.388\times 10^{-9}
    \end{array}\right)$ \end{tabular} \\
    \hline
    \begin{tabular}{l} PMNS matrix\\(output) \end{tabular} & $\left|V_{\mathrm{PMNS}}\right| \simeq
    \left(\begin{array}{lll}
    0.818 & 0.556 & 0.147 \\
    0.304 & 0.635 & 0.710 \\
    0.489 & 0.536 & 0.688
    \end{array}\right)$
    \\
    \hline
    \begin{tabular}{l} Mixing angles\\(output) \end{tabular}
    & $\left(\begin{array}{lll}
    \theta_{12} & \theta_{13} & \theta_{23}
    \end{array}\right)
    \simeq \left(\begin{array}{lll}
    0.190\pi & 0.047\pi & 0.255\pi
    \end{array}\right)$ \\
    \hline
\end{tabular}
}
\label{tab:lepton_15_4}
\end{table}

\vspace{\stretch{1}}
\newpage
\vspace*{\stretch{1}}

\begin{table}[H]
\centering
\scalebox{0.82}{
\begin{tabular}{l|c}
    \hline
    \begin{tabular}{l} Charges \end{tabular} &
    ${\cal Q}=\left(\begin{array}{ccc|ccc|ccc|cc}
        L_{1} & L_{2} & L_{3} & N_{1} & N_{2} & N_{3} & l_{1} & l_{2} & l_{3} & H & \phi \\ 
        \hline
        4 & 2 & 2 & -5 & -8 & 0 & -8 & -3 & -7 & 2 & 1 \\ 
    \end{array}\right)$ \\ 
    \hline
    \begin{tabular}{l} $\mathcal{O}\left(1\right)$ coeff. \end{tabular} & $y^{l} \simeq
    \left(\begin{array}{rrr}
        -0.993 & -1.019 & -0.765 \\
        1.091 & 1.168 & -1.515 \\
        -1.132 & 0.729 & 0.885
    \end{array}\right) \ ,\ 
    y^{\nu} \simeq
    \left(\begin{array}{rrr}
        1.556 & 0.977 & -0.725 \\
        1.116 & -0.906 & 1.181 \\
        0.395 & -1.202 & -0.935
    \end{array}\right)$ \\ 
    & $y^{N} \simeq
    \left(\begin{array}{rrr}
        1.288 & 0.347 & 1.142 \\
        0.347 & -1.248 & 0.900 \\
        1.142 & 0.900 & 0.967
    \end{array}\right)$ \\ 
    \hline
    \begin{tabular}{l} Intrinsic value \end{tabular} & $\mathcal{V}_{\mathrm{opt}}\simeq-1.144$ \\ 
    \hline
    \begin{tabular}{l} Masses\\(output) \end{tabular} & \begin{tabular}{l} $\left(\begin{array}{lll}
    m_{e}/\MeV & m_{\mu}/\GeV & m_{\tau}/\GeV\\
    m_{\nu_{1}}/\meV & m_{\nu_{2}}/\meV & m_{\nu_{3}}/\meV
    \end{array}\right)
    \simeq \left(\begin{array}{lll}
    0.41096 & 0.019511 & 3.97394 \\
    5.937\times 10^{-5} & 1.185\times 10^{-4} & 8.285\times 10^{-3}
    \end{array}\right)$ \end{tabular} \\
    \hline
    \begin{tabular}{l} PMNS matrix\\(output) \end{tabular} & $\left|V_{\mathrm{PMNS}}\right| \simeq
    \left(\begin{array}{lll}
    0.826 & 0.545 & 0.145 \\
    0.263 & 0.600 & 0.756 \\
    0.499 & 0.586 & 0.638
    \end{array}\right)$
    \\
    \hline
    \begin{tabular}{l} Mixing angles\\(output) \end{tabular}
    & $\left(\begin{array}{lll}
    \theta_{12} & \theta_{13} & \theta_{23}
    \end{array}\right)
    \simeq \left(\begin{array}{lll}
    0.186\pi & 0.046\pi & 0.277\pi
    \end{array}\right)$ \\
    \hline
\end{tabular}
}
\label{tab:lepton_15_14}
\end{table}

\begin{table}[H]
\centering
\scalebox{0.82}{
\begin{tabular}{l|c}
    \hline
    \begin{tabular}{l} Charges \end{tabular} &
    ${\cal Q}=\left(\begin{array}{ccc|ccc|ccc|cc}
        L_{1} & L_{2} & L_{3} & N_{1} & N_{2} & N_{3} & l_{1} & l_{2} & l_{3} & H & \phi \\ 
        \hline
        4 & 4 & 5 & -1 & 0 & -6 & -7 & -3 & -4 & 2 & 1 \\ 
    \end{array}\right)$ \\ 
    \hline
    \begin{tabular}{l} $\mathcal{O}\left(1\right)$ coeff. \end{tabular} & $y^{l} \simeq
    \left(\begin{array}{rrr}
        -1.029 & 1.092 & 1.035 \\
        1.138 & 0.861 & -0.925 \\
        -0.784 & -1.103 & -0.706
    \end{array}\right) \ ,\ 
    y^{\nu} \simeq
    \left(\begin{array}{rrr}
        1.088 & -1.299 & 0.974 \\
        -0.889 & -0.740 & 1.129 \\
        1.169 & 1.114 & 1.271
    \end{array}\right)$ \\ 
    & $y^{N} \simeq
    \left(\begin{array}{rrr}
        -0.972 & 0.902 & -1.050 \\
        0.902 & -0.775 & 1.136 \\
        -1.050 & 1.136 & -0.839
    \end{array}\right)$ \\ 
    \hline
    \begin{tabular}{l} Intrinsic value \end{tabular} & $\mathcal{V}_{\mathrm{opt}}\simeq-1.220$ \\ 
    \hline
    \begin{tabular}{l} Masses\\(output) \end{tabular} & \begin{tabular}{l} $\left(\begin{array}{lll}
    m_{e}/\MeV & m_{\mu}/\GeV & m_{\tau}/\GeV\\
    m_{\nu_{1}}/\meV & m_{\nu_{2}}/\meV & m_{\nu_{3}}/\meV
    \end{array}\right)
    \simeq \left(\begin{array}{lll}
    0.28375 & 0.065479 & 0.26595 \\
    1.923\times 10^{-6} & 2.699\times 10^{-6} & 6.037\times 10^{-4}
    \end{array}\right)$ \end{tabular} \\
    \hline
    \begin{tabular}{l} PMNS matrix\\(output) \end{tabular} & $\left|V_{\mathrm{PMNS}}\right| \simeq
    \left(\begin{array}{lll}
    0.839 & 0.524 & 0.145 \\
    0.452 & 0.525 & 0.721 \\
    0.302 & 0.671 & 0.678
    \end{array}\right)$
    \\
    \hline
    \begin{tabular}{l} Mixing angles\\(output) \end{tabular}
    & $\left(\begin{array}{lll}
    \theta_{12} & \theta_{13} & \theta_{23}
    \end{array}\right)
    \simeq \left(\begin{array}{lll}
    0.178\pi & 0.046\pi & 0.260\pi
    \end{array}\right)$ \\
    \hline
\end{tabular}
}
\label{tab:lepton_15_7}
\end{table}

\vspace{\stretch{1}}
\newpage
\vspace*{\stretch{1}}

\begin{table}[H]
\centering
\scalebox{0.82}{
\begin{tabular}{l|c}
    \hline
    \begin{tabular}{l} Charges \end{tabular} &
    ${\cal Q}=\left(\begin{array}{ccc|ccc|ccc|cc}
        L_{1} & L_{2} & L_{3} & N_{1} & N_{2} & N_{3} & l_{1} & l_{2} & l_{3} & H & \phi \\ 
        \hline
        5 & 5 & 6 & -1 & -6 & -5 & -5 & -2 & -3 & 2 & 1 \\ 
    \end{array}\right)$ \\ 
    \hline
    \begin{tabular}{l} $\mathcal{O}\left(1\right)$ coeff. \end{tabular} & $y^{l} \simeq
    \left(\begin{array}{rrr}
        1.165 & 1.350 & -0.629 \\
        -1.176 & -0.856 & -1.143 \\
        1.718 & 0.740 & -0.643
    \end{array}\right) \ ,\ 
    y^{\nu} \simeq
    \left(\begin{array}{rrr}
        -1.354 & -0.963 & 1.316 \\
        -0.626 & -1.025 & -0.999 \\
        -1.139 & 0.651 & -0.451
    \end{array}\right)$ \\ 
    & $y^{N} \simeq
    \left(\begin{array}{rrr}
        -0.712 & -0.744 & -1.013 \\
        -0.744 & 1.545 & 0.950 \\
        -1.013 & 0.950 & -0.926
    \end{array}\right)$ \\ 
    \hline
    \begin{tabular}{l} Intrinsic value \end{tabular} & $\mathcal{V}_{\mathrm{opt}}\simeq-1.237$ \\ 
    \hline
    \begin{tabular}{l} Masses\\(output) \end{tabular} & \begin{tabular}{l} $\left(\begin{array}{lll}
    m_{e}/\MeV & m_{\mu}/\GeV & m_{\tau}/\GeV\\
    m_{\nu_{1}}/\meV & m_{\nu_{2}}/\meV & m_{\nu_{3}}/\meV
    \end{array}\right)
    \simeq \left(\begin{array}{lll}
    0.90616 & 0.062690 & 0.30145 \\
    2.554\times 10^{-8} & 5.640\times 10^{-8} & 1.407\times 10^{-6}
    \end{array}\right)$ \end{tabular} \\
    \hline
    \begin{tabular}{l} PMNS matrix\\(output) \end{tabular} & $\left|V_{\mathrm{PMNS}}\right| \simeq
    \left(\begin{array}{lll}
    0.801 & 0.580 & 0.147 \\
    0.346 & 0.650 & 0.676 \\
    0.488 & 0.491 & 0.722
    \end{array}\right)$
    \\
    \hline
    \begin{tabular}{l} Mixing angles\\(output) \end{tabular}
    & $\left(\begin{array}{lll}
    \theta_{12} & \theta_{13} & \theta_{23}
    \end{array}\right)
    \simeq \left(\begin{array}{lll}
    0.199\pi & 0.047\pi & 0.240\pi
    \end{array}\right)$ \\
    \hline
\end{tabular}
}
\label{tab:lepton_15_9}
\end{table}

\begin{table}[H]
\centering
\scalebox{0.82}{
\begin{tabular}{l|c}
    \hline
    \begin{tabular}{l} Charges \end{tabular} &
    ${\cal Q}=\left(\begin{array}{ccc|ccc|ccc|cc}
        L_{1} & L_{2} & L_{3} & N_{1} & N_{2} & N_{3} & l_{1} & l_{2} & l_{3} & H & \phi \\ 
        \hline
        4 & 4 & 5 & -5 & -9 & -5 & -1 & -7 & -5 & 2 & 1 \\ 
    \end{array}\right)$ \\ 
    \hline
    \begin{tabular}{l} $\mathcal{O}\left(1\right)$ coeff. \end{tabular} & $y^{l} \simeq
    \left(\begin{array}{rrr}
        -0.725 & -0.861 & -0.907 \\
        1.049 & 0.620 & -1.094 \\
        -1.168 & 1.079 & -0.770
    \end{array}\right) \ ,\ 
    y^{\nu} \simeq
    \left(\begin{array}{rrr}
        0.844 & 0.937 & -0.705 \\
        -1.417 & 0.903 & -1.098 \\
        1.359 & -1.399 & -0.783
    \end{array}\right)$ \\ 
    & $y^{N} \simeq
    \left(\begin{array}{rrr}
        -1.024 & -0.748 & -1.253 \\
        -0.748 & -1.321 & 0.791 \\
        -1.253 & 0.791 & 1.241
    \end{array}\right)$ \\ 
    \hline
    \begin{tabular}{l} Intrinsic value \end{tabular} & $\mathcal{V}_{\mathrm{opt}}\simeq-1.335$ \\ 
    \hline
    \begin{tabular}{l} Masses\\(output) \end{tabular} & \begin{tabular}{l} $\left(\begin{array}{lll}
    m_{e}/\MeV & m_{\mu}/\GeV & m_{\tau}/\GeV\\
    m_{\nu_{1}}/\meV & m_{\nu_{2}}/\meV & m_{\nu_{3}}/\meV
    \end{array}\right)
    \simeq \left(\begin{array}{lll}
    0.43970 & 0.016993 & 3.77445 \\
    3.177\times 10^{-7} & 1.472\times 10^{-6} & 7.500\times 10^{-6}
    \end{array}\right)$ \end{tabular} \\
    \hline
    \begin{tabular}{l} PMNS matrix\\(output) \end{tabular} & $\left|V_{\mathrm{PMNS}}\right| \simeq
    \left(\begin{array}{lll}
    0.843 & 0.519 & 0.143 \\
    0.466 & 0.571 & 0.676 \\
    0.269 & 0.636 & 0.723
    \end{array}\right)$
    \\
    \hline
    \begin{tabular}{l} Mixing angles\\(output) \end{tabular}
    & $\left(\begin{array}{lll}
    \theta_{12} & \theta_{13} & \theta_{23}
    \end{array}\right)
    \simeq \left(\begin{array}{lll}
    0.176\pi & 0.046\pi & 0.239\pi
    \end{array}\right)$ \\
    \hline
\end{tabular}
}
\label{tab:lepton_15_21}
\end{table}

\vspace{\stretch{1}}
\newpage
\vspace*{\stretch{1}}

\begin{table}[H]
\centering
\scalebox{0.82}{
\begin{tabular}{l|c}
    \hline
    \begin{tabular}{l} Charges \end{tabular} &
    ${\cal Q}=\left(\begin{array}{ccc|ccc|ccc|cc}
        L_{1} & L_{2} & L_{3} & N_{1} & N_{2} & N_{3} & l_{1} & l_{2} & l_{3} & H & \phi \\ 
        \hline
        3 & 3 & 4 & -1 & -4 & -5 & -8 & -2 & -1 & 2 & 1 \\ 
    \end{array}\right)$ \\ 
    \hline
    \begin{tabular}{l} $\mathcal{O}\left(1\right)$ coeff. \end{tabular} & $y^{l} \simeq
    \left(\begin{array}{rrr}
        -0.995 & 0.891 & 0.892 \\
        -1.111 & 1.040 & 0.904 \\
        -0.835 & -1.267 & -1.245
    \end{array}\right) \ ,\ 
    y^{\nu} \simeq
    \left(\begin{array}{rrr}
        1.145 & 0.929 & 0.803 \\
        -1.055 & 0.881 & 0.641 \\
        -1.163 & -1.461 & 0.902
    \end{array}\right)$ \\ 
    & $y^{N} \simeq
    \left(\begin{array}{rrr}
        1.153 & -0.802 & 1.079 \\
        -0.802 & -1.238 & 1.087 \\
        1.079 & 1.087 & 1.255
    \end{array}\right)$ \\ 
    \hline
    \begin{tabular}{l} Intrinsic value \end{tabular} & $\mathcal{V}_{\mathrm{opt}}\simeq-1.433$ \\ 
    \hline
    \begin{tabular}{l} Masses\\(output) \end{tabular} & \begin{tabular}{l} $\left(\begin{array}{lll}
    m_{e}/\MeV & m_{\mu}/\GeV & m_{\tau}/\GeV\\
    m_{\nu_{1}}/\meV & m_{\nu_{2}}/\meV & m_{\nu_{3}}/\meV
    \end{array}\right)
    \simeq \left(\begin{array}{lll}
    0.43090 & 0.274049 & 15.33941 \\
    6.446\times 10^{-6} & 2.046\times 10^{-5} & 8.656\times 10^{-5}
    \end{array}\right)$ \end{tabular} \\
    \hline
    \begin{tabular}{l} PMNS matrix\\(output) \end{tabular} & $\left|V_{\mathrm{PMNS}}\right| \simeq
    \left(\begin{array}{lll}
    0.835 & 0.529 & 0.149 \\
    0.287 & 0.651 & 0.703 \\
    0.469 & 0.545 & 0.695
    \end{array}\right)$
    \\
    \hline
    \begin{tabular}{l} Mixing angles\\(output) \end{tabular}
    & $\left(\begin{array}{lll}
    \theta_{12} & \theta_{13} & \theta_{23}
    \end{array}\right)
    \simeq \left(\begin{array}{lll}
    0.180\pi & 0.048\pi & 0.252\pi
    \end{array}\right)$ \\
    \hline
\end{tabular}
}
\label{tab:lepton_15_23}
\end{table}

\begin{table}[H]
\centering
\scalebox{0.82}{
\begin{tabular}{l|c}
    \hline
    \begin{tabular}{l} Charges \end{tabular} &
    ${\cal Q}=\left(\begin{array}{ccc|ccc|ccc|cc}
        L_{1} & L_{2} & L_{3} & N_{1} & N_{2} & N_{3} & l_{1} & l_{2} & l_{3} & H & \phi \\ 
        \hline
        1 & 1 & 2 & -7 & -9 & 0 & -3 & -7 & -6 & 2 & 1 \\ 
    \end{array}\right)$ \\ 
    \hline
    \begin{tabular}{l} $\mathcal{O}\left(1\right)$ coeff. \end{tabular} & $y^{l} \simeq
    \left(\begin{array}{rrr}
        -1.394 & 1.351 & -1.084 \\
        -0.496 & 0.590 & 1.240 \\
        0.893 & -0.953 & -1.379
    \end{array}\right) \ ,\ 
    y^{\nu} \simeq
    \left(\begin{array}{rrr}
        -1.285 & -0.877 & 0.999 \\
        1.073 & 1.395 & -1.372 \\
        -0.755 & -0.588 & -0.539
    \end{array}\right)$ \\ 
    & $y^{N} \simeq
    \left(\begin{array}{rrr}
        -1.647 & -0.933 & 0.916 \\
    -0.933 & -0.810 & -0.952 \\
    0.916 & -0.952 & -1.052
    \end{array}\right)$ \\ 
    \hline
    \begin{tabular}{l} Intrinsic value \end{tabular} & $\mathcal{V}_{\mathrm{opt}}\simeq-1.572$ \\ 
    \hline
    \begin{tabular}{l} Masses\\(output) \end{tabular} & \begin{tabular}{l} $\left(\begin{array}{lll}
    m_{e}/\MeV & m_{\mu}/\GeV & m_{\tau}/\GeV\\
    m_{\nu_{1}}/\meV & m_{\nu_{2}}/\meV & m_{\nu_{3}}/\meV
    \end{array}\right)
    \simeq \left(\begin{array}{lll}
    0.60598 & 0.298396 & 16.92295 \\
    2.110\times 10^{-4} & 5.616\times 10^{-3} & 9.628\times 10^{-3}
    \end{array}\right)$ \end{tabular} \\
    \hline
    \begin{tabular}{l} PMNS matrix\\(output) \end{tabular} & $\left|V_{\mathrm{PMNS}}\right| \simeq
    \left(\begin{array}{lll}
    0.818 & 0.557 & 0.147 \\
    0.317 & 0.649 & 0.692 \\
    0.480 & 0.519 & 0.707
    \end{array}\right)$
    \\
    \hline
    \begin{tabular}{l} Mixing angles\\(output) \end{tabular}
    & $\left(\begin{array}{lll}
    \theta_{12} & \theta_{13} & \theta_{23}
    \end{array}\right)
    \simeq \left(\begin{array}{lll}
    0.190\pi & 0.047\pi & 0.246\pi
    \end{array}\right)$ \\
    \hline
\end{tabular}
}
\label{tab:lepton_15_19}
\end{table}

\vspace{\stretch{1}}
\newpage
\vspace*{\stretch{1}}

\begin{table}[H]
\centering
\scalebox{0.82}{
\begin{tabular}{l|c}
    \hline
    \begin{tabular}{l} Charges \end{tabular} &
    ${\cal Q}=\left(\begin{array}{ccc|ccc|ccc|cc}
        L_{1} & L_{2} & L_{3} & N_{1} & N_{2} & N_{3} & l_{1} & l_{2} & l_{3} & H & \phi \\ 
        \hline
        7 & 8 & 8 & -2 & -4 & -8 & 2 & -2 & -5 & 2 & 1 \\ 
    \end{array}\right)$ \\ 
    \hline
    \begin{tabular}{l} $\mathcal{O}\left(1\right)$ coeff. \end{tabular} & $y^{l} \simeq
    \left(\begin{array}{rrr}
        0.720 & 1.157 & 0.683 \\
        -1.058 & 1.405 & 0.757 \\
        1.298 & -0.971 & 1.274
    \end{array}\right) \ ,\ 
    y^{\nu} \simeq
    \left(\begin{array}{rrr}
        0.919 & -1.133 & 0.728 \\
        0.890 & 1.450 & 1.382 \\
        1.003 & -1.161 & -1.245
    \end{array}\right)$ \\ 
    & $y^{N} \simeq
    \left(\begin{array}{rrr}
        1.206 & -0.872 & -1.036 \\
        -0.872 & -0.978 & 0.591 \\
        -1.036 & 0.591 & -1.118
    \end{array}\right)$ \\ 
    \hline
    \begin{tabular}{l} Intrinsic value \end{tabular} & $\mathcal{V}_{\mathrm{opt}}\simeq-2.070$ \\ 
    \hline
    \begin{tabular}{l} Masses\\(output) \end{tabular} & \begin{tabular}{l} $\left(\begin{array}{lll}
    m_{e}/\MeV & m_{\mu}/\GeV & m_{\tau}/\GeV\\
    m_{\nu_{1}}/\meV & m_{\nu_{2}}/\meV & m_{\nu_{3}}/\meV
    \end{array}\right)
    \simeq \left(\begin{array}{lll}
    0.06487 & 0.011648 & 2.41311 \\
    8.151\times 10^{-11} & 1.627\times 10^{-10} & 4.678\times 10^{-10}
    \end{array}\right)$ \end{tabular} \\
    \hline
    \begin{tabular}{l} PMNS matrix\\(output) \end{tabular} & $\left|V_{\mathrm{PMNS}}\right| \simeq
    \left(\begin{array}{lll}
    0.806 & 0.573 & 0.147 \\
    0.278 & 0.586 & 0.761 \\
    0.522 & 0.573 & 0.632
    \end{array}\right)$
    \\
    \hline
    \begin{tabular}{l} Mixing angles\\(output) \end{tabular}
    & $\left(\begin{array}{lll}
    \theta_{12} & \theta_{13} & \theta_{23}
    \end{array}\right)
    \simeq \left(\begin{array}{lll}
    0.197\pi & 0.047\pi & 0.279\pi
    \end{array}\right)$ \\
    \hline
\end{tabular}
}
\label{tab:lepton_15_5}
\end{table}

\vspace{\stretch{1}}

\subsection{The case of $10^{16}\,\GeV$}
\label{app:16GeV}

We show benchmarks for the leptons at $M=10^{16}\,\GeV$.
Note that the VEV is the same as Table \ref{tab:benchmark_quark16}.
Following tables are ordered by the intrinsic values.

\vspace{\stretch{1}}

\begin{table}[H]
\centering
\scalebox{0.82}{
\begin{tabular}{l|c}
    \hline
    \begin{tabular}{l} Charges \end{tabular} &
    ${\cal Q}=\left(\begin{array}{ccc|ccc|ccc|cc}
        L_{1} & L_{2} & L_{3} & N_{1} & N_{2} & N_{3} & l_{1} & l_{2} & l_{3} & H & \phi \\ 
        \hline
        -4 & -4 & -4 & 3 & 8 & 1 & 6 & 3 & 2 & -3 & -1 \\ 
    \end{array}\right)$ \\ 
    \hline
    \begin{tabular}{l} $\mathcal{O}\left(1\right)$ coeff. \end{tabular} & $y^{l} \simeq
    \left(\begin{array}{rrr}
        -0.652 & -1.396 & 1.040 \\
        -1.175 & -1.376 & -1.075 \\
        1.171 & -1.295 & -0.925
    \end{array}\right) \ ,\ 
    y^{\nu} \simeq
    \left(\begin{array}{rrr}
        0.815 & 0.553 & -0.837 \\
        0.903 & -1.349 & 1.040 \\
        1.326 & -1.310 & -0.467

    \end{array}\right)$ \\ 
    & $y^{N} \simeq
    \left(\begin{array}{rrr}
        1.068 & -1.147 & 0.874 \\
        -1.147 & 1.003 & 1.487 \\
        0.874 & 1.487 & -0.885

    \end{array}\right)$ \\ 
    \hline
    \begin{tabular}{l} Intrinsic value \end{tabular} & $\mathcal{V}_{\mathrm{opt}}\simeq-0.595$ \\ 
    \hline
    \begin{tabular}{l} Masses\\(output) \end{tabular} & \begin{tabular}{l} $\left(\begin{array}{lll}
    m_{e}/\MeV & m_{\mu}/\GeV & m_{\tau}/\GeV\\
    m_{\nu_{1}}/\meV & m_{\nu_{2}}/\meV & m_{\nu_{3}}/\meV
    \end{array}\right)
    \simeq \left(\begin{array}{lll}
    0.46065 & 0.188688 & 1.00064 \\
    1.724\times 10^{-12} & 9.254\times 10^{-12} & 2.056\times 10^{-11}
    \end{array}\right)$ \end{tabular} \\
    \hline
    \begin{tabular}{l} PMNS matrix\\(output) \end{tabular} & $\left|V_{\mathrm{PMNS}}\right| \simeq
    \left(\begin{array}{lll}
    0.834 & 0.531 & 0.151 \\
    0.456 & 0.508 & 0.731 \\
    0.311 & 0.679 & 0.665
    \end{array}\right)$
    \\
    \hline
    \begin{tabular}{l} Mixing angles\\(output) \end{tabular}
    & $\left(\begin{array}{lll}
    \theta_{12} & \theta_{13} & \theta_{23}
    \end{array}\right)
    \simeq \left(\begin{array}{lll}
    0.180\pi & 0.048\pi & 0.265\pi
    \end{array}\right)$ \\
    \hline
\end{tabular}
}
\end{table}

\vspace{\stretch{1}}
\newpage
\vspace*{\stretch{1}}

\begin{table}[H]
\centering
\scalebox{0.82}{
\begin{tabular}{l|c}
    \hline
    \begin{tabular}{l} Charges \end{tabular} &
    ${\cal Q}=\left(\begin{array}{ccc|ccc|ccc|cc}
        L_{1} & L_{2} & L_{3} & N_{1} & N_{2} & N_{3} & l_{1} & l_{2} & l_{3} & H & \phi \\ 
        \hline
        -7 & -7 & -8 & 0 & 8 & 7 & -2 & 1 & -1 & -3 & -1 \\ 
    \end{array}\right)$ \\ 
    \hline
    \begin{tabular}{l} $\mathcal{O}\left(1\right)$ coeff. \end{tabular} & $y^{l} \simeq
    \left(\begin{array}{rrr}
        0.914 & 1.041 & 1.494 \\
        0.895 & -0.960 & 1.102 \\
        -0.626 & 1.497 & -0.701
    \end{array}\right) \ ,\ 
    y^{\nu} \simeq
    \left(\begin{array}{rrr}
        1.006 & -0.919 & 0.684 \\
        -0.766 & -0.547 & -0.671 \\
        1.029 & 1.115 & -1.081

    \end{array}\right)$ \\ 
    & $y^{N} \simeq
    \left(\begin{array}{rrr}
        1.238 & -0.823 & -0.964 \\
        -0.823 & 1.372 & 1.099 \\
        -0.964 & 1.099 & -1.332

    \end{array}\right)$ \\ 
    \hline
    \begin{tabular}{l} Intrinsic value \end{tabular} & $\mathcal{V}_{\mathrm{opt}}\simeq-0.852$ \\ 
    \hline
    \begin{tabular}{l} Masses\\(output) \end{tabular} & \begin{tabular}{l} $\left(\begin{array}{lll}
    m_{e}/\MeV & m_{\mu}/\GeV & m_{\tau}/\GeV\\
    m_{\nu_{1}}/\meV & m_{\nu_{2}}/\meV & m_{\nu_{3}}/\meV
    \end{array}\right)
    \simeq \left(\begin{array}{lll}
    0.80136 & 0.145047 & 5.01901 \\
    7.459\times 10^{-18} & 2.919\times 10^{-17} & 1.434\times 10^{-16}
    \end{array}\right)$ \end{tabular} \\
    \hline
    \begin{tabular}{l} PMNS matrix\\(output) \end{tabular} & $\left|V_{\mathrm{PMNS}}\right| \simeq
    \left(\begin{array}{lll}
    0.803 & 0.578 & 0.146 \\
    0.348 & 0.653 & 0.673 \\
    0.484 & 0.489 & 0.725
    \end{array}\right)$
    \\
    \hline
    \begin{tabular}{l} Mixing angles\\(output) \end{tabular}
    & $\left(\begin{array}{lll}
    \theta_{12} & \theta_{13} & \theta_{23}
    \end{array}\right)
    \simeq \left(\begin{array}{lll}
    0.199\pi & 0.047\pi & 0.238\pi
    \end{array}\right)$ \\
    \hline
\end{tabular}
}
\label{tab:lepton_16_2}
\end{table}

\begin{table}[H]
\centering
\scalebox{0.82}{
\begin{tabular}{l|c}
    \hline
    \begin{tabular}{l} Charges \end{tabular} &
    ${\cal Q}=\left(\begin{array}{ccc|ccc|ccc|cc}
        L_{1} & L_{2} & L_{3} & N_{1} & N_{2} & N_{3} & l_{1} & l_{2} & l_{3} & H & \phi \\ 
        \hline
        -4 & -4 & -4 & 3 & 1 & 7 & 5 & 1 & 2 & -3 & -1 \\ 
    \end{array}\right)$ \\ 
    \hline
    \begin{tabular}{l} $\mathcal{O}\left(1\right)$ coeff. \end{tabular} & $y^{l} \simeq
    \left(\begin{array}{rrr}
        -0.852 & -0.887 & 1.019 \\
        -1.008 & -1.517 & 1.229 \\
        -1.024 & -1.058 & 0.705
    \end{array}\right) \ ,\ 
    y^{\nu} \simeq
    \left(\begin{array}{rrr}
        -1.082 & -0.531 & -0.937 \\
        -1.280 & 0.913 & 0.857 \\
        -1.075 & 0.843 & 1.381

    \end{array}\right)$ \\ 
    & $y^{N} \simeq
    \left(\begin{array}{rrr}
        0.961 & 1.197 & 0.678 \\
        1.197 & 0.721 & 0.963 \\
        0.678 & 0.963 & 1.007

    \end{array}\right)$ \\ 
    \hline
    \begin{tabular}{l} Intrinsic value \end{tabular} & $\mathcal{V}_{\mathrm{opt}}\simeq-1.025$ \\ 
    \hline
    \begin{tabular}{l} Masses\\(output) \end{tabular} & \begin{tabular}{l} $\left(\begin{array}{lll}
    m_{e}/\MeV & m_{\mu}/\GeV & m_{\tau}/\GeV\\
    m_{\nu_{1}}/\meV & m_{\nu_{2}}/\meV & m_{\nu_{3}}/\meV
    \end{array}\right)
    \simeq \left(\begin{array}{lll}
    0.56125 & 0.187528 & 7.91476 \\
    2.236\times 10^{-12} & 6.171\times 10^{-12} & 6.840\times 10^{-11}
    \end{array}\right)$ \end{tabular} \\
    \hline
    \begin{tabular}{l} PMNS matrix\\(output) \end{tabular} & $\left|V_{\mathrm{PMNS}}\right| \simeq
    \left(\begin{array}{lll}
    0.820 & 0.554 & 0.143 \\
    0.487 & 0.545 & 0.683 \\
    0.300 & 0.630 & 0.716
    \end{array}\right)$
    \\
    \hline
    \begin{tabular}{l} Mixing angles\\(output) \end{tabular}
    & $\left(\begin{array}{lll}
    \theta_{12} & \theta_{13} & \theta_{23}
    \end{array}\right)
    \simeq \left(\begin{array}{lll}
    0.189\pi & 0.046\pi & 0.242\pi
    \end{array}\right)$ \\
    \hline
\end{tabular}
}
\label{tab:lepton_16_3}
\end{table}

\vspace{\stretch{1}}
\newpage
\vspace*{\stretch{1}}

\begin{table}[H]
\centering
\scalebox{0.82}{
\begin{tabular}{l|c}
    \hline
    \begin{tabular}{l} Charges \end{tabular} &
    ${\cal Q}=\left(\begin{array}{ccc|ccc|ccc|cc}
        L_{1} & L_{2} & L_{3} & N_{1} & N_{2} & N_{3} & l_{1} & l_{2} & l_{3} & H & \phi \\ 
        \hline
        -6 & -6 & -6 & 0 & 2 & 7 & 5 & 1 & 0 & -3 & -1 \\ 
    \end{array}\right)$ \\ 
    \hline
    \begin{tabular}{l} $\mathcal{O}\left(1\right)$ coeff. \end{tabular} & $y^{l} \simeq
    \left(\begin{array}{rrr}
        -0.904 & 0.448 & -0.970 \\
        1.449 & 0.821 & 0.675 \\
        0.780 & -0.970 & -1.099
    \end{array}\right) \ ,\ 
    y^{\nu} \simeq
    \left(\begin{array}{rrr}
        -1.044 & -0.849 & -0.503 \\
        0.940 & -1.495 & -0.779 \\
        -0.970 & 1.413 & 0.590

    \end{array}\right)$ \\ 
    & $y^{N} \simeq
    \left(\begin{array}{rrr}
        1.232 & -0.812 & 0.917 \\
        -0.812 & 0.809 & 1.028 \\
        0.917 & 1.028 & 0.990

    \end{array}\right)$ \\ 
    \hline
    \begin{tabular}{l} Intrinsic value \end{tabular} & $\mathcal{V}_{\mathrm{opt}}\simeq-1.124$ \\ 
    \hline
    \begin{tabular}{l} Masses\\(output) \end{tabular} & \begin{tabular}{l} $\left(\begin{array}{lll}
    m_{e}/\MeV & m_{\mu}/\GeV & m_{\tau}/\GeV\\
    m_{\nu_{1}}/\meV & m_{\nu_{2}}/\meV & m_{\nu_{3}}/\meV
    \end{array}\right)
    \simeq \left(\begin{array}{lll}
    0.07069 & 0.094949 & 0.91865 \\
    3.144\times 10^{-16} & 3.918\times 10^{-16} & 1.417\times 10^{-14}
    \end{array}\right)$ \end{tabular} \\
    \hline
    \begin{tabular}{l} PMNS matrix\\(output) \end{tabular} & $\left|V_{\mathrm{PMNS}}\right| \simeq
    \left(\begin{array}{lll}
    0.815 & 0.561 & 0.147 \\
    0.489 & 0.528 & 0.694 \\
    0.311 & 0.637 & 0.705
    \end{array}\right)$
    \\
    \hline
    \begin{tabular}{l} Mixing angles\\(output) \end{tabular}
    & $\left(\begin{array}{lll}
    \theta_{12} & \theta_{13} & \theta_{23}
    \end{array}\right)
    \simeq \left(\begin{array}{lll}
    0.192\pi & 0.047\pi & 0.247\pi
    \end{array}\right)$ \\
    \hline
\end{tabular}
}
\label{tab:lepton_16_4}
\end{table}

\begin{table}[H]
\centering
\scalebox{0.82}{
\begin{tabular}{l|c}
    \hline
    \begin{tabular}{l} Charges \end{tabular} &
    ${\cal Q}=\left(\begin{array}{ccc|ccc|ccc|cc}
        L_{1} & L_{2} & L_{3} & N_{1} & N_{2} & N_{3} & l_{1} & l_{2} & l_{3} & H & \phi \\ 
        \hline
        -4 & -4 & -4 & 9 & 9 & 8 & 4 & 1 & 2 & -3 & -1 \\ 
    \end{array}\right)$ \\ 
    \hline
    \begin{tabular}{l} $\mathcal{O}\left(1\right)$ coeff. \end{tabular} & $y^{l} \simeq
    \left(\begin{array}{rrr}
        1.309 & -0.891 & -1.023 \\
        0.774 & -1.253 & -1.132 \\
        -0.956 & -1.078 & -0.395
    \end{array}\right) \ ,\ 
    y^{\nu} \simeq
    \left(\begin{array}{rrr}
        0.843 & -0.959 & -0.725 \\
        0.890 & -0.690 & 1.089 \\
        1.230 & 1.485 & 0.862

    \end{array}\right)$ \\ 
    & $y^{N} \simeq
    \left(\begin{array}{rrr}
        1.069 & 1.671 & -0.779 \\
        1.671 & 1.767 & -0.738 \\
        -0.779 & -0.738 & -0.553

    \end{array}\right)$ \\ 
    \hline
    \begin{tabular}{l} Intrinsic value \end{tabular} & $\mathcal{V}_{\mathrm{opt}}\simeq-1.560$ \\ 
    \hline
    \begin{tabular}{l} Masses\\(output) \end{tabular} & \begin{tabular}{l} $\left(\begin{array}{lll}
    m_{e}/\MeV & m_{\mu}/\GeV & m_{\tau}/\GeV\\
    m_{\nu_{1}}/\meV & m_{\nu_{2}}/\meV & m_{\nu_{3}}/\meV
    \end{array}\right)
    \simeq \left(\begin{array}{lll}
    1.26735 & 0.325209 & 7.23693 \\
    3.831\times 10^{-12} & 2.887\times 10^{-11} & 8.116\times 10^{-11}
    \end{array}\right)$ \end{tabular} \\
    \hline
    \begin{tabular}{l} PMNS matrix\\(output) \end{tabular} & $\left|V_{\mathrm{PMNS}}\right| \simeq
    \left(\begin{array}{lll}
    0.825 & 0.544 & 0.154 \\
    0.468 & 0.504 & 0.726 \\
    0.317 & 0.670 & 0.671
    \end{array}\right)$
    \\
    \hline
    \begin{tabular}{l} Mixing angles\\(output) \end{tabular}
    & $\left(\begin{array}{lll}
    \theta_{12} & \theta_{13} & \theta_{23}
    \end{array}\right)
    \simeq \left(\begin{array}{lll}
    0.186\pi & 0.049\pi & 0.263\pi
    \end{array}\right)$ \\
    \hline
\end{tabular}
}
\label{tab:lepton_16_5}
\end{table}

\vspace{\stretch{1}}
\newpage
\vspace*{\stretch{1}}

\begin{table}[H]
\centering
\scalebox{0.82}{
\begin{tabular}{l|c}
    \hline
    \begin{tabular}{l} Charges \end{tabular} &
    ${\cal Q}=\left(\begin{array}{ccc|ccc|ccc|cc}
        L_{1} & L_{2} & L_{3} & N_{1} & N_{2} & N_{3} & l_{1} & l_{2} & l_{3} & H & \phi \\ 
        \hline
        -4 & -4 & -4 & 0 & 6 & 0 & 6 & 4 & 3 & -3 & -1 \\ 
    \end{array}\right)$ \\ 
    \hline
    \begin{tabular}{l} $\mathcal{O}\left(1\right)$ coeff. \end{tabular} & $y^{l} \simeq
    \left(\begin{array}{rrr}
        1.051 & -0.993 & 1.379 \\
        0.842 & 0.924 & -1.288 \\
        -1.070 & 1.056 & 1.029
    \end{array}\right) \ ,\ 
    y^{\nu} \simeq
    \left(\begin{array}{rrr}
        -1.075 & 0.901 & 0.903 \\
        1.154 & -1.072 & -1.325 \\
        -1.228 & -1.251 & -0.825

    \end{array}\right)$ \\ 
    & $y^{N} \simeq
    \left(\begin{array}{rrr}
        -1.126 & 1.229 & -0.679 \\
        1.229 & 0.914 & -0.885 \\
        -0.679 & -0.885 & -1.182

    \end{array}\right)$ \\ 
    \hline
    \begin{tabular}{l} Intrinsic value \end{tabular} & $\mathcal{V}_{\mathrm{opt}}\simeq-1.993$ \\ 
    \hline
    \begin{tabular}{l} Masses\\(output) \end{tabular} & \begin{tabular}{l} $\left(\begin{array}{lll}
    m_{e}/\MeV & m_{\mu}/\GeV & m_{\tau}/\GeV\\
    m_{\nu_{1}}/\meV & m_{\nu_{2}}/\meV & m_{\nu_{3}}/\meV
    \end{array}\right)
    \simeq \left(\begin{array}{lll}
    0.36609 & 0.019676 & 0.18129 \\
    6.045\times 10^{-13} & 1.651\times 10^{-12} & 3.870\times 10^{-11}
    \end{array}\right)$ \end{tabular} \\
    \hline
    \begin{tabular}{l} PMNS matrix\\(output) \end{tabular} & $\left|V_{\mathrm{PMNS}}\right| \simeq
    \left(\begin{array}{lll}
    0.831 & 0.537 & 0.146 \\
    0.495 & 0.593 & 0.635 \\
    0.255 & 0.600 & 0.758
    \end{array}\right)$
    \\
    \hline
    \begin{tabular}{l} Mixing angles\\(output) \end{tabular}
    & $\left(\begin{array}{lll}
    \theta_{12} & \theta_{13} & \theta_{23}
    \end{array}\right)
    \simeq \left(\begin{array}{lll}
    0.183\pi & 0.047\pi & 0.222\pi
    \end{array}\right)$ \\
    \hline
\end{tabular}
}
\label{tab:lepton_16_6}
\end{table}

\begin{table}[H]
\centering
\scalebox{0.82}{
\begin{tabular}{l|c}
    \hline
    \begin{tabular}{l} Charges \end{tabular} &
    ${\cal Q}=\left(\begin{array}{ccc|ccc|ccc|cc}
        L_{1} & L_{2} & L_{3} & N_{1} & N_{2} & N_{3} & l_{1} & l_{2} & l_{3} & H & \phi \\ 
        \hline
        -7 & -7 & -7 & 4 & 9 & 5 & 4 & 1 & 0 & -3 & -1 \\ 
    \end{array}\right)$ \\ 
    \hline
    \begin{tabular}{l} $\mathcal{O}\left(1\right)$ coeff. \end{tabular} & $y^{l} \simeq
    \left(\begin{array}{rrr}
        -0.893 & 1.493 & -0.743 \\
        -0.833 & -1.364 & -1.453 \\
        -1.026 & 1.150 & 1.274
    \end{array}\right) \ ,\ 
    y^{\nu} \simeq
    \left(\begin{array}{rrr}
        -1.235 & -0.730 & -1.313 \\
        1.024 & 0.990 & -1.357 \\
        -1.271 & -1.372 & 1.223

    \end{array}\right)$ \\ 
    & $y^{N} \simeq
    \left(\begin{array}{rrr}
        -1.364 & -1.711 & 0.874 \\
        -1.711 & -0.857 & -1.227 \\
        0.874 & -1.227 & -1.230

    \end{array}\right)$ \\ 
    \hline
    \begin{tabular}{l} Intrinsic value \end{tabular} & $\mathcal{V}_{\mathrm{opt}}\simeq-2.605$ \\ 
    \hline
    \begin{tabular}{l} Masses\\(output) \end{tabular} & \begin{tabular}{l} $\left(\begin{array}{lll}
    m_{e}/\MeV & m_{\mu}/\GeV & m_{\tau}/\GeV\\
    m_{\nu_{1}}/\meV & m_{\nu_{2}}/\meV & m_{\nu_{3}}/\meV
    \end{array}\right)
    \simeq \left(\begin{array}{lll}
    0.05448 & 0.025337 & 0.17501 \\
    5.824\times 10^{-18} & 1.987\times 10^{-17} & 2.256\times 10^{-16}
    \end{array}\right)$ \end{tabular} \\
    \hline
    \begin{tabular}{l} PMNS matrix\\(output) \end{tabular} & $\left|V_{\mathrm{PMNS}}\right| \simeq
    \left(\begin{array}{lll}
    0.841 & 0.519 & 0.151 \\
    0.451 & 0.520 & 0.726 \\
    0.298 & 0.678 & 0.671
    \end{array}\right)$
    \\
    \hline
    \begin{tabular}{l} Mixing angles\\(output) \end{tabular}
    & $\left(\begin{array}{lll}
    \theta_{12} & \theta_{13} & \theta_{23}
    \end{array}\right)
    \simeq \left(\begin{array}{lll}
    0.176\pi & 0.048\pi & 0.262\pi
    \end{array}\right)$ \\
    \hline
\end{tabular}
}
\label{tab:lepton_16_7}
\end{table}

\vspace{\stretch{1}}
\newpage

\bibliography{ref}{}

\providecommand{\href}[2]{#2}\begingroup\raggedright\begin{thebibliography}{10}

\bibitem{Ema:2016ops}
Y.~Ema, K.~Hamaguchi, T.~Moroi and K.~Nakayama, \emph{{Flaxion: a minimal extension to solve puzzles in the standard model}}, \href{https://doi.org/10.1007/JHEP01(2017)096}{\emph{JHEP} {\bfseries 01} (2017) 096} [\href{https://arxiv.org/abs/1612.05492}{{\ttfamily 1612.05492}}].

\bibitem{Calibbi:2016hwq}
L.~Calibbi, F.~Goertz, D.~Redigolo, R.~Ziegler and J.~Zupan, \emph{{Minimal axion model from flavor}}, \href{https://doi.org/10.1103/PhysRevD.95.095009}{\emph{Phys. Rev. D} {\bfseries 95} (2017) 095009} [\href{https://arxiv.org/abs/1612.08040}{{\ttfamily 1612.08040}}].

\bibitem{Davidson:1981zd}
A.~Davidson and K.C.~Wali, \emph{{MINIMAL FLAVOR UNIFICATION VIA MULTIGENERATIONAL PECCEI-QUINN SYMMETRY}}, \href{https://doi.org/10.1103/PhysRevLett.48.11}{\emph{Phys. Rev. Lett.} {\bfseries 48} (1982) 11}.

\bibitem{Wilczek:1982rv}
F.~Wilczek, \emph{{Axions and Family Symmetry Breaking}}, \href{https://doi.org/10.1103/PhysRevLett.49.1549}{\emph{Phys. Rev. Lett.} {\bfseries 49} (1982) 1549}.

\bibitem{Froggatt:1978nt}
C.D.~Froggatt and H.B.~Nielsen, \emph{{Hierarchy of Quark Masses, Cabibbo Angles and CP Violation}}, \href{https://doi.org/10.1016/0550-3213(79)90316-X}{\emph{Nucl. Phys. B} {\bfseries 147} (1979) 277}.

\bibitem{PhysRevLett.38.1440}
R.D.~Peccei and H.R.~Quinn, \emph{$\mathrm{CP}$ conservation in the presence of pseudoparticles}, \href{https://doi.org/10.1103/PhysRevLett.38.1440}{\emph{Phys. Rev. Lett.} {\bfseries 38} (1977) 1440}.

\bibitem{PhysRevD.16.1791}
R.D.~Peccei and H.R.~Quinn, \emph{Constraints imposed by $\mathrm{CP}$ conservation in the presence of pseudoparticles}, \href{https://doi.org/10.1103/PhysRevD.16.1791}{\emph{Phys. Rev. D} {\bfseries 16} (1977) 1791}.

\bibitem{Nishimura:2023nre}
S.~Nishimura, C.~Miyao and H.~Otsuka, \emph{{Exploring the flavor structure of quarks and leptons with reinforcement learning}}, \href{https://doi.org/10.1007/JHEP12(2023)021}{\emph{JHEP} {\bfseries 12} (2023) 021} [\href{https://arxiv.org/abs/2304.14176}{{\ttfamily 2304.14176}}].

\bibitem{DMRadio:2022pkf}
{\scshape DMRadio} collaboration, \emph{{Projected sensitivity of DMRadio-m3: A search for the QCD axion below 1\,\,\ensuremath{\mu}eV}}, \href{https://doi.org/10.1103/PhysRevD.106.103008}{\emph{Phys. Rev. D} {\bfseries 106} (2022) 103008} [\href{https://arxiv.org/abs/2204.13781}{{\ttfamily 2204.13781}}].

\bibitem{Preskill:1982cy}
J.~Preskill, M.B.~Wise and F.~Wilczek, \emph{{Cosmology of the Invisible Axion}}, \href{https://doi.org/10.1016/0370-2693(83)90637-8}{\emph{Phys. Lett. B} {\bfseries 120} (1983) 127}.

\bibitem{Abbott:1982af}
L.F.~Abbott and P.~Sikivie, \emph{{A Cosmological Bound on the Invisible Axion}}, \href{https://doi.org/10.1016/0370-2693(83)90638-X}{\emph{Phys. Lett. B} {\bfseries 120} (1983) 133}.

\bibitem{Dine:1982ah}
M.~Dine and W.~Fischler, \emph{{The Not So Harmless Axion}}, \href{https://doi.org/10.1016/0370-2693(83)90639-1}{\emph{Phys. Lett. B} {\bfseries 120} (1983) 137}.

\bibitem{Minkowski:1977sc}
P.~Minkowski, \emph{{$\mu \to e\gamma$ at a Rate of One Out of $10^{9}$ Muon Decays?}}, \href{https://doi.org/10.1016/0370-2693(77)90435-X}{\emph{Phys. Lett. B} {\bfseries 67} (1977) 421}.

\bibitem{Yanagida:1979as}
T.~Yanagida, \emph{{Horizontal gauge symmetry and masses of neutrinos}}, {\emph{Conf. Proc. C} {\bfseries 7902131} (1979) 95}.

\bibitem{Gell-Mann:1979vob}
M.~Gell-Mann, P.~Ramond and R.~Slansky, \emph{{Complex Spinors and Unified Theories}}, {\emph{Conf. Proc. C} {\bfseries 790927} (1979) 315} [\href{https://arxiv.org/abs/1306.4669}{{\ttfamily 1306.4669}}].

\bibitem{Mohapatra:1979ia}
R.N.~Mohapatra and G.~Senjanovic, \emph{{Neutrino Mass and Spontaneous Parity Nonconservation}}, \href{https://doi.org/10.1103/PhysRevLett.44.912}{\emph{Phys. Rev. Lett.} {\bfseries 44} (1980) 912}.

\bibitem{KIM19871}
J.E.~Kim, \emph{Light pseudoscalars, particle physics and cosmology}, \href{https://doi.org/https://doi.org/10.1016/0370-1573(87)90017-2}{\emph{Physics Reports} {\bfseries 150} (1987) 1}.

\bibitem{Hiramatsu:2012gg}
T.~Hiramatsu, M.~Kawasaki, K.~Saikawa and T.~Sekiguchi, \emph{{Production of dark matter axions from collapse of string-wall systems}}, \href{https://doi.org/10.1103/PhysRevD.85.105020}{\emph{Phys. Rev. D} {\bfseries 85} (2012) 105020} [\href{https://arxiv.org/abs/1202.5851}{{\ttfamily 1202.5851}}].

\bibitem{Ibe:2019yew}
M.~Ibe, S.~Kobayashi, M.~Suzuki and T.T.~Yanagida, \emph{{Dynamical solution to the axion domain wall problem}}, \href{https://doi.org/10.1103/PhysRevD.101.035029}{\emph{Phys. Rev. D} {\bfseries 101} (2020) 035029} [\href{https://arxiv.org/abs/1909.01604}{{\ttfamily 1909.01604}}].

\bibitem{NA62:2020pwi}
{\scshape NA62} collaboration, \emph{{Search for $\pi^0$ decays to invisible particles}}, \href{https://doi.org/10.1007/JHEP02(2021)201}{\emph{JHEP} {\bfseries 02} (2021) 201} [\href{https://arxiv.org/abs/2010.07644}{{\ttfamily 2010.07644}}].

\bibitem{NA62:2021zjw}
{\scshape NA62} collaboration, \emph{{Measurement of the very rare K$^{+}$\textrightarrow{}$ {\pi}^{+}\nu \overline{\nu} $ decay}}, \href{https://doi.org/10.1007/JHEP06(2021)093}{\emph{JHEP} {\bfseries 06} (2021) 093} [\href{https://arxiv.org/abs/2103.15389}{{\ttfamily 2103.15389}}].

\bibitem{PhysRevD.33.889}
M.S.~Turner, \emph{Cosmic and local mass density of ``invisible'' axions}, \href{https://doi.org/10.1103/PhysRevD.33.889}{\emph{Phys. Rev. D} {\bfseries 33} (1986) 889}.

\bibitem{Planck:2018jri}
{\scshape Planck} collaboration, \emph{{Planck 2018 results. X. Constraints on inflation}}, \href{https://doi.org/10.1051/0004-6361/201833887}{\emph{Astron. Astrophys.} {\bfseries 641} (2020) A10} [\href{https://arxiv.org/abs/1807.06211}{{\ttfamily 1807.06211}}].

\bibitem{Kofman:1995fi}
L.~Kofman, A.D.~Linde and A.A.~Starobinsky, \emph{{Nonthermal phase transitions after inflation}}, \href{https://doi.org/10.1103/PhysRevLett.76.1011}{\emph{Phys. Rev. Lett.} {\bfseries 76} (1996) 1011} [\href{https://arxiv.org/abs/hep-th/9510119}{{\ttfamily hep-th/9510119}}].

\bibitem{RL}
R.S.~Sutton and A.G.~Barto, \emph{Reinforcement learning: An introduction}, MIT press (2018).

\bibitem{DBLP:journals/corr/AbadiABBCCCDDDG16}
M.~Abadi, A.~Agarwal, P.~Barham, E.~Brevdo, Z.~Chen, C.~Citro et~al., \emph{Tensorflow: Large-scale machine learning on heterogeneous distributed systems}, {\emph{CoRR} {\bfseries abs/1603.04467} (2016) } [\href{https://arxiv.org/abs/1603.04467}{{\ttfamily 1603.04467}}].

\bibitem{Bergstra_2015}
J.~Bergstra, B.~Komer, C.~Eliasmith, D.~Yamins and D.D.~Cox, \emph{Hyperopt: a python library for model selection and hyperparameter optimization}, \href{https://doi.org/10.1088/1749-4699/8/1/014008}{\emph{Computational Science \& Discovery} {\bfseries 8} (2015) 014008}.

\bibitem{10.1145/3292500.3330701}
T.~Akiba, S.~Sano, T.~Yanase, T.~Ohta and M.~Koyama, \emph{Optuna: A next-generation hyperparameter optimization framework},  in \emph{Proceedings of the 25th ACM SIGKDD International Conference on Knowledge Discovery \& Data Mining}, KDD '19, (New York, NY, USA), p.~2623–2631, Association for Computing Machinery, 2019, \href{https://doi.org/10.1145/3292500.3330701}{DOI}.

\bibitem{Huang:2020hdv}
G.-y.~Huang and S.~Zhou, \emph{{Precise Values of Running Quark and Lepton Masses in the Standard Model}}, \href{https://doi.org/10.1103/PhysRevD.103.016010}{\emph{Phys. Rev. D} {\bfseries 103} (2021) 016010} [\href{https://arxiv.org/abs/2009.04851}{{\ttfamily 2009.04851}}].

\bibitem{ParticleDataGroup:2024prd}
{\scshape Particle Data Group} collaboration, \emph{{Review of Particle Physics}}, {\emph{to be published in Phys. Rev. D} {\bfseries 110} (2024) 030001}.

\bibitem{ParticleDataGroup:2022pth}
{\scshape Particle Data Group} collaboration, \emph{{Review of Particle Physics}}, \href{https://doi.org/10.1093/ptep/ptac097}{\emph{PTEP} {\bfseries 2022} (2022) 083C01}.

\bibitem{levine2018reinforcementlearningcontrolprobabilistic}
S.~Levine, \emph{Reinforcement learning and control as probabilistic inference: Tutorial and review},  2018.

\bibitem{10.5555/3104482.3104541}
M.P.~Deisenroth and C.E.~Rasmussen, \emph{Pilco: a model-based and data-efficient approach to policy search},  in \emph{Proceedings of the 28th International Conference on International Conference on Machine Learning}, ICML'11, (Madison, WI, USA), p.~465–472, Omnipress, 2011.

\bibitem{Brieden:2022lsd}
S.~Brieden, H.~Gil-Mar\'\i{}n and L.~Verde, \emph{{Model-agnostic interpretation of 10 billion years of cosmic evolution traced by BOSS and eBOSS data}}, \href{https://doi.org/10.1088/1475-7516/2022/08/024}{\emph{JCAP} {\bfseries 08} (2022) 024} [\href{https://arxiv.org/abs/2204.11868}{{\ttfamily 2204.11868}}].

\bibitem{Esteban:2020cvm}
I.~Esteban, M.C.~Gonzalez-Garcia, M.~Maltoni, T.~Schwetz and A.~Zhou, \emph{{The fate of hints: updated global analysis of three-flavor neutrino oscillations}}, \href{https://doi.org/10.1007/JHEP09(2020)178}{\emph{JHEP} {\bfseries 09} (2020) 178} [\href{https://arxiv.org/abs/2007.14792}{{\ttfamily 2007.14792}}].

\bibitem{PhysRevLett.104.041301}
S.J.~Asztalos, G.~Carosi, C.~Hagmann, D.~Kinion, K.~van Bibber, M.~Hotz et~al., \emph{Squid-based microwave cavity search for dark-matter axions}, \href{https://doi.org/10.1103/PhysRevLett.104.041301}{\emph{Phys. Rev. Lett.} {\bfseries 104} (2010) 041301}.

\bibitem{ADMX:2018ogs}
{\scshape ADMX} collaboration, \emph{{Piezoelectrically Tuned Multimode Cavity Search for Axion Dark Matter}}, \href{https://doi.org/10.1103/PhysRevLett.121.261302}{\emph{Phys. Rev. Lett.} {\bfseries 121} (2018) 261302} [\href{https://arxiv.org/abs/1901.00920}{{\ttfamily 1901.00920}}].

\bibitem{ADMX:2019uok}
{\scshape ADMX} collaboration, \emph{{Extended Search for the Invisible Axion with the Axion Dark Matter Experiment}}, \href{https://doi.org/10.1103/PhysRevLett.124.101303}{\emph{Phys. Rev. Lett.} {\bfseries 124} (2020) 101303} [\href{https://arxiv.org/abs/1910.08638}{{\ttfamily 1910.08638}}].

\bibitem{Salemi:2021gck}
C.P.~Salemi et~al., \emph{{Search for Low-Mass Axion Dark Matter with ABRACADABRA-10~cm}}, \href{https://doi.org/10.1103/PhysRevLett.127.081801}{\emph{Phys. Rev. Lett.} {\bfseries 127} (2021) 081801} [\href{https://arxiv.org/abs/2102.06722}{{\ttfamily 2102.06722}}].

\bibitem{PhysRevLett.127.261803}
{\scshape ADMX Collaboration} collaboration, \emph{Search for invisible axion dark matter in the $3.3--4.2\text{ }\text{ }\ensuremath{\mu}\mathrm{eV}$ mass range}, \href{https://doi.org/10.1103/PhysRevLett.127.261803}{\emph{Phys. Rev. Lett.} {\bfseries 127} (2021) 261803}.

\bibitem{Noordhuis:2022ljw}
D.~Noordhuis, A.~Prabhu, S.J.~Witte, A.Y.~Chen, F.~Cruz and C.~Weniger, \emph{{Novel Constraints on Axions Produced in Pulsar Polar-Cap Cascades}}, \href{https://doi.org/10.1103/PhysRevLett.131.111004}{\emph{Phys. Rev. Lett.} {\bfseries 131} (2023) 111004} [\href{https://arxiv.org/abs/2209.09917}{{\ttfamily 2209.09917}}].

\bibitem{Benabou:2025jcv}
J.N.~Benabou, C.~Dessert, K.C.~Patra, T.G.~Brink, W.~Zheng, A.V.~Filippenko et~al., \emph{{Search for Axions in Magnetic White Dwarf Polarization at Lick and Keck Observatories}},  \href{https://arxiv.org/abs/2504.12377}{{\ttfamily 2504.12377}}.

\bibitem{AxionLimits}
C.~O'Hare, ``cajohare/axionlimits: Axionlimits.'' \url{https://cajohare.github.io/AxionLimits/}, July, 2020.
\newblock 10.5281/zenodo.3932430.

\bibitem{Harvey:2021oue}
T.R.~Harvey and A.~Lukas, \emph{{Quark Mass Models and Reinforcement Learning}}, \href{https://doi.org/10.1007/JHEP08(2021)161}{\emph{JHEP} {\bfseries 08} (2021) 161} [\href{https://arxiv.org/abs/2103.04759}{{\ttfamily 2103.04759}}].

\end{thebibliography}\endgroup
\bibliographystyle{JHEP} 

\end{document}